\input psfig.tex
\def\ssp{\def\baselinestretch{1.0}\large\normalsize}
\documentclass[preprint]{aastex}
\slugcomment{}

\shortauthors{Matheson et al.}
\shorttitle{Type Ib/c Supernovae}

\begin{document}


\title{Optical Spectroscopy of Type Ib/c Supernovae}

\author{Thomas Matheson\footnote{Present Address: Harvard-Smithsonian
Center for Astrophysics, 60 Garden Street, Cambridge, MA 02138; tmatheson@cfa.harvard.edu}, Alexei V. Filippenko, Weidong Li, and Douglas
C. Leonard}
\affil{Department of Astronomy, University of California, Berkeley, CA
94720-3411}
\email{matheson, alex, weidong, leonard@astron.berkeley.edu}

\and

\author{Joseph C. Shields}
\affil{Department of Physics and Astronomy, Ohio University, Athens,
OH 45701}
\email{shields@phy.ohiou.edu}


\begin{abstract}

We present 84 spectra of Type Ib/c and Type IIb supernovae (SNe),
describing the individual SNe in detail.  The relative depths of the
helium absorption lines in the spectra of the SNe Ib appear to provide
a measurement of the temporal evolution of the SN, with \ion{He}{1}
$\lambda$5876 and \ion{He}{1} $\lambda$7065 growing in strength
relative to \ion{He}{1} $\lambda$6678 over time.  Light curves for
three of the SNe Ib provide a sequence for correlating the helium-line
strengths.  We find that some SNe Ic show evidence for weak helium
absorption, but most do not.  Aside from the presence or absence of
the helium lines, there are other spectroscopic differences between
SNe Ib and SNe Ic.  On average, the \ion{O}{1} $\lambda$7774 line is
stronger in SNe Ic than in SNe Ib.  In addition, the SNe Ic have
distinctly broader emission lines at late times, indicating either a
consistently larger explosion energy and/or lower envelope mass for
SNe Ic than for SNe Ib.  While SNe Ib appear to be basically
homogeneous, the SNe Ic are quite heterogeneous in their spectroscopic
characteristics.  Three SNe Ic that may have been associated with
gamma-ray bursts are also discussed; two of these have clearly
peculiar spectra, while the third seems fairly typical.
 
\end{abstract}


\keywords{gamma rays: bursts---stars: mass-loss---supernovae:
general}

\section{Introduction}

Since Minkowski (1941) first differentiated supernovae (SNe) into two
classes based on the absence (Type I) or presence (Type II) of
hydrogen lines in their spectra, SNe have traditionally been
classified by their spectroscopic characteristics near maximum
brightness.  The hydrogen-poor Type I SNe appeared initially to be a
very homogeneous class, but some subtle differences were gradually
noticed.  The Type I class was thus split into subtypes.  The most
distinctive feature of the prototypical Type I SNe---now referred to
as Type Ia SNe (or SNe Ia)---is a strong absorption line near 6150
\AA\ that is the result of blueshifted \ion{Si}{2}
$\lambda\lambda$6347, 6371 (often referred to jointly as
$\lambda$6355).  The other Type I subtypes lack this 6150 \AA\ feature
and are discriminated by the presence (SNe Ib) or absence (SNe Ic) of
\ion{He}{1} lines; other than this difference, Types Ib and Ic are
very similar.  The nebular-phase spectra of the various types are also
distinctive.  Type II SNe continue to be dominated by Balmer emission,
but also show lines of neutral and singly-ionized oxygen and calcium.
The late-time spectra of Type Ia SNe consist mainly of many
overlapping lines of iron and cobalt.  Types Ib and Ic have spectra
that are quite similar to those of late-time Type II SNe, but without
the hydrogen lines.  For a more thorough review of the history and
details of SN classification and spectroscopy, see, e.g., Harkness \&
Wheeler (1990), Branch (1990), and Filippenko (1997).

These spectroscopic differences reflect the nature of the progenitors
for each of the SN types.  The evolution of the spectra of Type Ia
SNe, along with their environments and homogeneous light curves, has
led to a general consensus that they are the result of the
thermonuclear disruption of a white dwarf (see, e.g., Branch et
al. 1995).  By a similar line of reasoning, Type II SNe arise from the
core collapse of massive ($\gtrsim 8-10$ M$_{\sun}$) stars (e.g.,
Woosley \& Weaver 1986; Arnett et al. 1989).  The progenitors of Type
Ib and Ic SNe were not as clear, as the basic data needed to
characterize the SNe themselves were limited (see Harkness \& Wheeler
1990 and Clocchiatti et al. 1997 for more details).

Evidence gradually began to accumulate that the Type Ib and Ic SNe
were related to Type II SNe.  Type Ib/c SNe and Type II SNe appear in
similar environments (Van Dyk, Hamuy, \& Filippenko 1996, and
references therein); in addition, Type Ib/c SNe occur almost
exclusively in late-type galaxies (Porter \& Filippenko 1987) and they
often exhibit radio emission that is thought to arise from
circumstellar interaction (e.g., Weiler et al. 1998, and references
therein).  As mentioned above, the late-time spectra of Type Ib/c SNe
are very similar to those of Type II SNe, but without the hydrogen.
Thus, it is likely that the progenitors of Type Ib/c SNe are also
massive stars, although white dwarf progenitor models have been
considered (e.g., Branch \& Nomoto 1986; Iben et al. 1987).

More direct evidence for the connection between Type Ib/c SNe and Type
II SNe is provided by several SNe that have transformed from one
spectral type into another.  The first of these was SN 1987K.  It was
initially classified as a Type II SN, but later observations showed no
hydrogen lines; the actual metamorphosis was not observed (Filippenko
1988).  Woosley et al. (1987) had already considered such a scenario
and had suggested the new name of ``Type IIb.''  SN 1993J was also a
Type II SN when first observed, but then developed the \ion{He}{1}
lines of a Type Ib (e.g., Filippenko, Matheson, \& Ho 1993; Wheeler \&
Filippenko 1996; Matheson et al. 2000a, and references therein).  Qiu
et al. (1999) show a similar transformation for SN 1996cb.  These
observed metamorphoses provide a clear link between the Type II SNe
and Type Ib/c SNe.  The hydrogen missing from the spectra of Type Ib/c
SNe is thought to have been lost from the progenitor through winds or
mass transfer to a companion (e.g., Woosley, Langer, \& Weaver 1993;
Yamaoka, Shigeyama, \& Nomoto 1993; Nomoto et al. 1994; Iwamoto et
al. 1994; for SN 1993J specifically, see Woosley et al. 1994; Wheeler
\& Filippenko 1996; Matheson et al. 2000a, and references therein).

The differences between progenitors that yield Type Ib SNe and Type Ic
SNe remain uncertain.  It may be that the same mechanism of mass
transfer invoked above to explain the disappearance of hydrogen can
also remove the helium from a star that then explodes as a Type Ic SN.
SN 1999cq, a Type Ic based on its spectrum, exhibited
intermediate-width \ion{He}{1} emission lines that might represent the
interaction of the ejecta with dense clumps of almost pure helium (no
Balmer lines were present; Matheson et al. 2000b).  This could be a
direct detection of the helium layer that was in the process of being
lost by the progenitor.  Some Type Ic SNe, however, still show broad
helium lines.  There have been claims of the \ion{He}{1}
$\lambda$10830 line in SN 1990W (Wheeler et al. 1994), SN 1994I
(Filippenko et al. 1995), and SN 1994ai (Benetti, quoted in
Clocchiatti et al. 1997).  Wheeler et al. (1994) suggest that the
presence of this line makes SN 1990W a Type Ib SN instead,
highlighting the difficulty in distinguishing between SNe Ib and SNe
Ic.  Reanalysis of other Type Ic SNe spectra has also shown some
indications of several helium lines (Clocchiatti et al. 1996; Munari
et al. 1998), although other studies have countered these claims
(Baron et al. 1999; Millard et al. 1999).

The difficulty of exciting helium to produce lines may provide an
alternative cause for the differences between Type Ib and Ic SNe.
Gamma rays emitted by newly synthesized $^{56}$Ni accelerate electrons
that then act as a source of nonthermal excitation for helium
(Harkness et al. 1987; Lucy 1991).  Thus, helium could still be
present in both Type Ib and Ic SNe, but lines are only observed if the
nickel is mixed into the the helium layer in quantities sufficient to
excite the helium (e.g., Wheeler et al. 1987; Shigeyama et al. 1990;
Hachisu et al. 1991).  

To help resolve the controversy over the nature of SNe Ib and SNe Ic,
we studied the aggregate collection of spectra of stripped-envelope
SNe (Types Ib, Ic, and IIb) that the SN group at U.C. Berkeley has
obtained in the past 12 years.  Some of these spectra, notably those
of SN 1994I and SN 1993J, have been presented in detail elsewhere
(Filippenko et al. 1995; Filippenko, Matheson, \& Ho 1993; Filippenko,
Matheson, \& Barth 1994; Matheson et al. 2000a).  Discounting them,
the sample consists of 84 spectra of SNe Ib, Ic, and IIb.  There are
also spectra of three SNe initially classified as SNe Ib/c, but whose
characteristics indicate that they are in fact SNe Ia.  In \S 2 the
observations are described in general; the individual SNe are
presented in detail in \S 3.  The evolution of the spectra of SNe Ib
is discussed in \S 4.1.  Helium lines in SNe Ic are considered in \S
4.2.  Some spectroscopic distinctions of SNe Ib and SNe Ic are shown
in \S 4.3.  The evolution and heterogeneity of SNe Ic are highlighted
in \S 4.4, with a special emphasis on two peculiar SNe Ic in \S 4.5
that may be associated with gamma-ray bursts.

\section{Observations}

Low-dispersion spectra of the SNe Ib/c considered here were obtained
with a variety of instruments.  The majority were observed at the
Cassegrain focus of the Shane 3-m reflector at Lick Observatory, using
either the original CCD spectrograph (the ``UV Schmidt''; Miller \&
Stone 1987) up to the 1992 January 09 data or the Kast double
spectrograph (Miller \& Stone 1993) from 1992 March to the present.
Several spectra were also obtained with the double spectrograph on the
Hale 5-m reflector at Palomar Observatory (Oke \& Gunn 1982) and with
the Low Resolution Imaging Spectrometer (LRIS; Oke et al. 1995) at the
Keck~II 10-m telescope.  A single spectrum was observed with the Kitt
Peak Mayall 4-m telescope using the RC spectrograph.  A long, narrow
(typically 2$\arcsec$, but smaller for the LRIS observations) slit was
used.  Various gratings and grisms were utilized, yielding resolutions
(full width at half maximum, FWHM) ranging from 6 \AA\ to 15 \AA.
Details of the exposures are given in Table 1.  Most of the spectra
were taken with the slit oriented at, or near (within 10$^{\circ}$),
the parallactic angle (Filippenko 1982).

Standard CCD processing and spectrum extraction were accomplished with
VISTA (Terndrup, Lauer, \& Stover 1984) or IRAF\footnote{IRAF is
distributed by the National Optical Astronomy Observatories, which are
operated by the Association of Universities for Research in Astronomy,
Inc., under cooperative agreement with the National Science
Foundation.}.  Optimal extraction was used for the IRAF reductions
(Horne 1986).  The wavelength scale was established using low-order
polynomial fits to spectra of calibration lamps.  The typical
root-mean-square (rms) deviation for the wavelength solution was
$0.1-0.5$ \AA, depending on the resolution for the particular
exposure.  Final adjustments to the wavelength scale were obtained by
using the background sky regions to provide an absolute scale.  We
employed our own routines to flux calibrate the data; comparison stars
are listed in Table 1.  Particular care was taken to remove telluric
absorption features through division by an intrinsically featureless
spectrum, where possible (Wade \& Horne 1988).  The flux standard was
routinely employed for this purpose.  While the relative
spectrophotometry is excellent (individual exceptions are noted in the
discussion below), no attempt has been made to calibrate the fluxes
to an absolute scale.

For four of the SNe (1998dt, 1998fa, 1999di, and 1999dn), we have
photometric observations obtained with the 0.75-m Katzman Automatic
Imaging Telescope at Lick Observatory (KAIT; Treffers et al. 1997;
Richmond, Treffers, \& Filippenko 1993; Li et al. 2000; Filippenko et
al. 2001).  The detector used by KAIT is a SITe $512 \times 512$ pixel
CCD with a field of view of 6.\arcmin8 $\times$ 6.\arcmin8.  For SN
1998dt, we used a point-spread function (PSF) fitting technique to
measure magnitudes.  SN 1998fa, SN 1999di, and SN 1999dn were in
complex regions of their host galaxies.  We therefore employed a
galaxy subtraction method to isolate the SNe; magnitudes were then
determined using PSF fitting.  Not all of the SNe were followed with
standard $UBVRI$ filters.  Their host galaxies, though, are part of
the Lick Observatory Supernova Search (LOSS; Li et al. 2000;
Filippenko et al. 2001) and thus are imaged regularly (every three to
four days).  These observations are taken without a filter, yielding
unfiltered magnitudes.  Photometric calibration of KAIT images
indicates that the unfiltered response is similar to (but broader
than) that of a standard $R$ filter (Riess et al. 1999; the KAIT $R$
filter is a Bessell $R$ [Bessell 1990]).  For SNe 1998dt and 1999dn,
there was photometric coverage with the $R$-band filter, and the
unfiltered observations match the $R$ light curve quite well.
Therefore, we will use both $R$-band and unfiltered magnitudes to
construct light curves for the SNe.  The differences between $R$-band
and unfiltered observations may increase as the spectrum becomes more
line-dominated, since the edges of the two passbands may include
different sets of lines.  The chief use of the light curves, however,
is to provide a temporal order for the spectra, so it is the epoch of
maximum light that is most important, and the measured magnitudes from
the two passbands roughly agree at this phase for the SNe with
$R$-band and unfiltered coverage.

\section{Description of Individual Supernovae}

Throughout this section, the magnitude of the SN at the time of
discovery is quoted directly from the I.A.U. Circulars.  The systems
used to detect the SNe are very heterogeneous, with most using
unfiltered observations.  None of the numbers reported is
well-calibrated on a standard system and some may have large
uncertainties.  They should only be used as rough indicators of
relative brightness.  All calendar dates reported are UT.  In all the
figures of the SNe spectra, the systemic heliocentric velocity has
been removed.  If a velocity could be determined from narrow emission
lines in the SN spectrum (from superposed \ion{H}{2} regions) then
that value was used; if not, then the velocity of the host galaxy
(i.e., of the nucleus) was removed.  If the velocity of the host
galaxy was found from our spectra, this is explicitly mentioned.
Otherwise, a reference to the source for the galactic velocity is
given or, for most NGC, UGC, or IC objects, we describe it as a
``listed'' velocity taken from the de Vaucouleurs et al. (1991) RC3
catalog.

\emph{SN 1988L}.---This supernova was discovered in the course of the
Berkeley Automated Supernova Search (BASS) on 1988 April 30 at a
magnitude of 16.5 (Perlmutter \& Pennypacker 1988) in NGC 5480.  The
narrow H$\alpha$ emission line from a superposed \ion{H}{2} region is
observed at 6608 \AA, yielding a recession velocity of 2050
km~s$^{-1}$; the velocity of NGC 5480 is listed as 1856 km~s$^{-1}$.
It was initially classified as Type I (possibly Ia or Ib; Graham,
Boroson, \& Oke 1988), and finally confirmed as Type Ib (Filippenko,
Spinrad, \& McCarthy 1988; Kidger 1988).  Some of these spectra have
already been published (Filippenko 1988).  Our first spectrum (1988
May 11; Figure \ref{sn1988l-mont}) is eleven days after discovery, and
still in the photospheric phase.  The later spectra exhibit the
typical nebular lines of Type Ib/c SNe, namely [\ion{O}{1}]
$\lambda\lambda$6300, 6364, [\ion{Ca}{2}] $\lambda\lambda$7291, 7324,
\ion{O}{1} $\lambda$7774, and the calcium near-infrared (near-IR)
triplet.

\emph{SN 1990B}.---The BASS also discovered this SN on 1990 January 20
in NGC 4568 at a magnitude of 16 (Perlmutter \& Pennypacker 1990).  We
find a recession velocity for the SN of 2200 km~s$^{-1}$ from narrow
H$\alpha$ emission from an \ion{H}{2} region, while the listed
galactic velocity for NGC 4568 is 2255 km~s$^{-1}$.  The lack of
hydrogen lines indicated that the SN was of Type I (Dopita \& Ryder
1990), and subsequent spectra confirmed that it was a SN Ib
(Filippenko, Spinrad, \& Dickinson 1990; Kirshner \& Leibundgut 1990;
Benetti, Cappellaro, \& Turatto 1990a).  Later analysis indicated that
SN 1990B was actually a Type Ic SN (Clocchiatti et al. 2001); at the
earlier stages, the existence of a separate subclass of helium-poor
Type I SNe (i.e., the SNe Ic) had yet to be demonstrated.  Radio
emission was also detected from SN 1990B (Sramek, Weiler, \& Panagia
1990; Van Dyk et al. 1993).  Some of the spectra shown here (Figure
\ref{sn1990b-mont}) are also presented by Clocchiatti et al. (2001).
Our first three spectra (1990 January 23, February 10, and February
27) show some absorption features, but they do not exhibit the full
set of helium lines that would indicate a SN Ib; thus SN 1990B is a SN
Ic according to our spectra; see, however, \S 4.2.  By 1990 March 25,
the SN had begun to enter the nebular phase.

\emph{SN 1990U}.---Another BASS discovery, SN 1990U was found on 1990
July 27 in NGC 7479 at magnitude 16 (Pennypacker, Perlmutter, \&
Marvin 1990).  Narrow H$\alpha$ emission from an \ion{H}{2} region
indicates a recession velocity of 2500 km~s$^{-1}$ with a listed
galactic velocity of 2378 km~s$^{-1}$.  It was initially classified as
a SN Ib (Filippenko \& Shields 1990a).  The relative weakness (or
absence) of the \ion{He}{1} lines led to a re-evaluation of the object
as a Type Ic (Filippenko \& Shields 1990b).  As can be seen in Figure
\ref{sn1990u-mont}, there are features in the photospheric spectra,
but the \ion{He}{1} lines are not prominent.

\emph{SN 1990aa}.---The BASS found SN 1990aa on 1990 September 4 at
magnitude 17 in UGC 540.  The recession velocity of the SN based on
the narrow H$\alpha$ emission line from an \ion{H}{2} region is 5020
km~s$^{-1}$ while the listed velocity for UGC 540 is 4982 km~s$^{-1}$.
The preliminary classification was as Type Ia (Della Valle 1990;
Benetti, Cappellaro, \& Turatto 1990b), but it was recognized as a SN
Ib, or possibly SN Ic, with fully calibrated spectra (Filippenko \&
Shields 1990b).  Our spectra (Figure \ref{sn1990aa-mont}) show that
this is a SN Ic; there are many features in the 1990 September 27
spectrum, but the \ion{He}{1} lines are not present at great strength,
if at all.

\emph{SN 1990aj}.---McNaught (1991a) discovered SN 1990aj on a plate
taken 1990 December 19 at magnitude 18 in NGC 1640.  Pre-discovery
images showed it on 1990 December 6.  There are no visible (\ion{H}{2}
region) narrow emission lines, but the centroid of the [\ion{O}{1}]
$\lambda\lambda$6300, 6364 line gives a velocity of 1900 km~s$^{-1}$;
the listed recession velocity of NGC 1640 is 1602 km~s$^{-1}$.  The
supernova's spectral features indicated that it was a late-time SN Ib
(Della Valle \& Pasquini 1991), but without photospheric spectra, it
is unclear whether this was a SN Ib or SN Ic.  Our spectrum only shows
the late nebular lines of a SN Ib/c (Figure \ref{sn1990aj-mont}).

\emph{SN 1991A}.---The BASS discovered SN 1991A on 1991 January 1 at
magnitude 18 in IC 2973 (Pennypacker et al. 1991).  Narrow H$\alpha$
emission from an \ion{H}{2} region gives a recession velocity of 3250
km~s$^{-1}$ while the listed galactic value is 3206 km~s$^{-1}$.  It
was classified as a Type Ic SN (Filippenko 1991a), with the
possibility of a weak H$\alpha$ component (Filippenko 1991b).  This
broad feature near H$\alpha$ is visible in our spectra (Figure
\ref{sn1991a-mont}).  The rest of the spectrum is relatively
featureless between 6000 and 7500 \AA.  The 1991 January 17 spectrum
shows an apparent double minimum between 5500 \AA\ and 5800 \AA.  The
redward minimum is most likely \ion{Na}{1}~D, while the blueward one
may be \ion{He}{1} $\lambda$5876; see the discussion of helium in SNe
Ic in \S 4.2

\emph{SN 1991D}.---SN 1991D was discovered by Remillard et al. (1991)
on 1991 February 6 at magnitude $\sim$ 16.5 in an anonymous galaxy.
This galaxy was actually number 23 of List 1 of the Cal\'an-Tololo
Seyfert Galaxies (Maza \& Ruiz 1989), also known as PGC/LEDA 84044.
There are no narrow (\ion{H}{2} region) emission lines in the spectrum
of the SN, but the redshift of the host galaxy is $\sim$ 12500
km~s$^{-1}$ (Maza \& Ruiz 1989).  The discoverers' own spectrum
indicated it was of Type I.  Later spectra showed that it was a SN Ib
or SN Ic (Kirshner, Huchra, \& McAfee 1991), but most likely a Type Ib
(Filippenko \& Shields 1991b).  Figure \ref{sn1991d-mont} shows the
spectrum obtained by Remillard et al. (1991; the 1991 February 7
spectrum), as well as our own.  The wavelength coverage of the 1991
February 7 spectrum is not extensive, but \ion{He}{1} $\lambda$7065
and $\lambda$7281 are visible, and there are hints of those lines in
the February 23 spectrum as well, lending credence to the
classification as a SN Ib.  The poor signal-to-noise (S/N) ratio of
the spectra does hamper a definitive classification.

\emph{SN 1991K}.---McNaught (1991b) discovered SN 1991K on 1991
February 21 in NGC 2851 at a magnitude of 18.  This SN shows no narrow
(\ion{H}{2} region) emission lines in its spectrum.  The recession
velocity of NGC 2851 is 5096 km~s$^{-1}$ (Phillips 1991a).
There was some confusion about the classification of this object.  The
first spectra were consistent with Type II or Type Ic (Filippenko \&
Shields 1991a), but subsequent reports indicated Type Ib (Phillips
1991a).  Late-time spectra ($\gtrsim$ 100 days), however, resemble
those of Type Ia SNe (G\'omez, L\'opez, \& S\'anchez 1996).  Our
spectra, one of which is seen in Figure \ref{comp-mont}, show a
striking similarity with a normal Type Ia SN at later times.

\emph{SN 1991L}.---Pollas (1991) discovered SN 1991L on plates taken
1991 February 24 at magnitude 18 in MCG +07-34-134.  While the
spectrum of SN 1991L shows no narrow (\ion{H}{2} region) emission
lines, the narrow H$\alpha$ emission from our spectrum of the galaxy
nucleus gives a recession velocity of 9050 km~s$^{-1}$.  Filippenko,
Shields, \& Nomoto (1991) classified the SN as a Type Ib or Ic.  The
fully calibrated spectrum (Figure \ref{sn1991l-mont}), however, does
not stand out as a clear example of any known type of SN.  The
features at the blue end are reminiscent of SNe Ia (cf. Figure
\ref{comp-mont}), but the iron features that dominate this region are
common to all SN types.  In addition, the lack of the \ion{Si}{2}
absorption belies a SN Ia claim.  The peculiar, overluminous Type Ia
SN 1991T had a very weak silicon line at early times (Filippenko et
al. 1992a; Phillips et al. 1992), but the rest of the spectrum does not
match that of SN 1991L.  With just this one spectrum, the exact nature
of SN 1991L will remain a mystery, but it is probably a SN Ib or SN
Ic.

\emph{SN 1991N}.---The BASS found SN 1991N on 1991 March 29 at
magnitude 15 in NGC 3310 (Perlmutter et al. 1991).  It was classified
as a Type Ic (possibly Type Ib; Filippenko 1991c).  This SN is
superposed on a very bright \ion{H}{2} region.  The recession velocity
derived from the narrow emission lines of the \ion{H}{2} region is
1190 km~s$^{-1}$, while that of NGC 3310 is listed as 980 km~s$^{-1}$.
Figure \ref{sn1991n-mont} shows our spectra of SN 1991N.  The
photospheric spectra are relatively smooth in the range 6000$-$7500
\AA\ where most of the distinctive \ion{He}{1} lines are found,
indicating that this is definitely a SN Ic.  The spectra of SN 1991N
are heavily contaminated by starlight, as indicated by the very blue
late-time spectra.

\emph{SN 1991ar}.---SN 1991ar was found by McNaught (1991c) on 1991
September 3 at magnitude 17 in IC 49.  The SN has a recession velocity
of 4520 km~s$^{-1}$ based on the narrow (\ion{H}{2} region) H$\alpha$
emission line.  The listed velocity of IC 49 is 4562 km~s$^{-1}$.
Phillips (1991b) classified it as a Type Ib; this was confirmed by
Filippenko \& Matheson (1991b), despite an earlier misidentification
as a Type II (Filippenko \& Matheson 1991a).  Our spectra (Figure
\ref{sn1991ar-mont}), although noisy, show distinctly the \ion{He}{1}
series.

\emph{SN 1995F}.---Lane \& Gray (1995) discovered SN 1995F on 1995
February 11 at a magnitude of $\sim$ 15 in NGC 2726.  The narrow
(\ion{H}{2} region) H$\alpha$ emission line in the spectra of this SN
indicates a recession velocity of 1440 km~s$^{-1}$, while the listed
galactic value for NGC 2726 is 1518 km~s$^{-1}$.  It was classified as
a Type Ic (possibly Type Ib; Filippenko \& Barth 1995).  The
photospheric spectra of SN 1995F (Figure \ref{sn1995f-mont}) show many
features, but they are not at the correct wavelengths to be helium, so
the object is a SN Ic.

\emph{SN 1995bb}.---During the course of a redshift survey, Tokarz \&
Garnavich (1995) found SN 1995bb in a spectrum of an anonymous galaxy
taken on 1995 November 29.  The spectrum indicated the nebular
features of a Type Ib/c SN.  We find a recession velocity for SN
1995bb of 1610 km~s$^{-1}$ based on the narrow (\ion{H}{2} region)
H$\alpha$ emission line in the spectrum, in contrast to the value of
1740 km~s$^{-1}$ reported by Tokarz \& Garnavich (1995).  Figure
\ref{sn1995bb-mont} displays our spectrum, which shows only the
nebular-phase lines of [\ion{O}{1}] $\lambda\lambda$6300, 6364 and
\ion{Mg}{1}] $\lambda$4571.

\emph{SN 1996cb}.---SN 1996cb was discovered by Nakano \& Aoki (1996)
and Qiao et al. (1996) on 1996 December 16 (December 19 for Qiao et
al.) at magnitude 16 in NGC 3510.  The narrow (\ion{H}{2} region)
H$\alpha$ emission line in our spectra gives a recession velocity of
800 km~s$^{-1}$, while NGC 3510 has a listed velocity of 705
km~s$^{-1}$.  Early spectra indicated that this was a Type II SN
(Garnavich \& Kirshner 1996), but Garnavich \& Kirshner (1997) soon
noticed that SN 1996cb began to show \ion{He}{1} lines.  It thus
became another example of the SNe that change from SNe II to objects
that resemble SNe Ib such as SN 1987K and SN 1993J (see, e.g. Matheson
et al. 2000a; Qiu et al. 1999).  Our spectra (Figure
\ref{sn1996cb-mont}) are from later times, and do not show the P-Cygni
profiles of the helium lines.  They do, however, retain a relatively
weak H$\alpha$ line on what would normally seem to be a spectrum of a
SN Ib, just as in the later-time spectra of SN 1993J (see Figure 5 of
Matheson et al. 2000a).

\emph{SN 1997C}.---The Beijing Astronomical Observatory (BAO)
Supernova Survey discovered SN 1997C on 1997 January 14 at magnitude
18 in NGC 3160 (Li et al. 1997b).  There are no narrow (\ion{H}{2}
region) emission lines in our spectrum of SN 1997C, but the redshift
of the host galaxy is 6795 km~s$^{-1}$ (Falco et al. 1999).  Li et
al. (1997b) classified the SN as a Type Ic.  Our spectrum, though,
appears to resemble that of a peculiar SN Ia at later times (Figure
\ref{comp-mont}) and we believe that SN 1997C was misclassified.

\emph{SN 1997dc}.---The BAO Supernova Survey also discovered SN 1997dc
on 1997 August 5 at magnitude 18 in NGC 7678 (Qiao et al. 1997).
Pre-discovery observations by the LOSS showed that the SN was present
as early as 1997 July 26 (Peng et al. 1997).  Our spectrum of SN
1997dc shows no narrow (\ion{H}{2} region) emission lines, but narrow
H$\alpha$ emission from the galaxy nucleus indicates a recession
velocity of 3480 km~s$^{-1}$; the velocity of NGC 7678 in the catalog
is listed as 3487 km~s$^{-1}$.  While initially classified as a SN Ic
(Li et al. 1997a), Piemonte, Benetti, \& Turatto (1997) reported the
presence of helium lines and concluded that SN 1997dc was a Type Ib.
Our spectrum (Figure \ref{sn1997dc-mont}) also shows the optical
\ion{He}{1} series of a typical SN Ib, although $\lambda$7281 is
relatively weak.

\emph{SN 1997dd}.---Nakano \& Aoki (1997a) communicated the discovery
of SN 1997dd on 1997 August 21 at magnitude 16 in NGC 6060.  Our
spectrum shows no narrow (\ion{H}{2} region) emission lines, but the
recession velocity of NGC 6060 is 4554 km~s$^{-1}$ (Huchra et
al. 1983).  A spectrum obtained by Suntzeff, Maza, \& Phillips (1997)
showed that SN 1997dd was a peculiar Type II, but they felt that it
would evolve into a SN IIb, such as SN 1987K, SN 1993J, or SN 1996cb
(see above).  Garnavich, Jha, \& Kirshner (1997) later reported that
helium lines had begun to appear in the spectrum of SN 1997dd,
confirming the prediction.  Our spectrum is from an early stage in the
transformation, before the helium lines have fully developed (Figure
\ref{sn1997dd-mont}).  Nevertheless, a distinct notch in the emission
component of H$\alpha$ shows the \ion{He}{1} $\lambda$6678 line.  This
is very similar to the early spectra of SN 1993J (see, e.g., Figure 1
of Matheson et al. 2000a).

\emph{SN 1997dq}.---Nakano \& Aoki (1997b) also discovered SN 1997dq,
but on 1997 November 2 at magnitude 15 in NGC 3810.  Spectra of SN
1997dq did not show any narrow (\ion{H}{2} region) emission lines,
though the listed recession velocity of NGC 3810 is 994 km~s$^{-1}$.
Jha et al. (1997) classified it as a Type Ib SN.  The \ion{He}{1}
features are not particularly strong (Figure \ref{sn1997dq-mont}) and
the overall impression is not that of a typical SN Ib.  In fact, there
are some similarities to the peculiar SN Ib/c 1997ef (see below).

\emph{SN 1997ef}.---Nakano \& Sano (1997) reported the discovery of SN
1997ef on 1997 November 26 at magnitude 17 in UGC 4107.  The narrow
(\ion{H}{2} region) H$\alpha$ emission line in the spectra of SN
1997ef gives a recession velocity for the SN of 3500 km~s$^{-1}$,
while the listed velocity of the galaxy is 3504 km~s$^{-1}$.  Early
spectra indicated an unusual object with broad features, but not
obviously those of a SN (Garnavich et al. 1997a; Hu et al. 1997;
Filippenko 1997b).  Further study yielded no definitive answer, but it
was thought to be probably a peculiar Type Ib/c (Garnavich et
al. 1997c; Wang, Howell, \& Wheeler 1998b).  Our spectra (Figure
\ref{sn1997ef-mont}) also show this to be a peculiar SN Ib/c.  The
fact that SN 1997ef had extremely broad emission lines at early stages
led some to consider it in relation to SNe that may have been
associated with gamma-ray bursts (see, e.g., Nomoto et al. 1999;
Woosley, Eastman, \& Schmidt 1999, and references therein).

\emph{SN 1997ei}.---Nakano \& Aoki (1997c) found SN 1997ei on 1997
December 23 at a magnitude of 18 in NGC 3963.  The velocity of SN
1997ei as determined from the narrow (\ion{H}{2} region) H$\alpha$
emission line is 3160 km~s$^{-1}$, with a listed velocity for the
galaxy of 3186 km~s$^{-1}$.  Initial classification was as a Type Ia
SN (Garnavich et al. 1997b; Ayani \& Yamaoka 1997), but subsequent
observations described it as Type Ic (Wang, Howell, \& Wheeler 1998a;
Filippenko \& Moran 1998a).  Our spectra (Figure \ref{sn1997ei-mont})
show many lines, but they are not at the proper positions to be the
\ion{He}{1} series, so the object is a SN Ic.

\emph{SN 1998T}.---SN 1998T was found on 1998 March 3 at a magnitude
of 15 (Li, Li, \& Wan 1998a) as part of the BAO Supernova Survey.  The
narrow emission line of H$\alpha$ from a superposed \ion{H}{2} region
indicates a recession velocity of 3080 km~s$^{-1}$ for SN 1998T, while
the listed velocity for the galaxy is 3033 km~s$^{-1}$.  Although the
first report located the SN in IC 694, the SN is actually in NGC 3690,
a nearby pair of interacting galaxies (all three comprise Arp 299;
Yamaoka et al. 1998).  Filippenko \& Moran (1998b) classified it as a
Type Ib SN.  The strong absorptions caused by \ion{He}{1} are visible
in our spectra (Figure \ref{sn1998t-mont}), despite contamination by a
very bright, superposed \ion{H}{2} region.

\emph{SN 1998dt}.---The LOSS located SN 1998dt on 1998 September 1 at
magnitude 17 in NGC 945 (Shefler et al. 1998).  The recession velocity
of SN 1998dt measured from narrow (\ion{H}{2} region) H$\alpha$
emission is 4580 km~s$^{-1}$, with a listed value for the galaxy of
4484 km~s$^{-1}$.  It was classified as a Type Ib (Jha et al. 1998a).
Figure \ref{sn1998dt-mont} shows our spectra of SN 1998dt; it is
unmistakably a Type Ib.

\emph{SN 1998fa}.---The LOSS discovered SN 1998fa on 1998 December 25
at magnitude 18 in UGC 3513 (Li et al. 1998b).  The recession velocity
of SN 1998fa measured from narrow (\ion{H}{2} region) H$\alpha$
emission is 7460 km~s$^{-1}$, with a listed value for the galaxy of
7316 km~s$^{-1}$.  Early spectra showed broad hydrogen emission,
leading to its classification as a Type II SN (Jha et al. 1998b).  The
onset of P-Cygni profiles of \ion{He}{1} at later times (Filippenko,
Leonard, \& Riess 1999) meant that this was actually a Type IIb SN,
such as SN 1987K or SN 1993J (cf. Figure 2 of Matheson et al. 2000a).
Our spectra (Figure \ref{sn1998fa-mont}) show a Type II SN with
relatively weak hydrogen lines on 1999 January 10, but a small notch
develops in H$\alpha$ by January 19 that is identified with
\ion{He}{1} $\lambda$6678.  The other helium lines appear as well.
Two spectra listed in Table 1 (1999 February 23 and March 12) are not
shown as the SN was not detectable by those dates.

\emph{SN 1999P}.---The High-Z Supernova Search Team (Schmidt et
al. 1998) discovered SN 1999P on 1999 January 13-14 at magnitude 21.8
in an anonymous galaxy (Garnavich et al. 1999).  From a narrow
(\ion{H}{2} region) H$\alpha$ line in the spectrum of SN 1999P, we
derive a recession velocity of 17700 km~s$^{-1}$.  In the course of
the follow-up for the High-Z work, it was found that SN 1999P was a SN
Ib/c (Filippenko, Riess, \& Leonard 1999).  The spectrum (Figure
\ref{sn1999p-mont}) indicates that the SN is beginning to enter the
nebular phase; thus, a definitive classification (SN Ib or Ic) is not
possible.

\emph{SN 1999bv}.---SN 1999bv was found by Schwartz (1999) on 1999
April 19 at magnitude 18 in MCG +10-25-14.  Narrow (\ion{H}{2} region)
H$\alpha$ emission gives a recession velocity of 5490 km~s$^{-1}$,
very similar to that of the host galaxy (MCG +10-25-14, $cz = $5510
km~s$^{-1}$; Jha et al. 1999a).  Jha et al. (1999a) described it as a
probable SNe Ib/c similar to SN 1988L.  This was confirmed by Hill et
al. (1999).  Our spectrum (Figure \ref{comp-mont}), though, is
actually quite similar to that of a late-time SN Ia.  We believe that
SN 1999bv was misclassified.

\emph{SN 1999di}.---Puckett \& Langoussis (1999) discovered SN 1999di
on 1999 August 8 at magnitude 18 in NGC 776.  The narrow (\ion{H}{2}
region) H$\alpha$ emission line in the spectra of SN 1999di indicates
a recession velocity of 4920 km~s$^{-1}$, effectively identical with
the listed value for the host galaxy, NGC 776 (4921 km~s$^{-1}$).
Filippenko et al. (1999) classified it as a Type Ib.  Spectra of SN
1999di (Figure \ref{sn1999di-mont}) show the \ion{He}{1} lines as well
as a strong absorption at $\sim$ 6300 \AA\ that may be \ion{Si}{2}
$\lambda$6355 or \ion{C}{2} $\lambda$6580 (cf. \S 4.2).  It is
possible that this absorption is due to H$\alpha$, but the
corresponding H$\beta$ line is not present.

\emph{SN 1999dn}.---The BAO Supernova Survey found SN 1999dn on 1999
August 20 at a magnitude of 16 in NGC 7714 (Qiu, Qiao, \& Hu 1999).
The recession velocity of SN 1999dn is 2700 km~s$^{-1}$, determined
from the narrow (\ion{H}{2} region) H$\alpha$ emission line.  The
listed velocity of NGC 7714 is 2799 km~s$^{-1}$.  There was confusion
about the classification of this SN.  While Ayani et al. (1999) and
Turatto et al. (1999) described it as a Type Ic SN, Qiu et al. (1999)
reported it as a Type Ia SN.  The helium lines were visible
(Pastorello et al. 1999), so it was also listed as a Type Ib/c SN.
Our spectra (Figure \ref{sn1999dn-mont}) unambiguously indicate that
it is a SN Ib.  This was also the conclusion of Deng et al. (2000),
who used a synthetic-spectrum code to model SN 1999dn.  They also
suggest that there is evidence for a weak hydrogen line, implying that
SN 1999dn might have had an extremely low-mass hydrogen envelope.

\emph{SN 1999ec}.---The LOSS discovered SN 1999ec on 1999 October 3 at
magnitude 18 in NGC 2207 (Modjaz \& Li 1999).  Narrow (\ion{H}{2}
region) H$\alpha$ emission in the spectrum of SN 1999ec yields a
recession velocity of 2800 km~s$^{-1}$, while NGC 2207 has a listed
velocity of 2746 km~s$^{-1}$.  It was classified as a SN Ib by Jha et
al. (1999b), who described it as having prominent helium lines in a
spectrum taken on 1999 October 5.  Our single spectrum (obtained 1999
October 8, Figure \ref{sn1999ec-mont}) does not appear to fit
obviously into any category.  While there is an absorption line near
5700 \AA, this could simply be \ion{Na}{1}~D alone, not \ion{Na}{1}~D
and \ion{He}{1} $\lambda$5876; note that the other helium lines are
invisible.  It could still be a SN Ib, but at late times.  The
[\ion{O}{1}] $\lambda\lambda$6300, 6364 lines, however, are not
present at their usual strength for a SN Ib.  We believe that this is
a peculiar SN I event, but the type is uncertain.  It may be related
to SN 1993R, an object that showed an unusual blend of SN Ib/c and SN
Ia features (Filippenko \& Matheson 1993; Matheson et al. 2001).

\section{Discussion}

To characterize our data on SNe Ib and SNe Ic, we will include
comparisons with several of the best-observed examples of each class.
One of the first well-studied SNe Ib was SN 1984L; we will use the
data from Harkness et al. (1987).  A late-time spectrum of SN Ib 1983N
is taken from Gaskell et al. (1986).  We will also use spectra of SN
1985F (Filippenko \& Sargent 1986).  As there was no spectrum of this
SN near maximum light, the type is uncertain.  The light curve
(Filippenko et al. 1986; Tsvetkov 1986), however, is more consistent
with those of SNe Ib than SNe Ic (Wheeler \& Harkness 1990), and we
will treat it as such.  By far the most extensively studied SN Ic is
SN 1994I; the spectra from Filippenko et al. (1995) will be used for
comparison.  In addition, we will consider the SN Ic 1987M
(Filippenko, Porter, \& Sargent 1990).

\subsection{Evolution of SNe Ib}

The set of spectra of SNe Ib in our sample is quite heterogeneous;
there are few epochs for any given SN.  For three of the SNe Ib,
though, we have a light curve and can place the spectra relative to
maximum light.  The $R$-band and unfiltered-magnitude light curves of
these three SNe are shown in Figure \ref{iblightcurve}.  The light
curves for the individual SNe were shifted in time and magnitude to
match one another with 0 mag indicating maximum brightness.  The
derived $R$-band maxima\footnote{While there may be discrepancies
between the $R$-band and unfiltered magnitudes for our SNe, the
correspondence between the $R$-band and unfiltered values for the SNe
that were observed in both bands indicates that the epoch of maximum
light derived from unfiltered observations is approximately coincident
with the time of $R$-band maximum light, especially within the
uncertainty described in the text.  Therefore, for SNe with few or no
$R$-band observations, we will refer to the date of unfiltered maximum
as the epoch of $R$-band maximum light for the purposes of this
paper.} for the SNe are as follows: SN 1998dt, 1998 September 12; SN
1999di, 1999 July 27; SN 1999dn, 1999 August 31.  Given the sparse
nature of the points, an exact match was not possible.  Shifting any
one of the three curves by a day or two in either direction yields a
similar result; thus, in all subsequent discussion of the relative
phases of the spectra, an uncertainty of one or two days is
appropriate.  For the sake of completeness, we also show a partial
light curve of the SN IIb 1998fa offset from the SNe Ib in Figure
\ref{iblightcurve}; its derived $R$-band maximum is 1999 January 2.

With the knowledge of the dates of the $R$-band maxima for the three
SNe Ib, we can register the spectra and assign them a phase, as listed
in Table 2.  When these spectra are arranged in phase order (Figure
\ref{ibmontage}), it is apparent that there is variation in the
relative strengths of the \ion{He}{1} P-Cygni absorptions.  It is
possible that this is just evidence of the differences among the SNe,
but the variation seems to follow the temporal order even within the
set of spectra for an individual object.  There may be variations in
the abundance of helium or the level of mixing, but we will ignore
those effects for the present time to see whether changes in the
helium lines do indicate a standard evolutionary behavior for the
spectra of SNe Ib.

To make a comparison of the lines among different spectra, we had to
develop a technique for characterizing the line strength.  Traditional
line-measuring methods are not applicable here.  The equivalent width
of the lines is dependent on a well-defined continuum, and this is not
available as the spectra consist of a pseudo-continuum of many
overlapping lines.  We attempted to determine the line strength using
local parameters for each line.  We defined an ``effective''
equivalent width by setting a local continuum, integrating the flux in
the line, and dividing by the value of the continuum.  To do this, we
chose relatively smooth regions on either side of the absorption to
define the local continuum (Figure \ref{depthfig} illustrates the
technique).  The median value of each region was evaluated and a
straight line fit through the two points (each point having the
wavelength of the center of its continuum region and the flux of the
median value).  We then chose the edges of the line (generally where
the continuum regions began, but with occasional subjective
excursions).  The flux within the line was evaluated in two ways.
First, we summed the flux directly.  Second, we fit a Gaussian to the
line and summed the flux it represented.  Both methods gave
effectively the same results.  As can be seen in Figure
\ref{depthfig}, the Gaussian fit the absorption rather well, although
for some lines other than helium, it was more problematic.  The
``equivalent widths'' determined with this technique did not yield any
pattern indicating a temporal sequence for our spectra of SNe Ib.
Complicating factors such as differing expansion velocity and
placement of the continuum introduced a large uncertainty into each
value.

Another possibility for measuring the lines came from considering only
the depth of the line to eliminate some of the impacts from these
potential sources of error.  The depth was not dependent on the
definition of the precise edges of the lines---the most subjective
aspect of the measurement.  One only had to find the minimum of the
line, determine the flux of the spectrum at that point, evaluate the
``continuum'' flux at that wavelength from the line fit to the regions
on either side of the absorption, and then use those fluxes to
calculate the fractional line depth.  The value for the line depth is
measured from these fluxes using the formula
\begin{displaymath}
{\rm{Fractional~Line~Depth}} = \frac{\it{F}_{\rm{cont}} - \it{F}_{\rm{min}}}{\it{F}_{\rm{cont}}},
\end{displaymath}
where ${F_{\rm{cont}}}$ and ${F_{\rm{min}}}$ represent the flux values
of the continuum and of the line minimum at the wavelength of the
minimum, respectively.  For noisy spectra, the region around the
minimum was smoothed by a boxcar chosen to match approximately three
times the spectral resolution.  This smoothed subset determined the
location and value for the minimum flux.  The value of the fractional
line depth thus gave a less subjective measurement for the scale of a
given absorption.  Larger values of the fractional line depth indicate
a deeper and, presumably, stronger absorption.  This is somewhat
related to the technique used by Nugent et al. (1995) to estimate the
strengths of absorptions in the spectra of SNe Ia, although their
continuum was determined by two points, not a fit to a continuum
region as we do.

We then took the line depths and normalized them to the fractional
depth of the \ion{He}{1} $\lambda$6678 line.  In the SNe Ic, we still
scaled to the feature closest to the expected position of this line,
but, as discussed below, \emph{it is likely not the helium line}.
Thus, for each spectrum we have a fractional line depth scaled to a
particular line.  If there are helium abundance variations, then the
normalization by the $\lambda$6678 line should help to lessen any
impact.  These values are listed in Table 3.

When the line depths for the SNe Ib that can be put in temporal order
are compared, a pattern appears for the helium lines (see Figure
\ref{linedepthrat}).  Both $\lambda$5876 (probably contaminated by
\ion{Na}{1}~D) and $\lambda$7065 grow in strength relative to
$\lambda$6678.  The values are not exactly monotonic, but the trend is
there within the scatter.  The $\lambda$7281 line was either not
present at great strength or too difficult to isolate in many of our
spectra; for the ones in which it could be measured, a trend was not
obvious.

To test this pattern, we examined two spectra of SN 1984L that have
the three relevant \ion{He}{1} lines.  For the 1984 September 23
spectrum, the line ratios indicate that it is probably between our day
21 and day 38 spectra (see Figure \ref{ibmontage}).  The date of
maximum for SN 1984L is not well known, but $B$ maximum was
approximately 1984 August 20\footnote{Note that Filippenko (1997; his
Figure 9) used an erroneous date for the $B$ maximum of SN 1984L to
assign epochs to displayed spectra (some relative separations are also
incorrect).  His epochs of $-8$, $-4$, 16, 20, 25, 45, and 58 days
relative to $B$ maximum should be 10, 14, 30, 34, 40, 59, and 72
days.}  (August 20 $\pm$ 4, Tsvetkov 1985; Schlegel \& Kirshner 1989).
That puts this spectrum at day 34, modulo differences between $B$ and
$R$ maximum, which falls within our (admittedly broad) window.  The
second spectrum (1984 September 29) fits in days 17-33, slightly
\emph{earlier} than the September 23 spectrum, so the scheme clearly
is not complete.  The first spectrum, though, could be day 21 and the
second thus day 27, thereby still fitting within our model, although
the date of maximum would have to be incorrect by almost two weeks,
and that is not likely.  If the September 29 spectrum is day 33, then
the September 23 spectrum would be day 27 (consistent with our model's
predictions) and the date of maximum would be off by six days.  As the
date of maximum is estimated from later points on the light curve,
this may be reasonable.

For SN 1991ar, the line-depth values for the 1991 September 16
spectrum indicate a phase of $21-38$ days, although this is more
uncertain than most.  The 1991 October 2 spectrum seems to be from a
later phase, probably greater than 52 days.  The 16 days between
observations is consistent with our predicted ages.  Discovery was on
September 2, so it was at least 14 days past explosion by September 16
and 30 days past explosion by October 2.  Our predicted ages are
relative to the date of maximum brightness, so we would expect that
the actual explosion date was much earlier than September 2.  Phillips
(1991b) called it ``a few weeks old'' on September 14, again consistent
with our numbers.

The line-depth values for the 1997 September 6 spectrum of SN 1997dc
are almost identical with those of the 1999 August 19 spectrum of SN
1999di.  This would make its age about 21 days past maximum and most
likely less than 33 days.  Discovery was on 1997 August 5, with
pre-discovery images confirming its presence on 1997 July 26 (see
discussion of the individual SNe above).  An explosion date near July
26 and a rise to maximum of approximately 15 days (cf. Figure
\ref{iblightcurve}) would make Sept 6 roughly 27 days past $R$-band
maximum.  Within the uncertainty of our age estimates, this is
consistent.

The model does not work at all with the values measured from SN 1998T.
Considering only the $\lambda$5876 line, both the 1998 March 6 and
March 27 spectra fit in the $21-38$ day range.  Discovery was on March
3 (Li et al. 1998a), but the March 6 spectrum is evidently older than
just a few days past maximum (compare with the early spectra of SN
1984L in Harkness et al. 1987).  This SN is superposed on an extremely
bright \ion{H}{2} region, and contamination from the narrow H$\alpha$
line made measurements of the $\lambda$6678 line problematic.  In
addition, while the ratios of the line depths are consistent with
those of other SNe Ib, the scale of the fractional line depth of the
individual lines was much less.  As can be seen in Figure
\ref{sn1998t-mont}, the spectra are very blue, indicating a high level
of starlight contamination.  It is possible that the combination of
measurement errors and dilution render our line-depth values useless
in this case.  SN 1998T is a clear exception to the otherwise high
level of homogeneity among SNe Ib.

When we try to determine the relative ages of the seven spectra of SNe
Ib (SNe 1984L, 1991ar, 1997dc, and 1998T) that were not used to
develop the pattern of helium-line depths, the result is not
completely successful.  Figure \ref{ibmontage} shows the training set
of spectra in the proper order with the six test spectra inserted in
the approximately correct positions.  The spectra of SN 1984L fit to a
certain degree, but the 1984 September 29 spectrum does not quite
match the model.  SN 1991ar and SN 1997dc seem to work well, although
it is difficult to be certain with only a single spectrum for SN
1997dc.  The results with SN 1998T are in conflict with the model.
The apparent contamination in the spectra may be the cause of this
problem, but it may also indicate that there is not a simple sequence
of temporal evolution of the helium lines for SNe Ib or that
helium-line strength is intrinsically variable.  The rest of the SNe
Ib in the sample, however, are remarkably homogeneous.

\subsection{Helium Lines in SNe Ib \emph{and} SNe Ic?}

One other result of the line-fitting technique is that the expansion
velocity of the minimum of the line is measured along with an estimate
of the velocity width of the absorption\footnote{For these values, and
all subsequent discussions of velocities, the relativistic formula was
used to convert Doppler shifts to velocities.}.  These values are
listed in Table 4.  The SNe Ib all have fairly consistent expansion
velocities in the range $7000-9500$ km~s$^{-1}$.  The SNe Ic have
similar velocities for the line referred to as \ion{He}{1}
$\lambda$5876, but this is probably \ion{Na}{1}~D\footnote{For the SNe
Ic, velocities are still computed relative to a rest wavelength of
5876 \AA.  If the line is \ion{Na}{1}~D, then the expansion velocities
are $\sim$ 800 km~s$^{-1}$ larger.}.  The line referred
to as \ion{He}{1} $\lambda$6678, though, has fairly large values
($16100-18900$ km~s$^{-1}$, with one at 12200 km~s$^{-1}$) for the SNe
Ic---\emph{so large that this line is probably not helium}, but rather
some other species that appears in most SNe Ic.

Clocchiatti et al. (1996) claim that helium lines, including
$\lambda$5876 and $\lambda$6678, \emph{are} present in the spectra of
many SNe Ic at a high velocity.  (Munari et al. [1998] made similar
claims for SN 1997X, but inspection of their spectra indicates that
this is a less certain identification.)  Figure \ref{ichelines} shows
two of the SNe that Clocchiatti et al. (1996) considered (SNe 1994I
and 1987M) along with several SNe Ic in our sample and a SN Ib for
comparison.  The expected positions of the \ion{He}{1} lines at
expansion velocities of $-15000$ to $-19000$ km~s$^{-1}$ are marked.
For SNe 1994I and 1987M, a broad feature is at the expected location
of high-velocity \ion{He}{1} $\lambda$6678, as well as a narrower line
at the expected high-velocity position of \ion{He}{1} $\lambda$5876.
The other helium lines may appear in SNe 1994I and 1987M, but they are
weak.  The other SNe Ic from our sample do not seem to exhibit this
same pattern, although, as mentioned above, in the 1991 January 17
spectrum, SN 1991A does have a double-minimum structure between 5300
\AA\ and 5800 \AA\ that may be \ion{Na}{1}~D and high-velocity
\ion{He}{1} $\lambda$5876.  There is a feature near 6300 \AA\ in some
of the SNe Ic that would match the \ion{He}{1} $\lambda$6678 line, but
the other lines are not there.  If anything, SN 1990B seems to show
weak \emph{low-velocity} helium lines that match the SN Ib, although
this identification is tentative.

It may be that the small features Clocchiatti et al. (1996)
interpreted as high-velocity helium are actually other species at low
velocities.  The small dip near 5550 \AA\ that they claimed as
high-velocity \ion{He}{1} $\lambda$5876 also appears in the spectrum
of the SN Ib 1999dn as a shoulder on the blue edge of the \ion{He}{1}
$\lambda$5876 + \ion{Na}{1}~D line, but the SN Ib does not have any
unexpected features at the locations of the other high-velocity helium
lines (the low-velocity position of \ion{He}{1} $\lambda$7065 is
coincident with the high-velocity position of \ion{He}{1}
$\lambda$7281 for SN 1999dn).

The identification of the line that appears near 6300 \AA\ remains
uncertain.  Millard et al. (2000) suggested \ion{C}{2} $\lambda$6580
was present in SN 1994I at a high, detached velocity of 16000
km~s$^{-1}$.  (Deng et al. [2000] come to similar conclusions for the
SN Ib 1999dn.)  This line would still have a fairly high velocity
($12000-13000$ km~s$^{-1}$) in our spectra, but seems more likely than
helium.  As discussed below, carbon and oxygen lines might stand out
more in SN Ic spectra because they are not diluted by a helium layer
as they would be in spectra of SNe Ib.  It would be at a very low
velocity if the line were \ion{Si}{2} $\lambda$6355, but that remains
a possibility.  Careful spectral synthesis modeling may be able to
finalize the identification of this line.

Clocchiatti et al. (1996) found that there was \emph{not} a continuum
of helium-line strength that would imply a gradual transition from SNe
Ib to SNe Ic.  This is also apparent from our spectra, especially in
light of the uncertain line near 6300 \AA.  For all of the objects
listed in Tables 3 and 4, if helium lines are present,
then they are distinct in the spectra.  The SNe Ib shown in Figure
\ref{ibmontage} all have fairly similar helium-line strengths, except
for SN 1998T.  It, though, appears to suffer a high level of
contamination that is diluting the helium lines.  In fact, the SNe Ib
in Figure \ref{ibmontage} are similar overall, except for the
appearance of the 6300 \AA\ line in SN 1999di (see Deng et al. 2000
for a further discussion of such lines in a SN Ib).

Other claims of detection of helium in SNe Ic have been put forth.
Filippenko et al. (1995) showed an apparent P-Cygni profile of
\ion{He}{1} $\lambda$10830 (a minimum at $\sim$ 10250 \AA) in SN
1994I.  The $\lambda$10830 line, though, should be much stronger than
the other optical helium lines (Swartz et al. 1993b) and can be
produced with very small amounts of helium (Wheeler, Swartz, \&
Harkness 1993; Baron et al. 1996).  Therefore, the \ion{He}{1}
$\lambda$10830 line could be present without requiring evidence for
the other optical helium lines.  More detailed spectral synthesis
models indicate that even the $\lambda$10830 identification may be
questionable (Baron et al. 1999; Millard et al. 1999).  These models
have found that \ion{C}{1} (multiplet 1, $\lambda$10695) and
\ion{Si}{1} multiplets 4 ($\lambda$12047), 5 ($\lambda$10790), 6
($\lambda$10482), and 13 ($\lambda$10869) may contribute to the
near-IR feature observed in SN 1994I.  In fact, \ion{Si}{1} may
contribute near 7000 \AA, explaining the features seen there in some
SNe Ic (Millard et al. 1999).  The SN Ic 1999cq showed
intermediate-width helium emission lines in its spectra, but these
were more likely the result of an interaction with circumstellar
material, not evidence for helium in the ejecta of the SN itself
(Matheson et al. 2000b).

It may be that the transition objects between SNe Ib and SNe Ic are
rare or have been overlooked.  For example, only a few objects have
been observed in the SN IIb class out of the hundreds of normal SNe II
and SNe Ib.  If this ratio holds for the SNe Ib with weak helium, then
perhaps an object bridging SNe Ib and SNe Ic has been missed.  Weak
lines in some SNe Ic are at the wrong positions to be helium, unless
the velocities are very large, and then only a subset of the lines
appears.  It is possible that the lines in SNe 1987M and 1994I are
helium, but the rest of our SN Ic spectra do not show the same
patterns.  If this is the case, then SNe 1987M and 1994I (and,
perhaps, 1990B) are the transition objects between SNe Ib and Ic,
while the true Type Ic SNe show no helium at all, with the possible
exception of \ion{He}{1} $\lambda$10830.

In addition to the controversy over the presence of helium, there have
been some claims of \emph{hydrogen} in SNe Ic.  Jeffery et al. (1991)
suggested that there was evidence for hydrogen in the spectra of SN
1987M.  Filippenko (1992) concurred on SN 1987M, as well as claiming
that SN 1991A (and perhaps SN 1990aa) also revealed weak hydrogen
lines.  If some SNe Ic do retain enough hydrogen to produce observable
lines, then it is likely that there is helium present as well.  In
this scenario, mixing of radioactive nickel into a helium layer would
be the more important factor in determining whether or not helium
lines appear.  As discussed below, though, the SNe Ic in general seem
to have less massive envelopes than the SNe Ib.  If that is the case,
the presence of hydrogen in the spectra of SNe Ic is quite mysterious,
although mixing at earlier phases in the star's evolution may have
introduced hydrogen into the carbon-oxygen or (low-mass) helium layer
of the star.  Detailed spectral modeling may indicate if this
identification is correct.

\subsection{Spectroscopic Distinction of SNe Ib and SNe Ic}

\subsubsection{Expansion Velocity from Late-Time Spectra}

If the difference between progenitors of SNe Ib and SNe Ic is due to
the absence (or at least partial absence) of a helium envelope in the
latter, then it may be possible to see an impact of this in the
late-time, nebular-phase spectra as well as at early times.  For a
given input energy, a lower-mass envelope will be expelled with a
higher velocity.  The apparent increase in velocity width from Types
II through IIb, Ib, and Ic may indicate a decreasing envelope mass
(Figure \ref{linewidthfig}).  Whether or not the explosion energy is the
same for all core-collapse events is questionable, but we will assume
that it does not vary much for the purposes of this analysis.

In a study of late-time spectra of SNe Ib, Schlegel \& Kirshner (1989)
found the velocity widths of [\ion{O}{1}] $\lambda\lambda$6300, 6364
and [\ion{Ca}{2}] $\lambda\lambda$7291, 7324 to be FWHM = 4500 $\pm$
600 km~s$^{-1}$.  Filippenko et al. (1995) determined larger values
for two SNe Ic (SN 1994I, FWHM = 7700 km~s$^{-1}$ for [\ion{O}{1}] and
FWHM = 9200 km~s$^{-1}$ for [\ion{Ca}{2}]; SN 1987M, FWHM = 7500
km~s$^{-1}$ for [\ion{O}{1}] and FWHM = 6200 km~s$^{-1}$ for
[\ion{Ca}{2}]).  They interpreted this as evidence for smaller
envelope mass and/or larger explosion energy in SNe Ic compared with
SNe Ib.

We measured the velocity width of nebular emission lines with a
Gaussian fit to the line after a continuum fit to the spectrum by hand
had been subtracted.  The continuum placement was subjective, but the
late-time spectra are simple and have fairly well-defined continua;
errors in the location of the continuum have little impact on the
value of the FWHM of the fit line.  The subset of our data with
measurable nebular-phase emission lines (along with SNe 1983N, 1985F,
1987M, and 1994I) are listed in Table 5 with the values from the
Gaussian fit for the FWHM of the lines.

Unfortunately, despite the large number of SNe Ib in our sample, none
of them is at a sufficiently late stage to provide clean, nebular
emission lines.  For the SNe Ib, we will rely on the values of
Schlegel \& Kirshner (1989) and the SN Ib 1983N.  There are many good
examples of late-time SNe Ic spectra.  Not all of the spectra are
useful though.  The values for SN Ic 1995F are very large.  The lines
appear to be contaminated (on the blue edge of [\ion{O}{1}]
$\lambda\lambda$6300, 6364 and red edge of [\ion{Ca}{2}]
$\lambda\lambda$7319, 7324---cf. Figure \ref{sn1995f-mont}), and we
will not use it in the determination of the mean values.  SNe Ic
1997dq and 1997ef seem to be spectroscopically peculiar (discussed
below), and thus will not be included for a general estimate of the
velocity widths of later-time SN Ic emission lines either.  We also
exclude SN 1991aj and SN 1995bb, as their exact classification is
uncertain.  The velocities of the lines for these two SNe are very
similar to the range listed for the other SNe Ic, and are clearly less
than the average Schlegel \& Kirshner (1989) reported for SNe Ib, but
without a definite type, they should not be used.  Adding them to the
SN Ic set of velocities, though, does not alter the values reported
below significantly.  Note that we do include SNe Ic 1994I and 1987M.
Our values for [\ion{O}{1}] $\lambda\lambda$6300, 6364 and
[\ion{Ca}{2}] $\lambda\lambda$7319 for the 1994 September 2 spectrum
of SN 1994I and the 1988 February 25 spectrum of SN 1987M are
virtually identical to the numbers reported by Filippenko et
al. (1995), despite using slightly different techniques to determine
the line widths.

For each of the relatively normal SNe that remains, we will consider
only the latest possible value for any given line.  To calculate the
mean value for the emission lines, each SN will contribute one number;
otherwise, a single SN with spectroscopic coverage over many late
epochs could bias the mean value.  In this way the heterogeneity of
the SNe Ic is reflected in the computed standard deviation.  For
[\ion{O}{1}] $\lambda\lambda$6300, 6364, the mean velocity
width\footnote{For this, and all subsequent discussion of line widths,
the velocity reported is FWHM.} for Type Ic SNe is 7600 $\pm$ 2100
km~s$^{-1}$ (median = 7300 km~s$^{-1}$) with nine data points.  The
mean is 8700 $\pm$ 2700 km~s$^{-1}$ (median = 8500 km~s$^{-1}$) for
[\ion{Ca}{2}] $\lambda\lambda$7319, 7324 with eight data points.
Combining both [\ion{O}{1}] $\lambda\lambda$6300, 6364 and
[\ion{Ca}{2}] $\lambda\lambda$7319, 7324 as has been done with the
Schlegel \& Kirshner (1989) data for SNe Ib, the mean velocity width
is 8100 $\pm$ 2400 km~s$^{-1}$ (median 7400 km~s$^{-1}$).

The calcium near-IR triplet is more problematic to measure than the
other lines discussed.  As it has three components, the line profile
is not as Gaussian-shaped as it is for [\ion{O}{1}]
$\lambda\lambda$6300, 6364 and [\ion{Ca}{2}] $\lambda\lambda$7319,
7324 and the total width of the line does not represent only the
expansion velocity.  The intrinsic width of the near-IR triplet in
velocity space is $\sim$ 5700 km~s$^{-1}$, significantly larger than
the velocity separation of the two [\ion{O}{1}] components of $\sim$
3000 km~s$^{-1}$.  With these caveats, the mean value for the near-IR
triplet must be viewed with caution.  The velocity width for the
triplet in SNe Ic is 10800 $\pm$ 1600 km~s$^{-1}$ (median = 10800
km~s$^{-1}$) with eight points contributing.  Table 5 also lists
\ion{O}{1} $\lambda$7774, but this line is not present in many of the
spectra and its mean value is not reliable.

Table 5 also lists values for two SNe Ib (1983N and 1985F).  The
average of these two for the [\ion{O}{1}] $\lambda\lambda$6300, 6364
lines is 5600 km~s$^{-1}$ while it is 5200 km~s$^{-1}$ for
[\ion{Ca}{2}] $\lambda\lambda$7319, 7324 (standard deviations for only
two measurements are not meaningful and are not reported).  When both
lines are considered together, the mean velocity width is 5400 $\pm$
300 km~s$^{-1}$.  With only two points, the mean values are much less
certain.  SN 1985F was part of the Schlegel \& Kirshner (1989) data
set.  If we combine the numbers from Schlegel \& Kirshner (1989) for
SNe 1984L, 1985F, and 1987K\footnote{Although SN 1987K is a Type IIb,
its velocity widths are comparable to those of the SNe Ib in the
Schlegel \& Kirshner (1989) sample.} with our values for SN 1983N
(considering only the latest possible value for each line in each SN),
the mean velocity width of [\ion{O}{1}] $\lambda\lambda$6300, 6364 and
[\ion{Ca}{2}] $\lambda\lambda$7319, 7324 for SNe Ib is 4900 $\pm$ 800
km~s$^{-1}$.  As mentioned above, [\ion{O}{1}] $\lambda$6300 and
[\ion{O}{1}] $\lambda$6364 are separated by $\sim$ 3000 km~s$^{-1}$.
This blending may artificially broaden the line as measured, yield
exaggerated expansion velocities.  The values for [\ion{O}{1}]
$\lambda\lambda$6300, 6364 in Table 5, however, are comparable to
those of [\ion{Ca}{2}] $\lambda\lambda$7319, 7324, whose two
components are just $\sim$ 200 km~s$^{-1}$ apart.  This implies that
$\lambda$6300 dominates the flux of doublet and the velocity as
measured is a good indication of the actual expansion.

The lines of [\ion{O}{1}] $\lambda\lambda$6300, 6364 and
[\ion{Ca}{2}] $\lambda\lambda$7319, 7324 at late times are fairly
representative of the velocity width of the expansion.  All of the
measurements for SNe Ic in Table 5 are larger than the
mean value for SNe Ib.  Only SN Ic 1990U has a velocity width
approaching that of the SNe Ib.  A Kolmogorov-Smirnov (KS) test
indicates that the distributions of line widths for SNe Ib and SNe Ic
are different at the 99\% confidence level.

It seems unlikely that every SN Ic has a higher explosion energy than
SNe Ib.  Thus, the more probable cause is that the SNe Ic have
lower-mass envelopes, lending credence to the argument that the
progenitors of SNe Ic have lost most or all of their helium layer.  If
a lack of mixing were the sole reason that SNe Ic do not show helium
lines, then one would expect that the envelope masses of SNe Ib and
SNe Ic would be comparable.  It may be that mixing still plays a role,
but the velocity widths of the late-time emission lines indicate that
there appears to be a distinct difference between the envelope masses
for SNe Ib and Ic.

\subsubsection{Permitted Oxygen Lines}

Another test of the lost-envelope scheme for producing SNe Ic is to
examine a feature that might appear more prominently if the helium
were removed.  A massive star that has lost its hydrogen and helium
(or at least most of the helium) would be left with a carbon-oxygen
core.  This new envelope could produce lines of carbon or oxygen at
relatively greater strength than could one in which those species are
diluted by the presence of a large amount of helium (or hydrogen).  If
this is the case for SNe Ic, then the oxygen lines present in their
spectra might appear to be relatively stronger than they do in the
spectra of SNe Ib.

To test this hypothesis, we chose to study the \ion{O}{1}
$\lambda$7774 line.  It is in a region of the spectrum that is
relatively uncontaminated by other lines and is readily accessible in
most of our spectra (unlike \ion{O}{1} $\lambda$9266).  Figure
\ref{oidepth} illustrates the apparent increase of \ion{O}{1}
$\lambda$7774 line strength from Types II through IIb, Ib, and Ic.
This increase could be the result of decreasing envelope mass.  In
each of our spectra that contained a relatively distinct P-Cygni
profile of \ion{O}{1} $\lambda$7774, we employed our fractional line
depth technique described above in \S 4.1.  The results are listed in
Table 6.  For this discussion, we will not include the values for the
peculiar SN Ic 1997dq.  The fractional line depth of the oxygen line
is the most useful of the measurements listed.  The velocity at the
minimum of the P-Cygni absorption is probably fairly accurate, but the
velocity width of the absorption measured by the fit Gaussian is not
particularly representative.  Unlike the \ion{He}{1} lines discussed
above, the Gaussian did not appear to match the profile very well; we
report the number for completeness.

The mean value for the line depth for all measurements of SNe Ib is
0.27 $\pm$ 0.11 (median = 0.25) while that for SNe Ic is 0.38 $\pm$
0.091 (median = 0.37).  If we consider only the values from the
earliest possible spectrum for each SN (i.e., only counting one value
for each SN, where we are examining the outermost layers of the
expanding ejecta), the mean for SNe Ib becomes 0.22 $\pm$ 0.10 (median
= 0.20) and that of the SNe Ic becomes 0.41 $\pm$ 0.099 (median =
0.39).  To see if there is a difference between the values for the SNe
Ib and SNe Ic, we compared them using a KS test.  The KS test
indicates that the two distributions (\ion{O}{1} line depths for SNe
Ib and Ic) are different at a confidence level of 97\%.\footnote{The
standard deviation listed for each set of line depth values appears to
indicate that the means are not significantly different.  The spread
of values in each distribution could be broad with a distinctly
different mean for each, though, so the KS test is a more rigorous
indicator of difference between the two distributions, even for a
small number of values (e.g., Press et al. 1986).}  The differences
suggest that the SNe Ic do have a relatively stronger \ion{O}{1}
$\lambda$7774 line, implying that they may have a lower-mass (or even
non-existent) helium envelope overlying the oxygen layer of their
progenitors.

The expansion velocity implied by the minimum of the \ion{O}{1}
$\lambda$7774 line is slightly larger for SNe Ic than for SNe Ib
(mean of 6900 km~s$^{-1}$ for SNe Ib vs. 8300 km~s$^{-1}$ for SNe
Ic).  This is consistent with the results described above for the
nebular-phase line widths, but the difference is smaller and probably
less significant.  The minimum for the \ion{O}{1} $\lambda$7774 line
was more difficult to define than for the helium lines due to its
generally broader shape and its position on a changing part of the overall
spectral profile.

\subsection{Are the SNe Ic a Uniform Class?}

Aside from SN 1998T, the spectra of the SNe Ib in Figure
\ref{ibmontage} are all fairly similar.  What about the SNe Ic?  There
are measured light curves for two of the SNe Ic in our sample, SN
1990B (Clocchiatti et al. 2001) and SN 1990U (Richmond, Filippenko, \&
Galisky 1998).  Combining these (and ignoring for now the possibility
of two different types of light curves for SNe Ic [Clocchiatti et
al. 1997; see below]), we can define the time of maximum light for SN
1990B and SN 1990U and register the two sets of spectra to a common
phase (Figure \ref{icmontage}).  For the case of the SNe Ib, the
relative strengths of the helium lines provided a possible indicator
for the relative age of the SNe. Unfortunately, there is not a
comparable measurement that shows the evolution of the SNe Ic.

Despite the lack of phase information, the spectra of SNe Ic in our
sample do allow some generalizations to be made.  One of the
spectroscopic differences among the various objects is the relative
smoothness of the spectrum in the range $6000-7500$ \AA.  For some
SNe, P-Cygni profiles appear with absorption minima near 6300 \AA\
and/or 6800 \AA.  The apparent \ion{He}{1} $\lambda$7065 minimum is
often near 6800 \AA, but the 6300 \AA\ line is probably not helium
(see discussion above).  The point is that some SNe Ic have one or
both of these features, and the others do not.  Considering only the
relatively normal SNe Ic that have spectra during the photospheric
phase (or at least show some photospheric features), we have three SNe
Ic that are relatively smooth in this region (SNe 1991A, 1991N, and
1994I) and five that show many distinct lines between $6000-7500$ \AA\
(SNe 1988L, 1990B, 1990U, 1990aa, and 1995F).

The expansion velocities from the late-time nebular line velocity
widths listed in Table 5 can be compared for the two subsets of SNe
Ic.  There is no correlation.  The two lowest-velocity SNe are SN
1990U and SN 1991N, yet their spectra do not look very similar.  In
fact, the spectra of SN 1991N and SN 1994I do look similar, but they
have disparate line widths.

Clocchiatti et al. (1997; see also Clocchiatti \& Wheeler 1997)
introduced a light-curve element into the classification of the SNe
Ic.  They found that there are two types of light curves, ``fast'' and
``slow.''  The light curve of SN 1990B is in the ``slow'' class
(Clocchiatti et al. 2001).  The comparison of this light curve with
that of SN 1990U (Richmond et al. 1998) shows that SN 1990U also has a
``slow'' light curve.  The only other SN Ic in our sample with a
measured light curve is SN 1994I (e.g., Richmond et al. 1996); it is
in the ``fast'' class.  As mentioned above, there do appear to be some
spectroscopic differences between SN 1994I and the two ``slow''
examples, SNe 1990B and 1990U.  The speed of the light curve is
apparently related to the overall mass of the ejecta, implying that SN
1994I had a smaller-mass envelope than did SN 1990B or SN 1990U.  SN
1990U had narrower nebular-phase emission lines than SN 1994I
(cf. Table 5), but SN 1990B was comparable to SN 1994I.  Without a
larger number of examples of the ``fast'' and ``slow'' classes having
spectroscopic coverage, it is not clear if any correlation exists.

There does not appear to be a method for subclassifying our spectra of
SNe Ic in a simple scheme.  The class as a whole is distinct from the
SNe Ib (and SNe II and SNe Ia), but further subdivision seems
unwarranted at this time.  Within the definable type of SN Ic, though,
there is a substantial level of heterogeneity.

\subsection{The Peculiar SNe Ic 1997dq and 1997ef}

The possible association of the peculiar SN Ic 1998bw with gamma-ray
burst (GRB) 980425 (e.g., Galama et al. 1998; Iwamoto et al. 1998;
Woosley, Eastman, \& Schmidt 1999) has helped to fuel a flurry of work
postulating that the GRBs are the result of some form of stellar
collapse (see, e.g., MacFadyen \& Woosley 1999; Nomoto et al. 2000;
Wheeler 2000; Lamb 2000, and references therein).  The temporal
appearance and location of SN 1998bw were approximately coincident
with GRB 980425.  The optical luminosity was significantly larger than
in a typical SN Ic (e.g., Galama et al. 1998; Iwamoto et al. 1998;
Woosley, Eastman, \& Schmidt 1999; for a comparison with other GRBs
and SNe, see Bloom et al. 1998).  Moreover, radio observations
exhibited a prompt turn-on and provided evidence of relativistic
motions (Kulkarni et al. 1998).  Finally, the spectral features were
broader than normal (see Nomoto et al. 2000; Kulkarni et al. 1998),
indicating a smaller mass and/or higher explosion energy.  SN 1997ef
(cf. Figure \ref{sn1997ef-mont}) also had broader than usual emission
and absorption features, causing Nomoto et al. (1999) to consider it
in the same class as SN 1998bw (see also Nomoto et al. 2000; Iwamoto
et al. 2000).  It was also possibly associated with a GRB (971115,
Wang \& Wheeler 1998).

The late-time spectra of SN 1997dq and SN 1997ef appear quite similar
(Figure \ref{dq-ef-comp}), but the early-time spectra of SN 1997dq do
not show lines as broad as those in SN 1997ef.  This probably just
reflects the lack of sufficiently early spectra for SN 1997dq.  The
GRB 971013 may be associated with SN 1997dq (Wang \& Wheeler 1998), so
it is possible that these two objects do represent a different class
than the normal SNe Ic presented above.  They may be examples of the
``hypernovae'' (Nomoto et al. 2000; Iwamoto et al. 2000) or
``collapsars'' (MacFadyen \& Woosley 1999) that are postulated as the
sources of GRBs.  Another SN in our sample, SN 1997ei, may also be
associated with a GRB (971120, Kippen et al. 1998; Wang \& Wheeler
1998), but its spectra do not seem especially unusual (cf. Figures
\ref{sn1997ei-mont} and \ref{dq-ef-comp}).  We do not know the
relative phases of all of our spectra, but none of our SN 1997dq or SN
1997ef spectra (which cover months of evolution) is a match for either
of the spectra of SN 1997ei.  Of course, apparent coincidental
association between SNe and GRBs is insufficient evidence for a direct
link between the objects.  On the other hand, if SN 1997ei and GRB
971120 are indeed physically linked, this may imply that the GRBs that
are also observed as SNe are a heterogeneous lot.  Models that produce
them must be able to account for the SNe 1998bw, 1997dq, and 1997ef
types as well as the more normal-looking SN 1997ei.

\section{Conclusions}

We present 84 spectra of SNe Ib, Ic, and IIb, as well as $R$-band and
unfiltered-magnitude light curves for three SNe Ib and one SN IIb.  We
find that three SNe in our sample that had originally been classified
as SNe Ib/c (SNe 1991K, 1997C, and 1999bv) have spectra that are more
consistent with those of late-time SNe Ia.  An examination of the
fractional depths of the helium lines in the SNe Ib reveals an
apparent progression of relative line strengths as the SN evolved.
The training set of SNe Ib for which we have light curves provides a
model for the development of the helium lines.  Other SNe Ib are fit
into the sequence of ages as determined by their helium-line
strengths, but with only partial success.  Examination of all the SNe
Ib spectra, though, does indicate that the SNe Ib have fairly
consistent helium-line strengths.  There are no clear examples of
weak, but distinct, helium lines in the SNe Ib.

In considering the possible presence of helium lines in SNe Ic, we
come to a different conclusion than Clocchiatti et al. (1996).  They
claim that spectra of various SNe Ic show helium lines at high
velocity.  The helium lines may be present in some of the objects that
Clocchiatti et al. (1996) studied, but the SNe Ic in our sample show
no consistent pattern of high-velocity helium lines.  Other lines,
such as \ion{C}{2} $\lambda$6580, may explain the features of the SN
Ic spectra.  SN 1990B does appear to show some weak helium lines, but
at low velocity.  This may represent the transition between SNe Ib and
SNe Ic, but the lines are not definitive.

We demonstrate other methods for distinguishing SNe Ib and SNe Ic
spectroscopically.  The emission lines in late-time, nebular-phase
spectra of SNe Ic have average line widths (FWHM) of 8100 $\pm$ 2400
km~s$^{-1}$, while incorporating the data of Schlegel \& Kirshner
(1989) gives a mean value of 4900 $\pm$ 800 km~s$^{-1}$ for the SNe
Ib.  They are distinctly different, with the SNe Ic having clearly
higher expansion velocities.  Unless the SNe Ic have consistently
higher explosion energies, this implies that SNe Ib have larger
envelope masses in general than do SNe Ic.

Another indicator of the difference between SNe Ib and SNe Ic is the
strength of the \ion{O}{1} $\lambda$7774 line.  If SNe Ic have had
their helium envelopes removed, then the exposed oxygen core could
yield relatively stronger lines as they are no longer diluted by the
helium.  We find that our fractional line depth measurements show
the SNe Ic to have stronger \ion{O}{1} lines (0.41 vs. 0.22 for SNe
Ib).  The KS test suggests that the difference between the SNe Ic and
SNe Ib is real, once again implying that the SNe Ic do have smaller
envelope masses than the SNe Ib.

Within the SN Ic class, there is a substantial heterogeneity.  We can
separate our SN Ic spectra into two classes based on the relative
smoothness in the range $6000-7500$ \AA, but the distinction is not
definitive.  In addition, no correlations with other line measurements
are found.  These two subsets may be related to the ``fast'' and
``slow'' photometric classes for SNe Ic that are described by
Clocchiatti et al. (1997), but, with only limited examples, this is
highly speculative.  While it is apparent from our results discussed
above that SNe Ic are distinct from SNe Ib (as well as from SNe II and
SNe Ia), there is no obvious method yet for spectroscopically
subclassifying them.

Finally, we explore two peculiar SNe Ic, 1997dq and 1997ef.  Both of
these SNe are possibly associated with GRBs.  SN 1997ef has extremely
broad emission lines at early times and has been modeled as a
``hypernova'' (e.g., Iwamoto et al. 2000).  Features in early-time
spectra of SN 1997dq are not as broad, but our temporal sampling is
sparse.  Later spectra of SN 1997dq match very well those of SN
1997ef, implying that it, too, is unusual in the same way as SN
1997ef.  SN 1997ei also may be positionally and temporally coincident
with a GRB, yet its spectra seem fairly normal for a SN Ic and do not
resemble SNe 1997dq or 1997ef at any phase.  It could be that the GRB
association with SN 1997ei is incorrect (as may be the case for all of
them), but if correct, then the models of ``hypernovae'' and
``collapsars'' must be able to produce typical SNe Ic such as SN
1997ei as well as the peculiar SN 1997ef.

The spectra presented in this paper, as well as the spectra of SN
1993J published by Matheson et al. (2000a), are available for analysis
by other researchers.  Please contact A.V.F. or T.M. if interested.

\acknowledgments 

This research was supported by NSF grants AST-8957063, AST-9115174,
AST-9417213, and AST-9987438 to A.V.F., as well as by NASA through
grants GO-7434, GO-7821 and GO-8243 from the Space Telescope Science
Institute, which is operated by AURA, Inc., under NASA contract NAS
5-26555.  T.M. acknowledges the support of an NSF Graduate Fellowship.
The Katzman Automatic Imaging Telescope was made possible by generous
donations from Sun Microsystems Inc. (Academic Equipment Grant
Program), the Hewlett-Packard Company, AutoScope Corporation, Lick
Observatory, the National Science Foundation, the University of
California, and the Sylvia and Jim Katzman Foundation.  We are
grateful to the staffs of the Lick, Keck, and Palomar Observatories
for their assistance with the observations.  The W. M. Keck
Observatory is operated as a scientific partnership among the
California Institute of Technology, the University of California, and
NASA; it was made possible by the generous financial support of the
W. M. Keck Foundation.  We wish to thank Aaron Barth, Ryan Chornock,
Alison Coil, Arjun Dey, Mark Dickinson, Mike Eracleous, Andrea
Gilbert, Jules Halpern, Luis Ho, Robert Kirshner, Pat McCarthy, Maryam
Modjaz, Ed Moran, Adam Riess, Wal Sargent, Jon Schacter, David
Schlegel, Hy Spinrad, and Chuck Steidel for providing assistance with
observations and data reduction.

\appendix

\ssp
\clearpage
\newpage
\begin{figure}[ht!]

\rotatebox{180}{
        \plotone{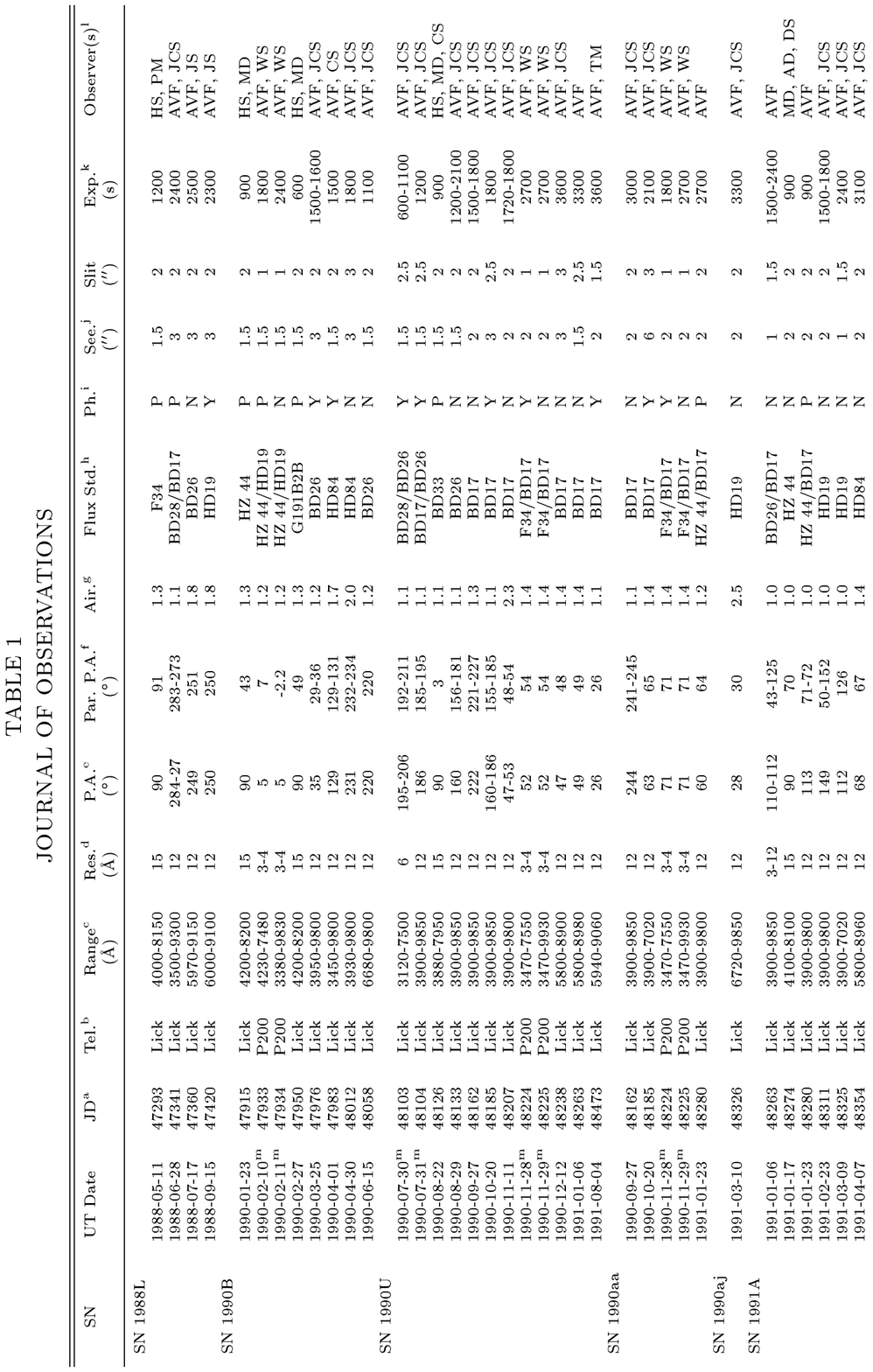}
  }

\end{figure}
 
\newpage
\begin{figure}[ht!]
\rotatebox{180}{
        \plotone{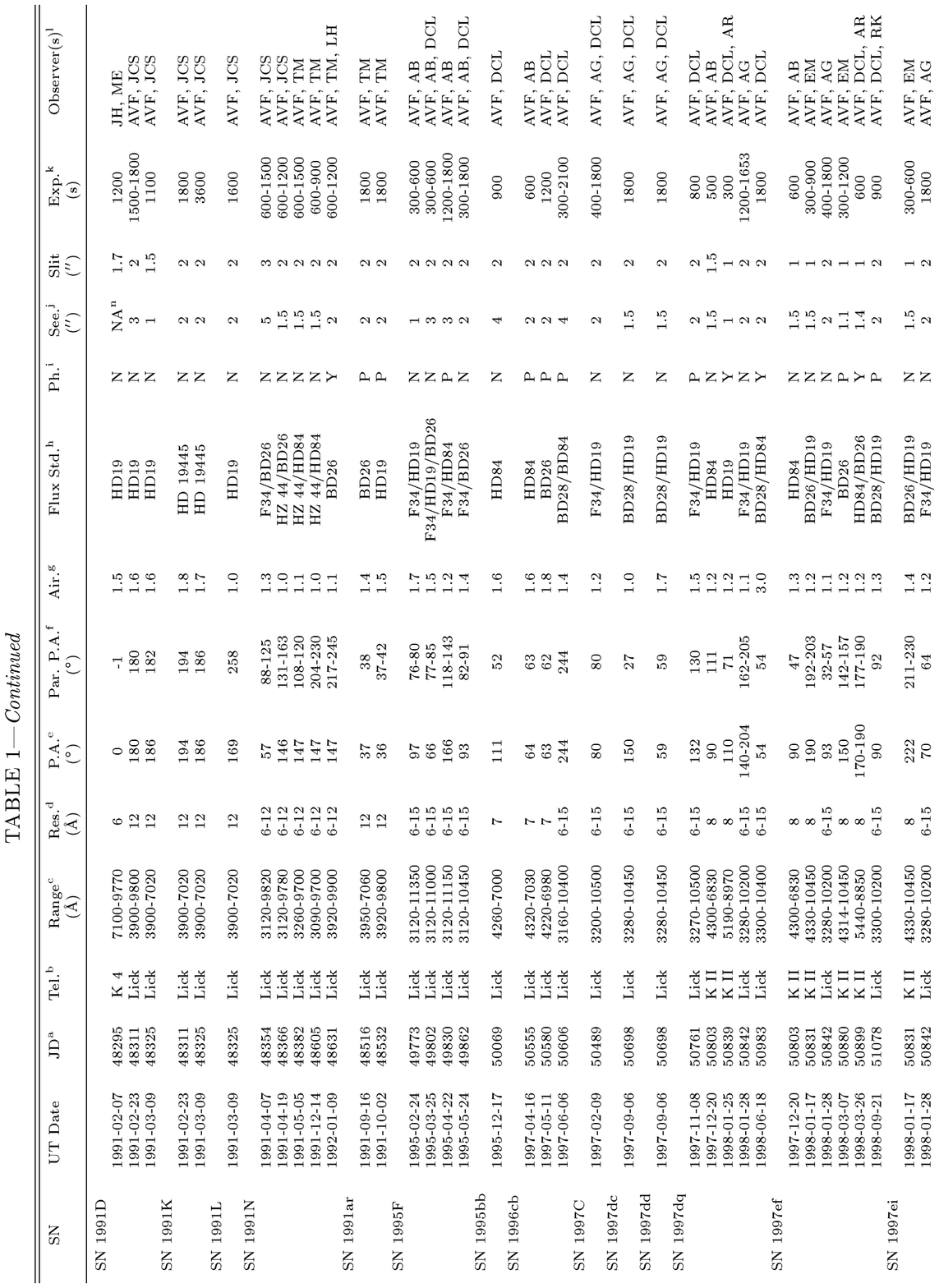}
        }

\end{figure}
\newpage
\begin{figure}[ht!]

\rotatebox{180}{
        \plotone{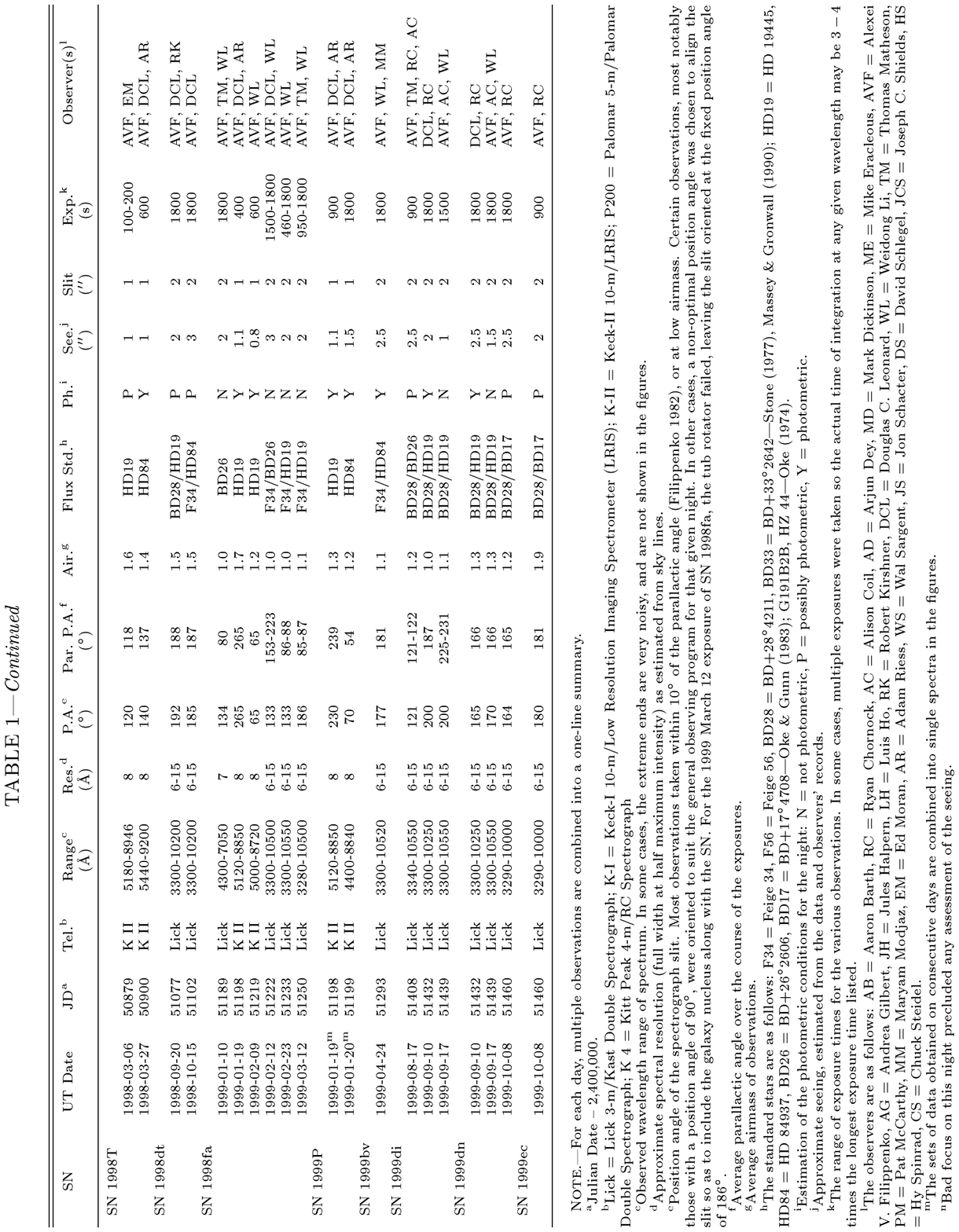}
}
\end{figure}
\newpage
\begin{figure}[ht!]

        \plotone{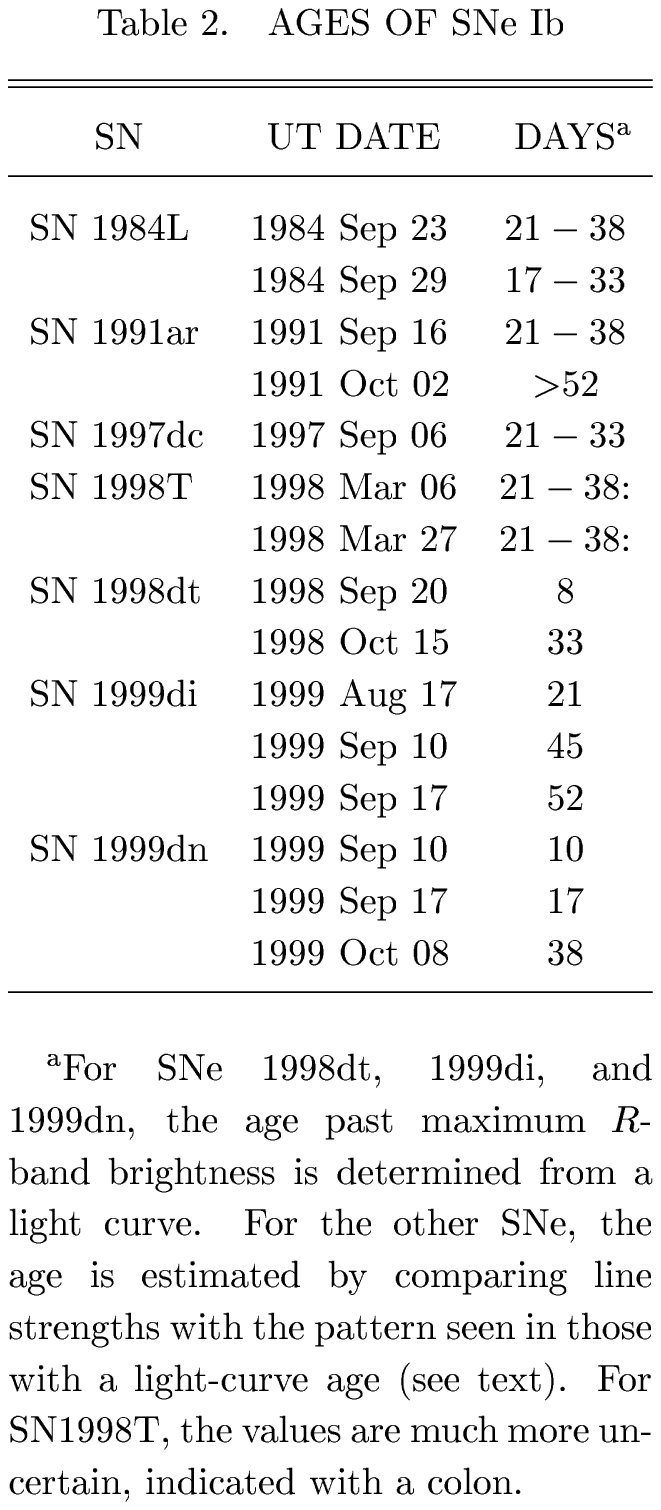}

\end{figure}
\newpage
\begin{figure}[ht!]

        \plotone{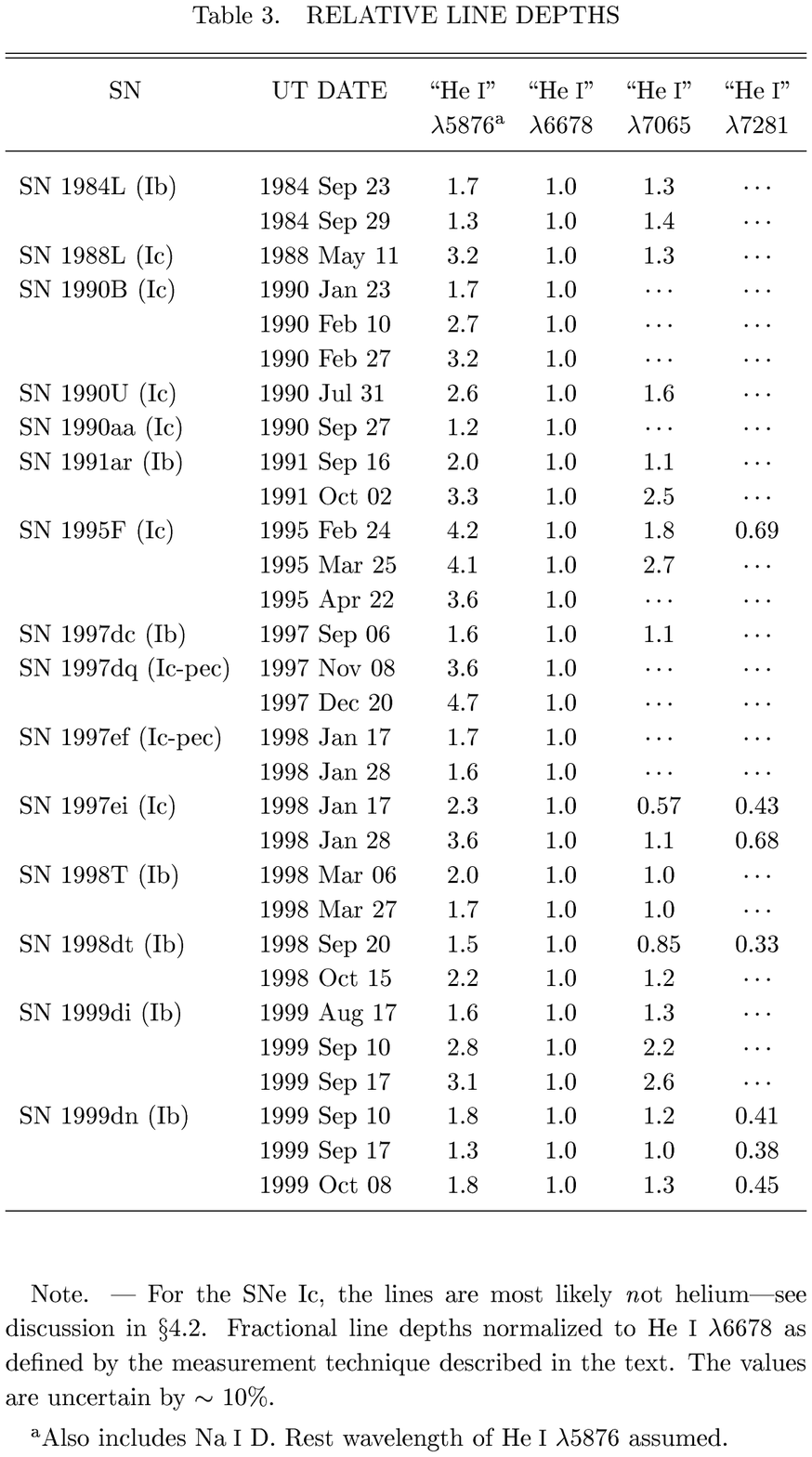}

\end{figure}
\newpage
\begin{figure}[ht!]

        \plotone{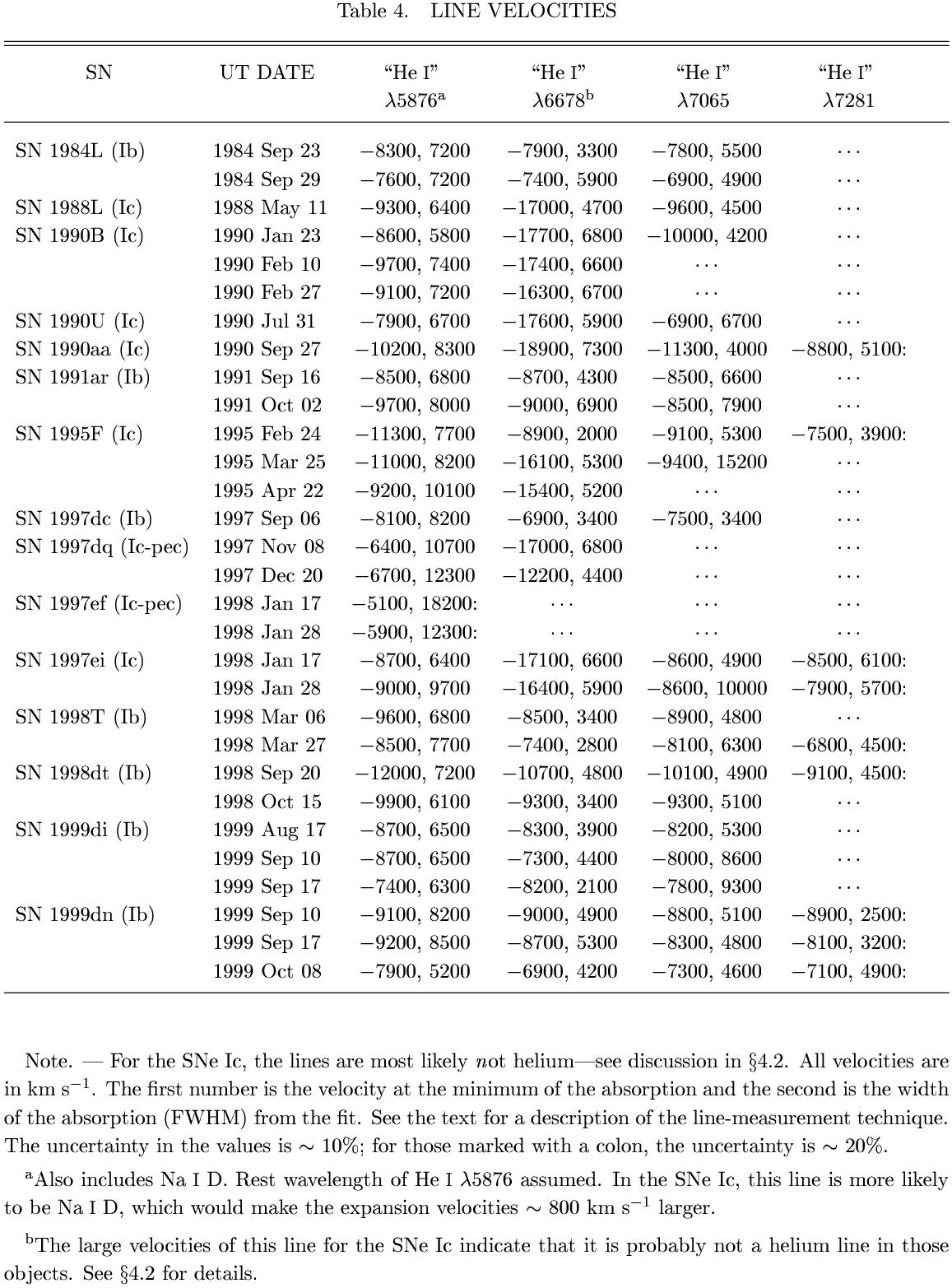}

\end{figure}
\newpage
\begin{figure}[ht!]

        \plotone{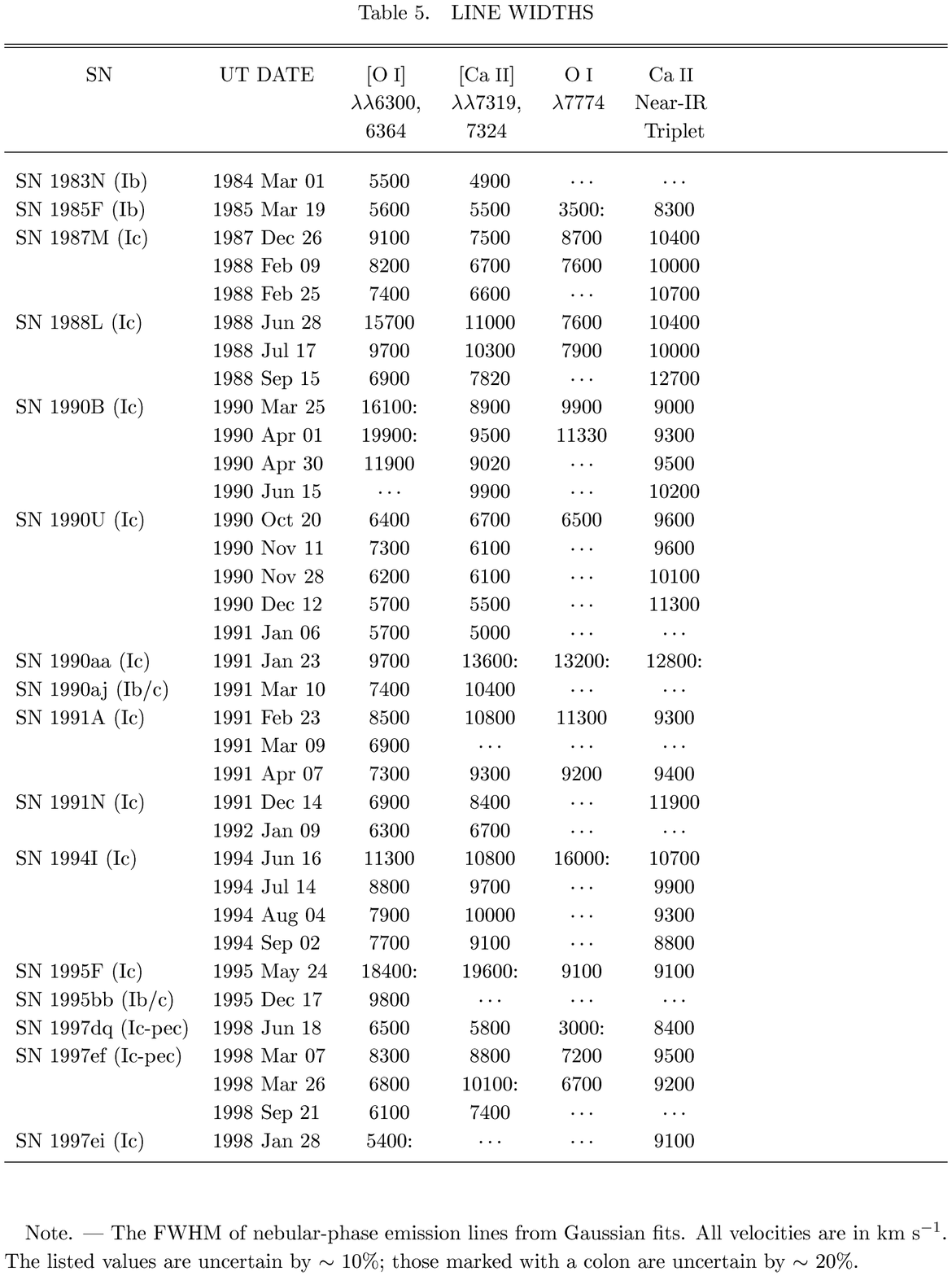}

\end{figure}
\newpage
\begin{figure}[ht!]

        \plotone{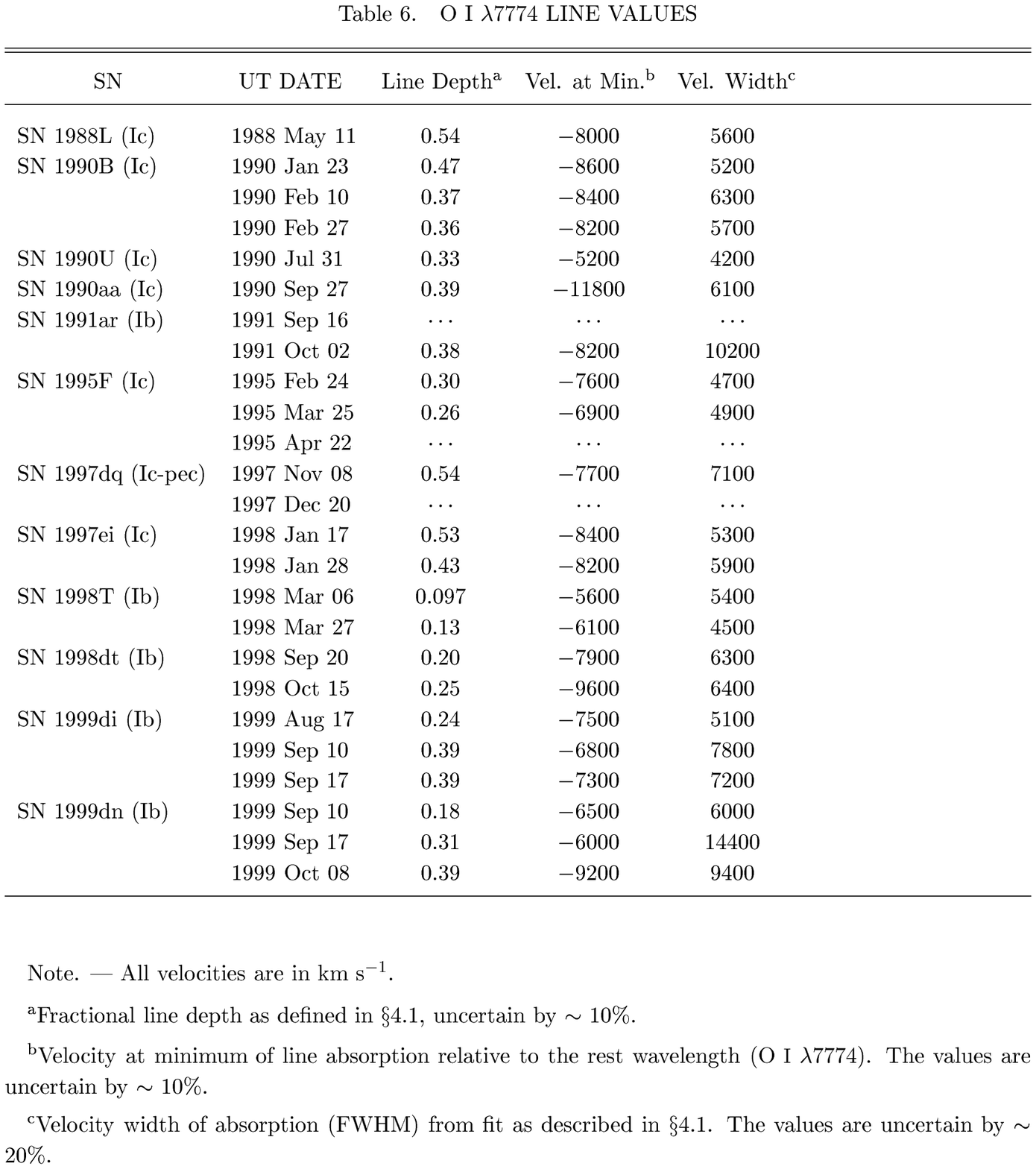}

\end{figure}
\clearpage

%% Generally speaking, only the figure captions, and not the figures
%% themselves, are included in electronic manuscript submissions.
%% Use \figcaption to format your figure captions. They should begin on a
%% new page.

\clearpage

%% No more than seven \figcaption commands are allowed per page,
%% so if you have more than seven captions, insert a \clearpage

%% There must be a \figcaption command for each legend. Key the text of the
%% legend and the optional \label in curly braces. If you wish, you may
%% include the name of the corresponding figure file in square brackets.
%% The label is for identification purposes only. It will not insert the
%% figures themselves into the document.
%% If you want to include your art in the paper, use \plotone.
%% Refer to the on-line documentation for details.

\clearpage
\newpage
\begin{figure}[ht!]
\plotone{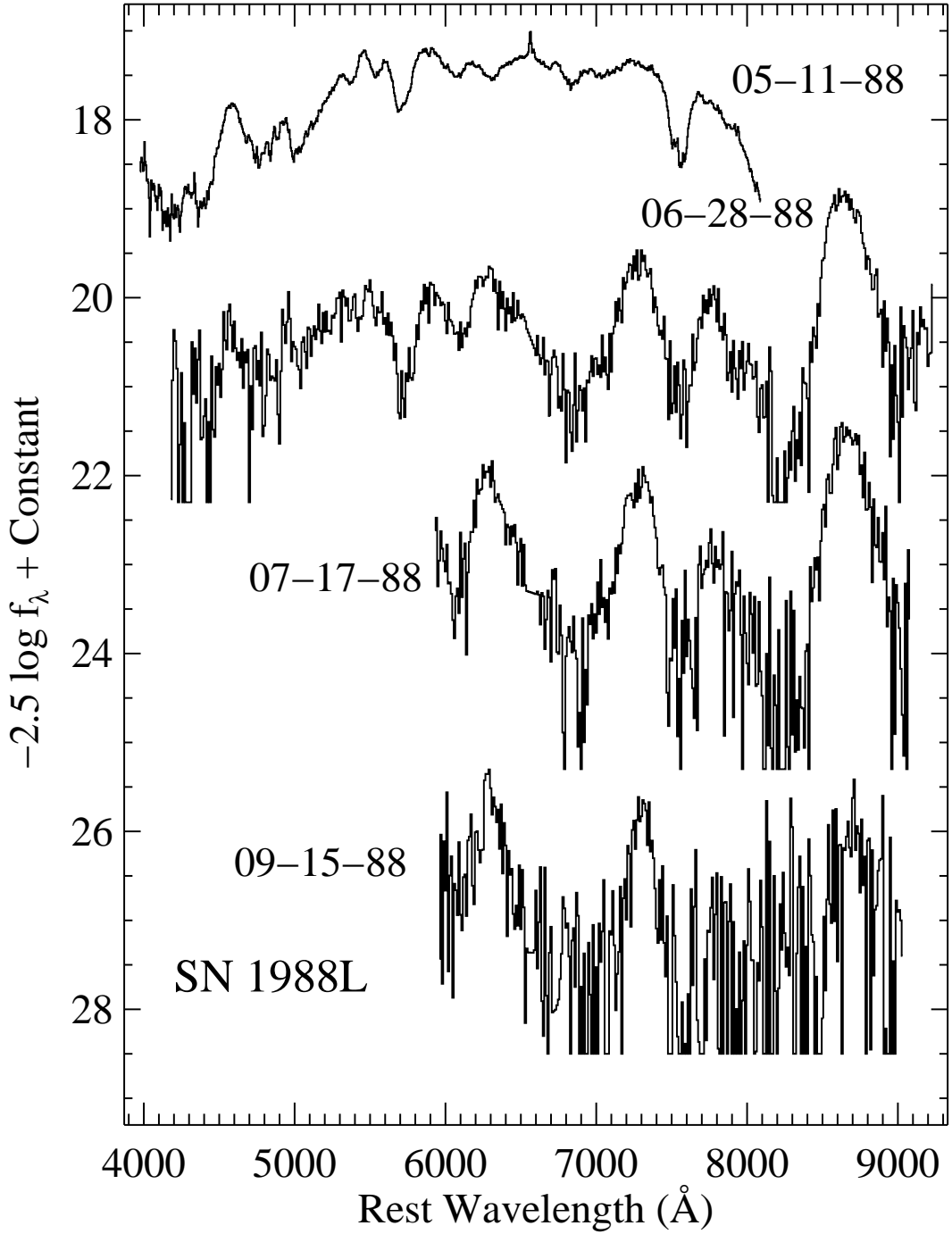}
\caption{Spectra of SN Ib/c 1988L; see also Filippenko (1988).  The
flux units are $-2.5$ log $f_{\lambda} - 21.10$, following the
definition of Space Telescope (ST) magnitudes (e.g., Koorneef et
al. 1986).  ST magnitudes are analogous to AB magnitudes ($-2.5$ log
$f_{\nu} - 48.60$; Oke \& Gunn 1983), with the zero point yielding
monochromatic magnitudes for Vega in the Johnson $B, V,$ and $R$
passbands of $\sim$ 0.  For clarity, the following constants have been
added to the individual spectra (from top to bottom): 0.0, 1.0, 3.0,
and 6.0.  The recession velocity of the SN has been removed as
described in the introduction to \S 3.\label{sn1988l-mont}}
\end{figure}
\clearpage
\begin{figure}[ht!]

        \plotone{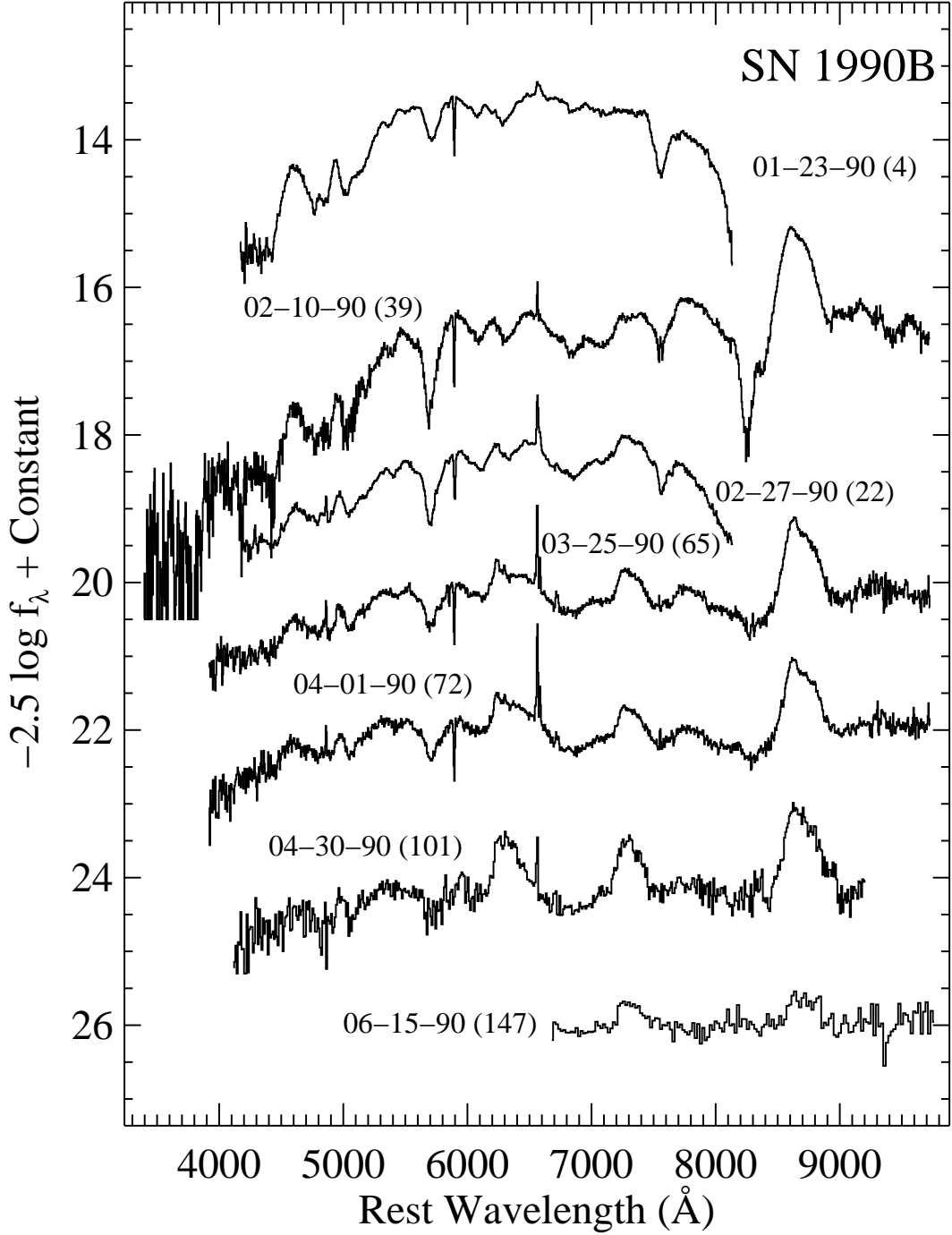}

\caption{Spectra of SN Ic 1990B with flux units as in Figure
\ref{sn1988l-mont}; see also Clocchiatti et al. (2001).  The following
constants have been added to the individual spectra (from top to
bottom): $-$2.0, $-$1.0, 1.5, 3.5, 4.5, 6.0, and 8.0.  The recession
velocity of the SN has been removed as described in the introduction
to \S 3.  The numbers after the date of observation indicate the
approximate number of days past $R$-band maximum brightness based upon
the light curve of Clocchiatti et al. (2001).\label{sn1990b-mont}}
\end{figure}
\clearpage
\begin{figure}[ht!]
\scalebox{0.9}{
        \plotone{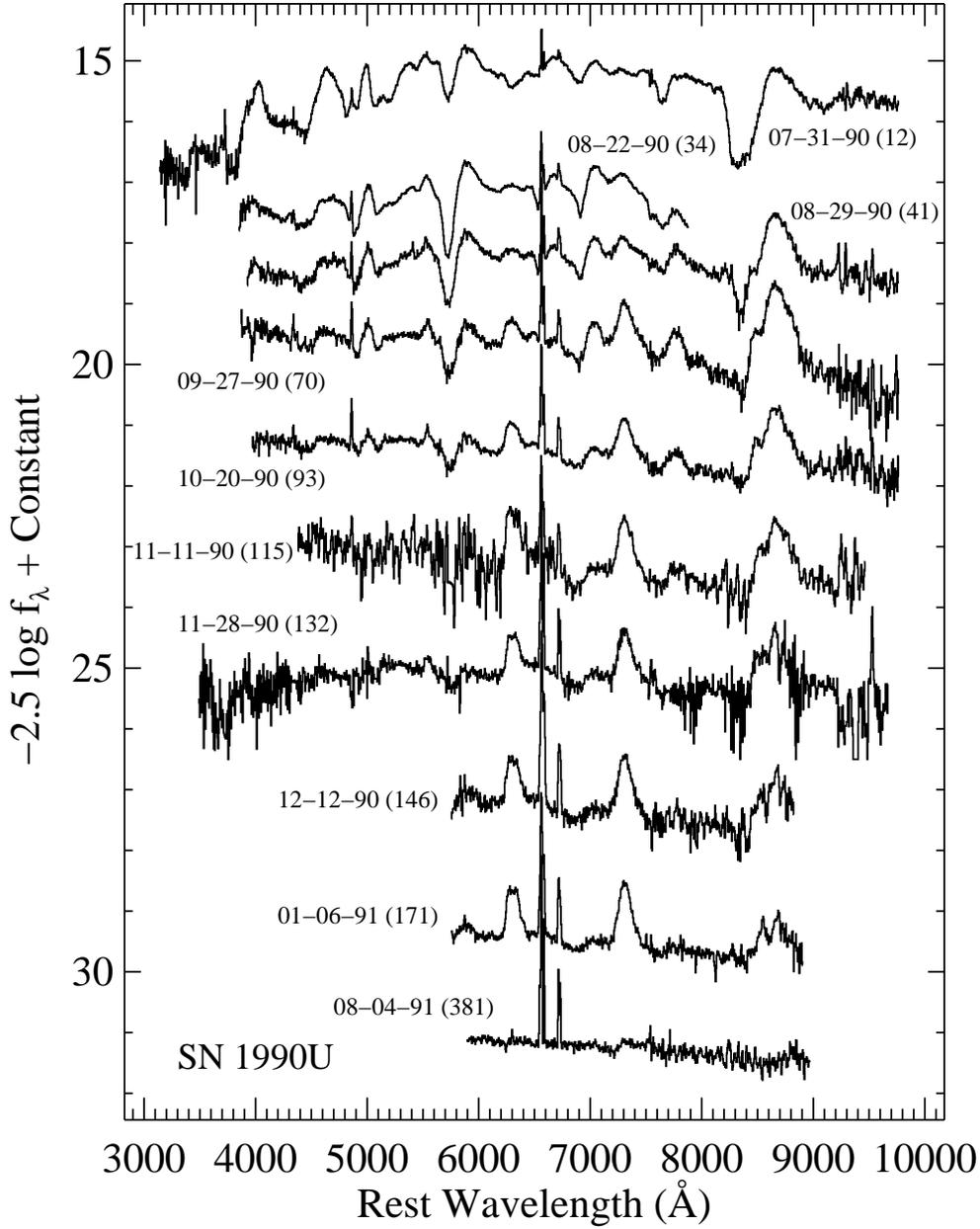}
}
\caption{Spectra of SN Ic 1990U with flux units as in Figure
\ref{sn1988l-mont}.  The following constants have been added to the
individual spectra (from top to bottom): $-$1.0, 0.0, 0.5, 1.5, 4.0,
4.0, 7.0, 9.0, 11.0, and 12.5.  The recession velocity of the SN has
been removed as described in the introduction to \S 3.  The numbers
after the date of observation indicate the approximate number of days
past $R$-band maximum brightness based upon the light curve of
Richmond, Filippenko, \& Galisky (1998).  As the light curve for SN
1990U does not contain points near maximum brightness, it was
temporally matched with the light curve for SN 1990B (Clocchiatti et
al. 2001) to estimate the date of maximum.\label{sn1990u-mont}}
\end{figure}
\clearpage
\begin{figure}[ht!]

        \plotone{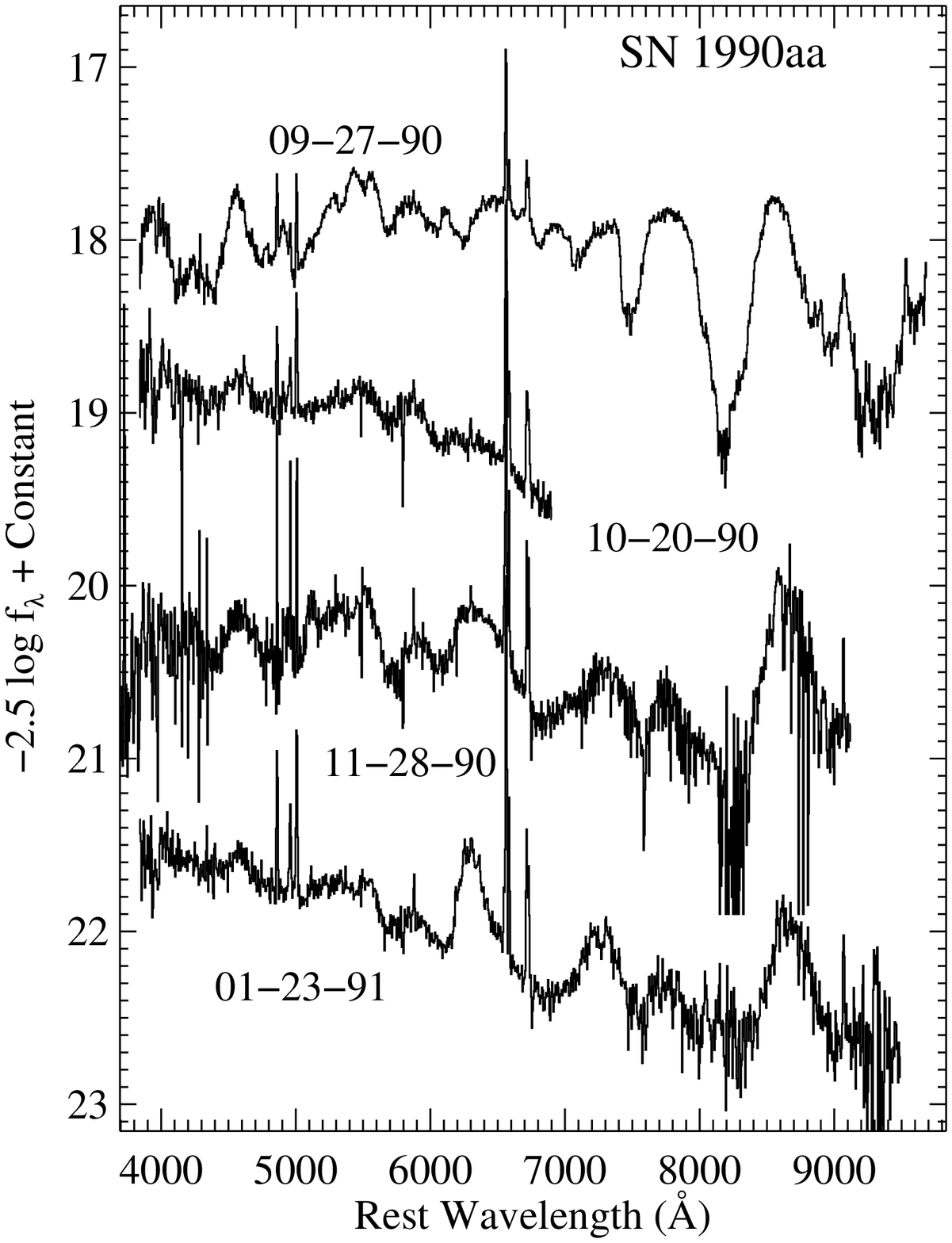}

\caption{Spectra of SN Ic 1990aa with flux units as in Figure
\ref{sn1988l-mont}; see also Filippenko (1992).  The following
constants have been added to the individual spectra (from top to
bottom): 0.0, 1.5, 2.5, and 4.0.  The recession velocity of the SN has
been removed as described in the introduction to \S
3.\label{sn1990aa-mont}}
\end{figure}
\clearpage
\begin{figure}[ht!]
\rotatebox{180}{
        \plotone{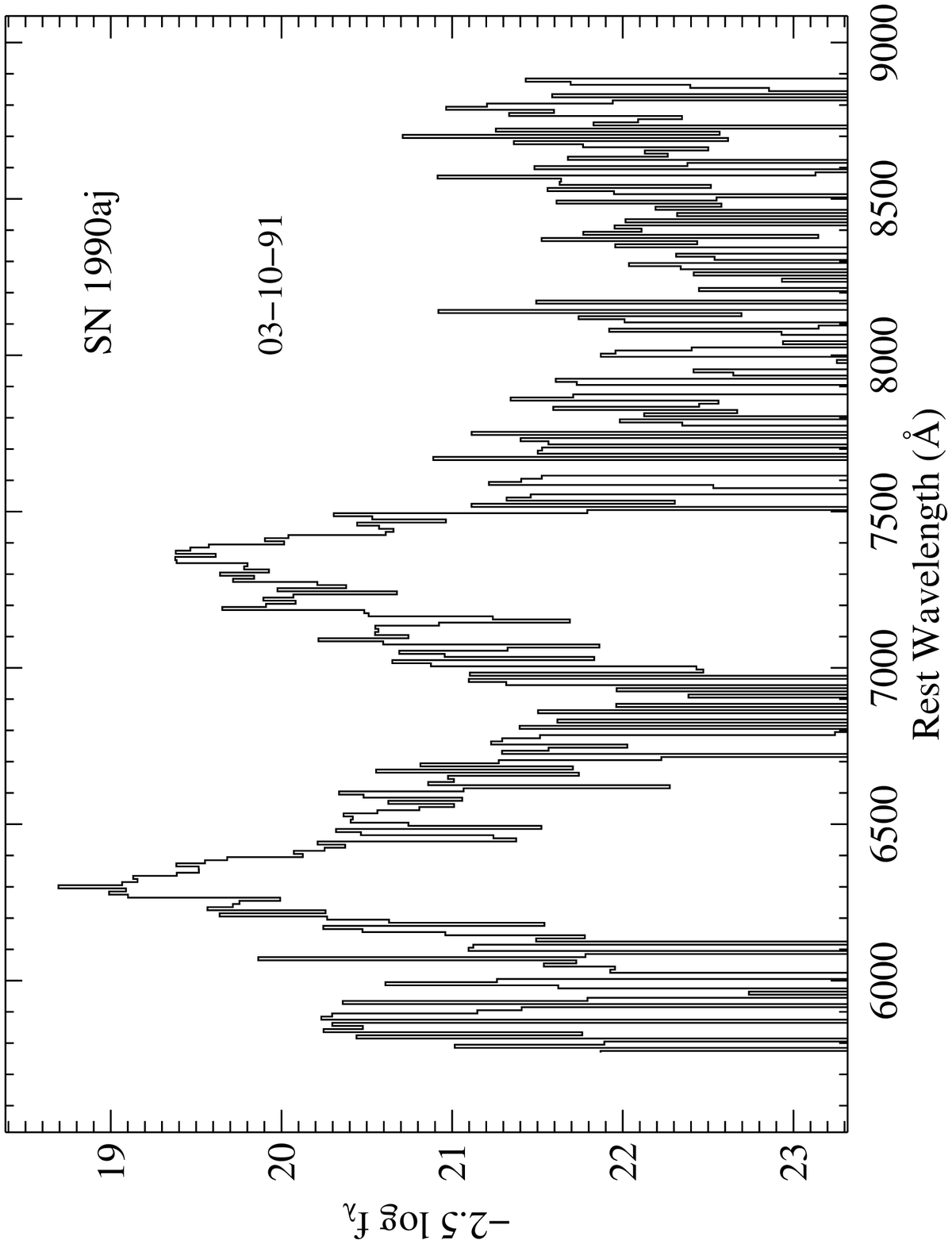}
 }
\caption{Spectrum of SN Ib/c 1990aj with flux units as in Figure
\ref{sn1988l-mont}.  The recession velocity of the SN has been removed
as described in the introduction to \S 3.\label{sn1990aj-mont}}
\end{figure}
\clearpage
\begin{figure}[ht!]

        \plotone{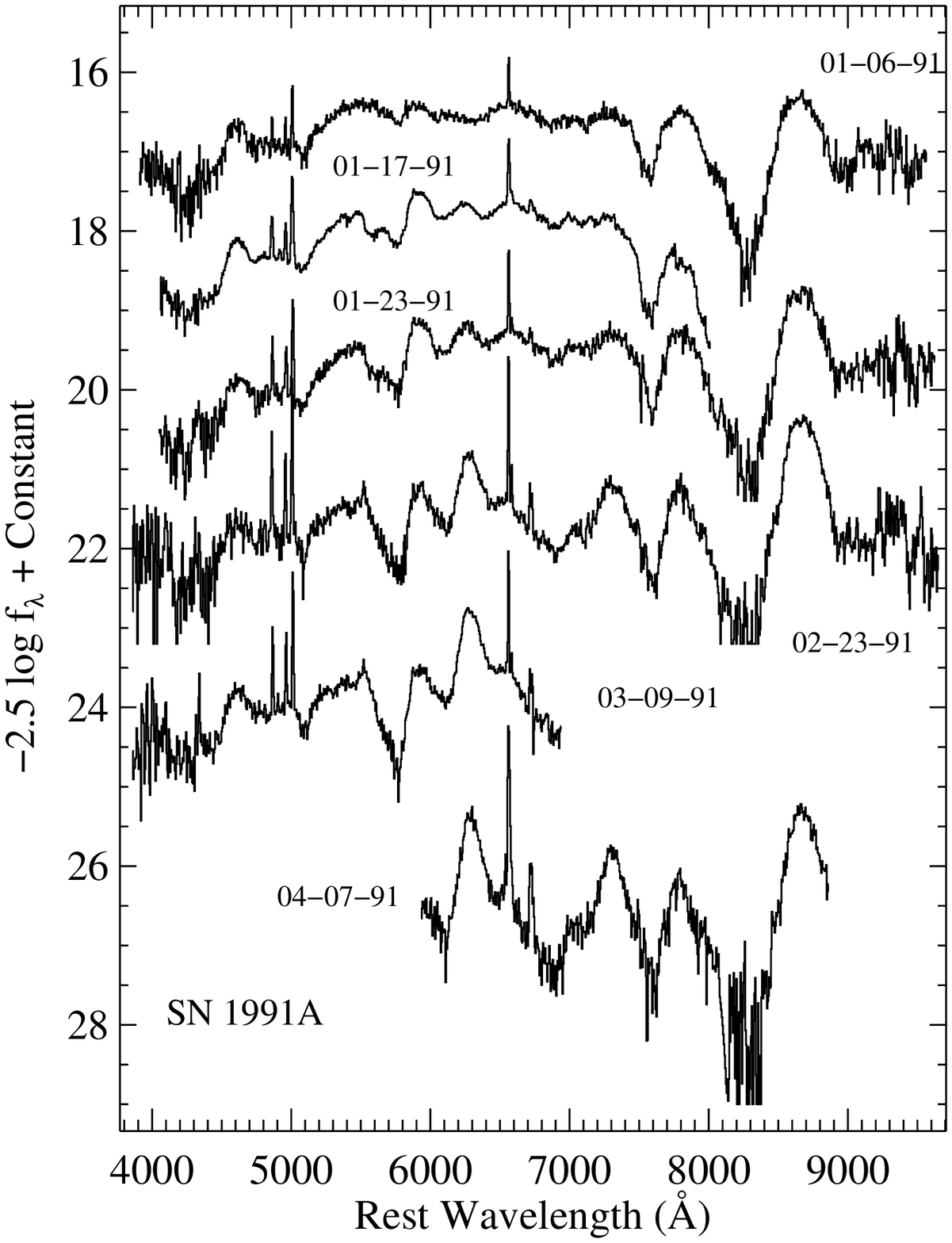}

\caption{Spectra of SN Ic 1991A with flux units as in Figure
\ref{sn1988l-mont}; see also Filippenko (1992).  The following
constants have been added to the individual spectra (from top to
bottom): $-$1.0, 0.5, 2.0, 3.5, 5.5, and 8.0.  The recession velocity
of the SN has been removed as described in the introduction to \S
3.\label{sn1991a-mont}}
\end{figure}
\clearpage
\begin{figure}[ht!]

        \plotone{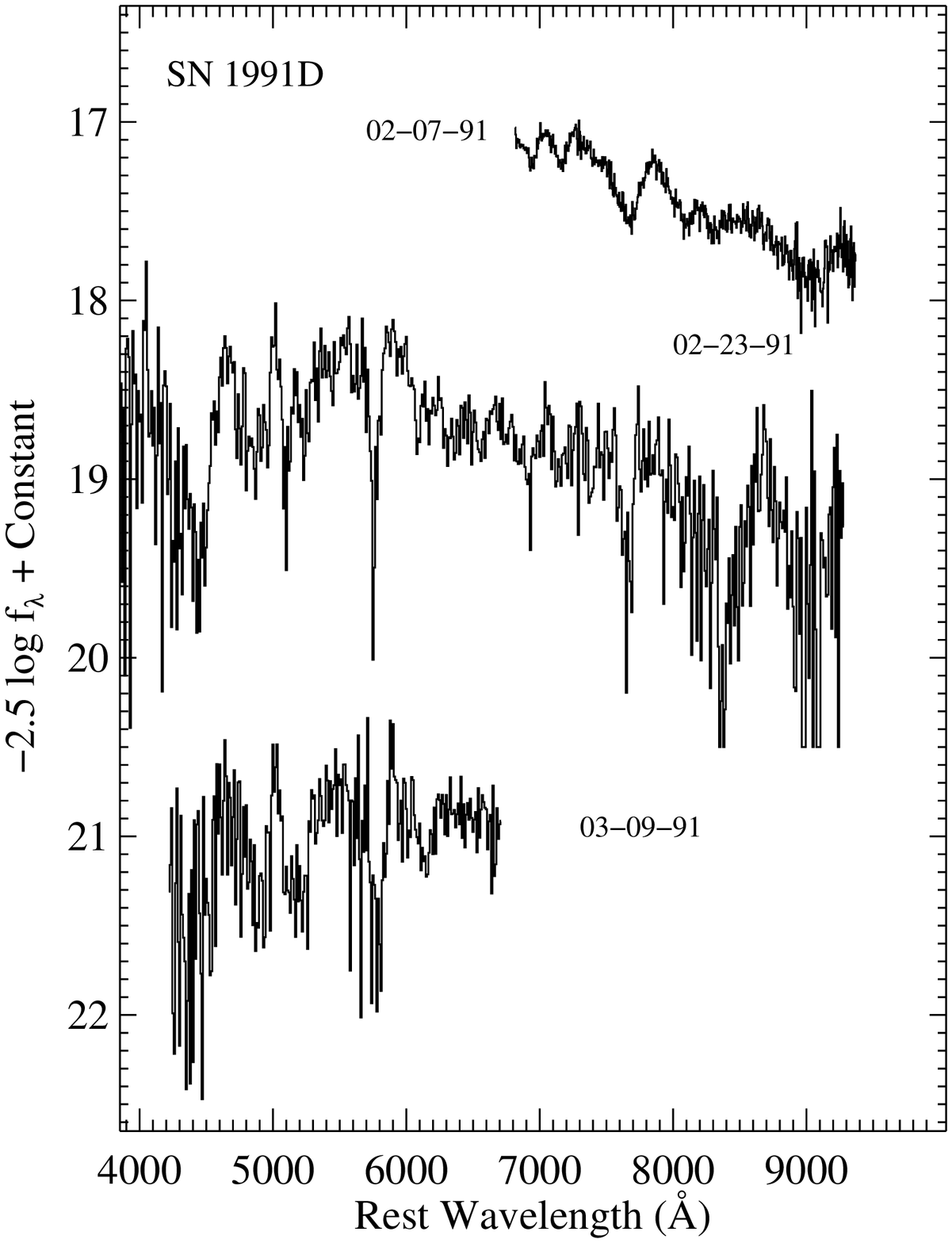}

\caption{Spectra of SN Ib 1991D with flux units as in Figure
\ref{sn1988l-mont}.  The following constants have been added to the
individual spectra (from top to bottom): $-$1.0, 0.0, and 1.0.  The
recession velocity of the SN has been removed as described in the
introduction to \S 3.\label{sn1991d-mont}}
\end{figure}
\clearpage
\begin{figure}[ht!]
\scalebox{0.9}{
        \plotone{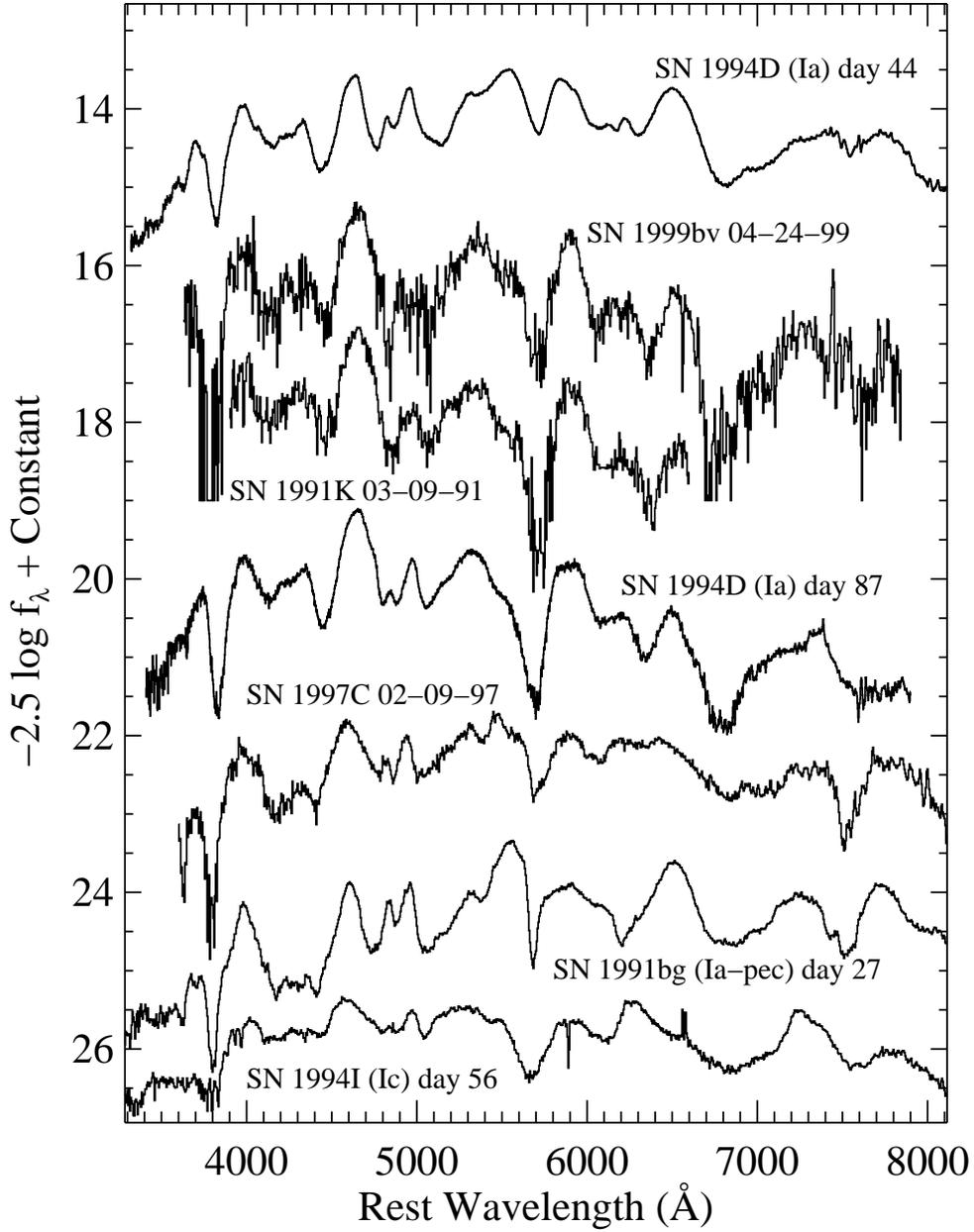}
}

\caption{Spectra of SN 1991K, SN 1997C, SN 1999bv, the Type Ia SN
1994D, the peculiar Type Ia SN 1991bg (Filippenko et al. 1992b), and
the Type Ic SN 1994I with flux units as in Figure \ref{sn1988l-mont},
offset by arbitrary amounts.  The recession velocities of the SNe have
been removed as described in the introduction to \S 3.  Note the
similarity of SN 1991K and SN 1999bv to SN 1994D, especially the day
87 spectrum.  They do not seem to match the spectrum of SN 1994I (at
any phase, cf. Filippenko et al. 1995).  SN 1997C is also similar to
SN 1994D, but shares more characteristics with the spectrum of the
sub-luminous Type Ia SN 1991bg.  We believe this indicates that SN
1991K, SN 1997C, and SN 1999bv were originally misclassified and were
actually SNe Ia caught past maximum.  Without spectra near maximum
brightness, though, there is still some uncertainty.\label{comp-mont}}
\end{figure}
\clearpage
\begin{figure}[ht!]
\rotatebox{180}{
        \plotone{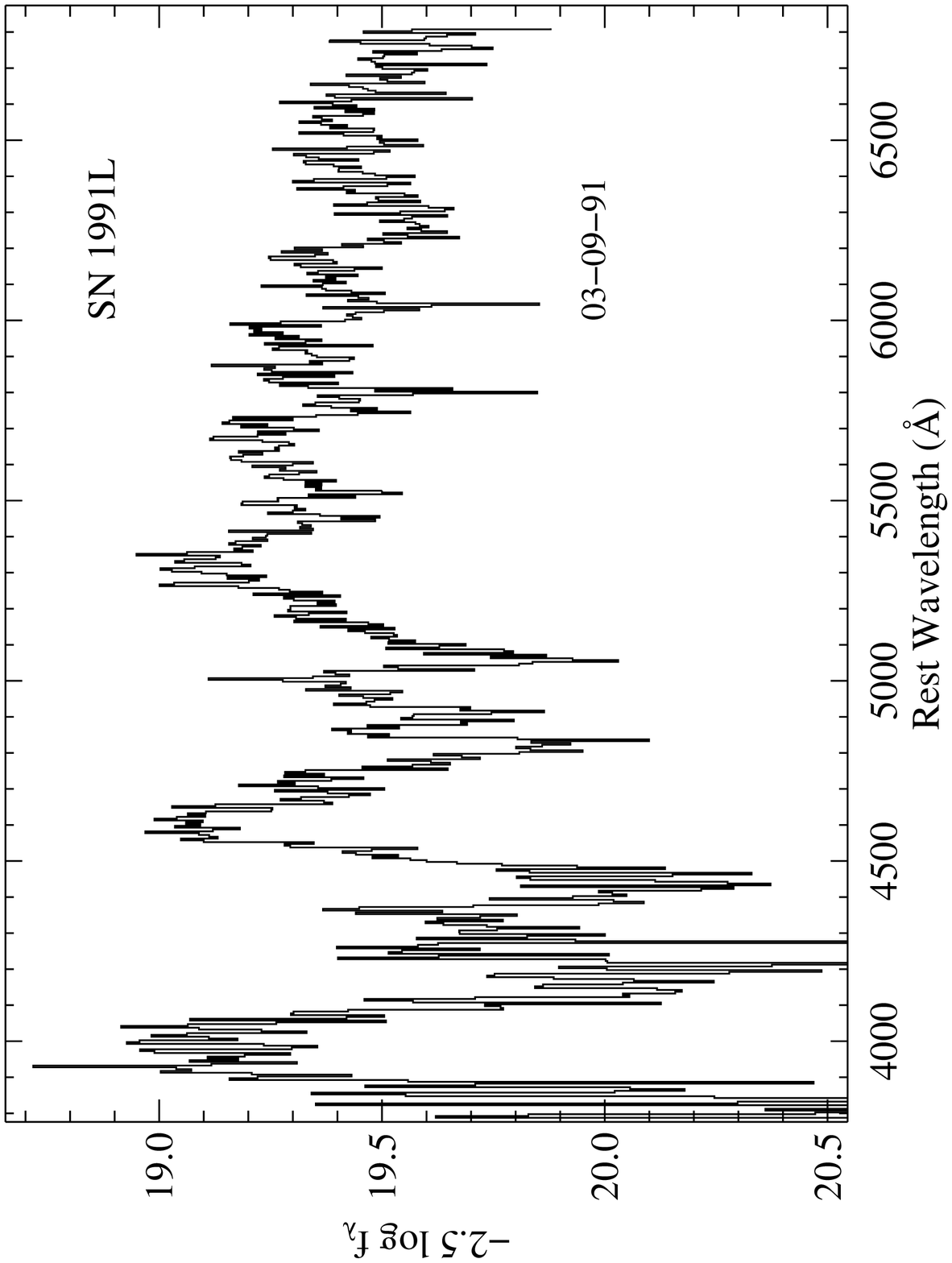}
}
\caption{Spectrum of SN Ib/c 1991L with flux units as in Figure
\ref{sn1988l-mont}.  The recession velocity of the SN has been removed
as described in the introduction to \S 3.\label{sn1991l-mont}}
\end{figure}
\clearpage
\begin{figure}[ht!]

        \plotone{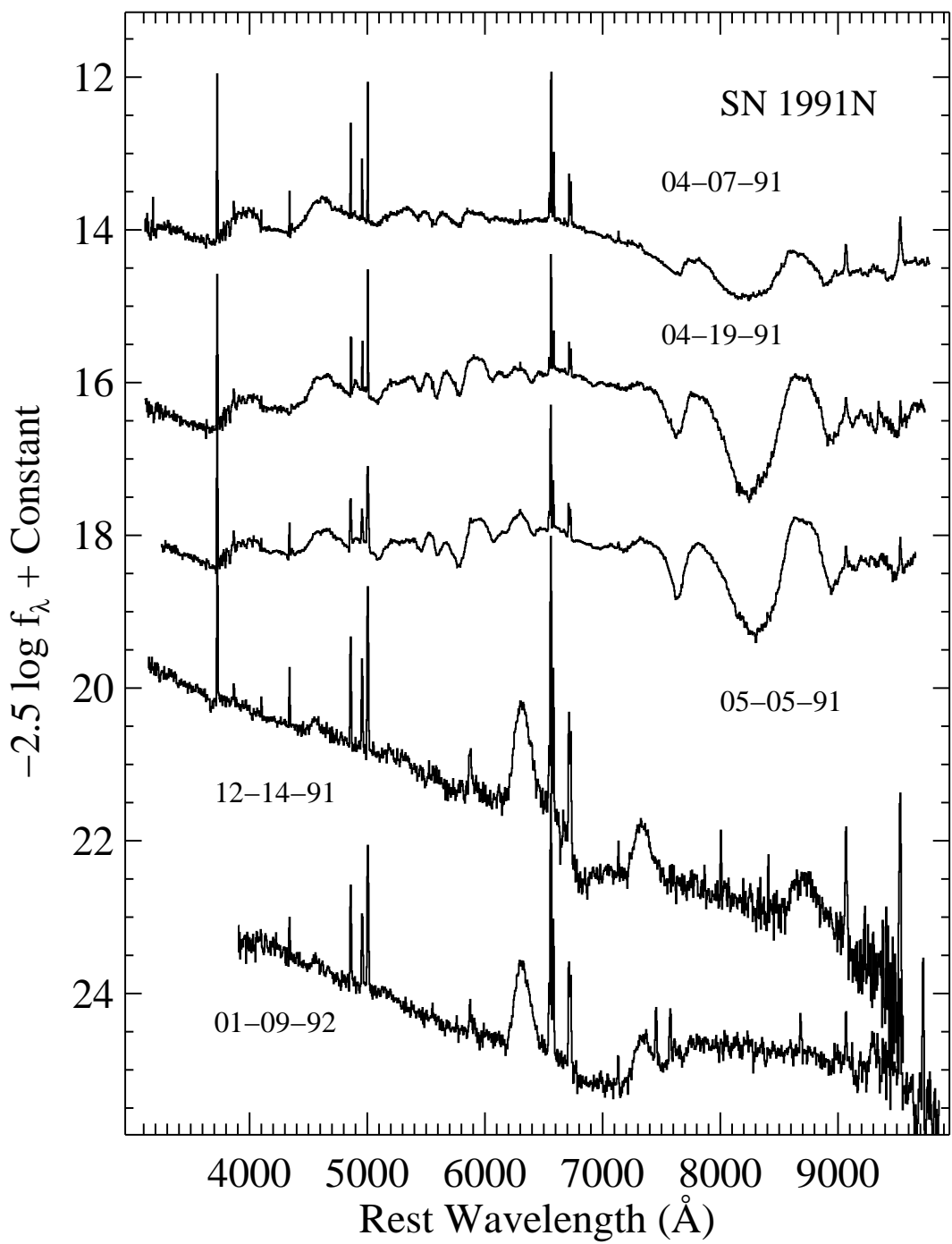}

\caption{Spectra of SN Ic 1991N with flux units as in Figure
\ref{sn1988l-mont}.  The continuum is heavily contaminated by
superposed stars; the strong, narrow emission lines are from a
superposed \ion{H}{2} region.  An apparent calibration error affects
the red half of the 1992 January 9 spectrum.  The following constants
have been added to the individual spectra (from top to bottom): 0.0,
1.5, 3.0, 5.5, and 8.0.  The recession velocity of the SN has been
removed as described in the introduction to \S 3.\label{sn1991n-mont}}
\end{figure}
\clearpage
\begin{figure}[ht!]

        \plotone{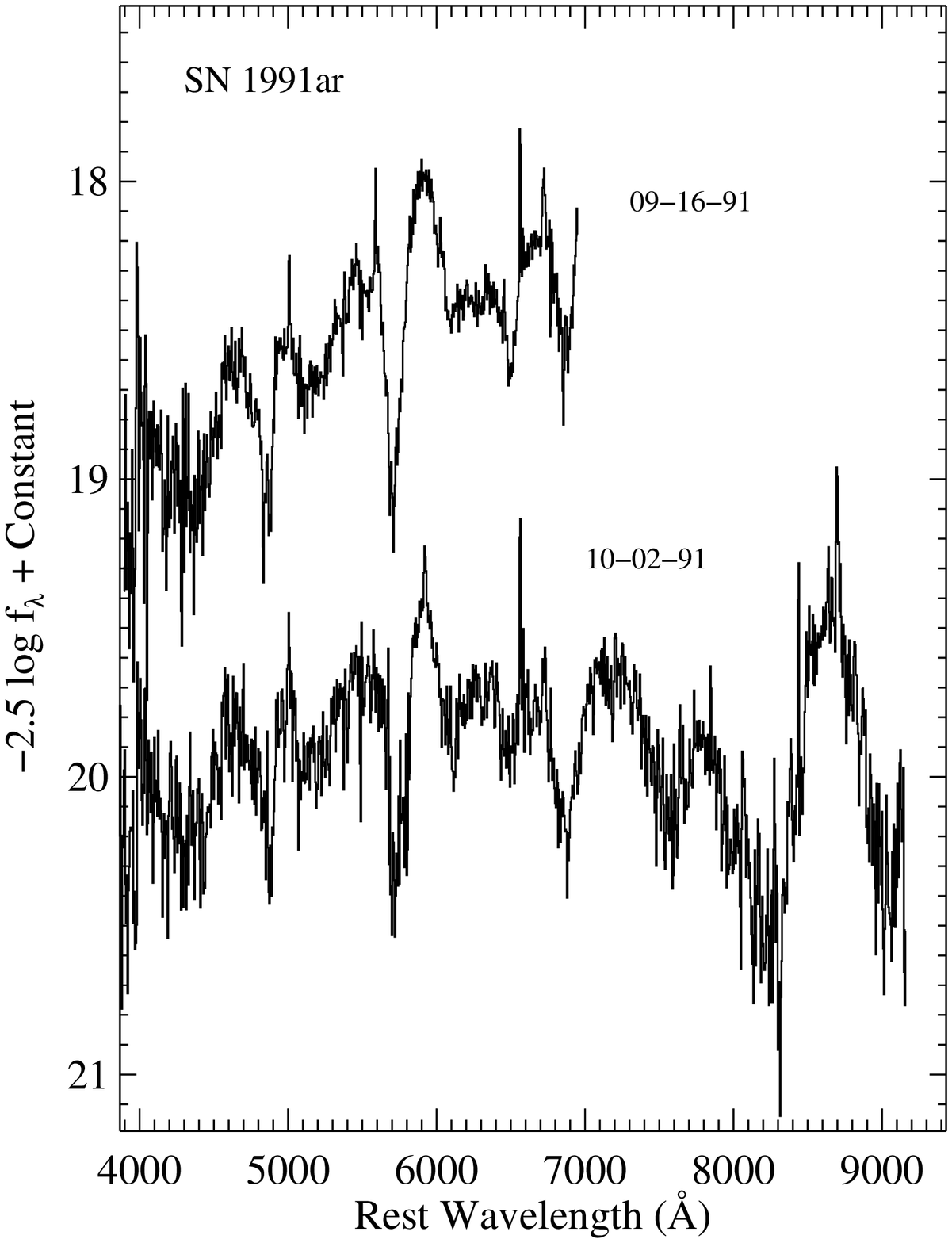}

\caption{Spectra of SN Ib 1991ar with flux units as in Figure
\ref{sn1988l-mont}.  The following constants have been added to the
individual spectra (from top to bottom): 0.0 and 1.0.  The recession
velocity of the SN has been removed as described in the introduction
to \S 3.\label{sn1991ar-mont}}
\end{figure}
\clearpage
\begin{figure}[ht!]

        \plotone{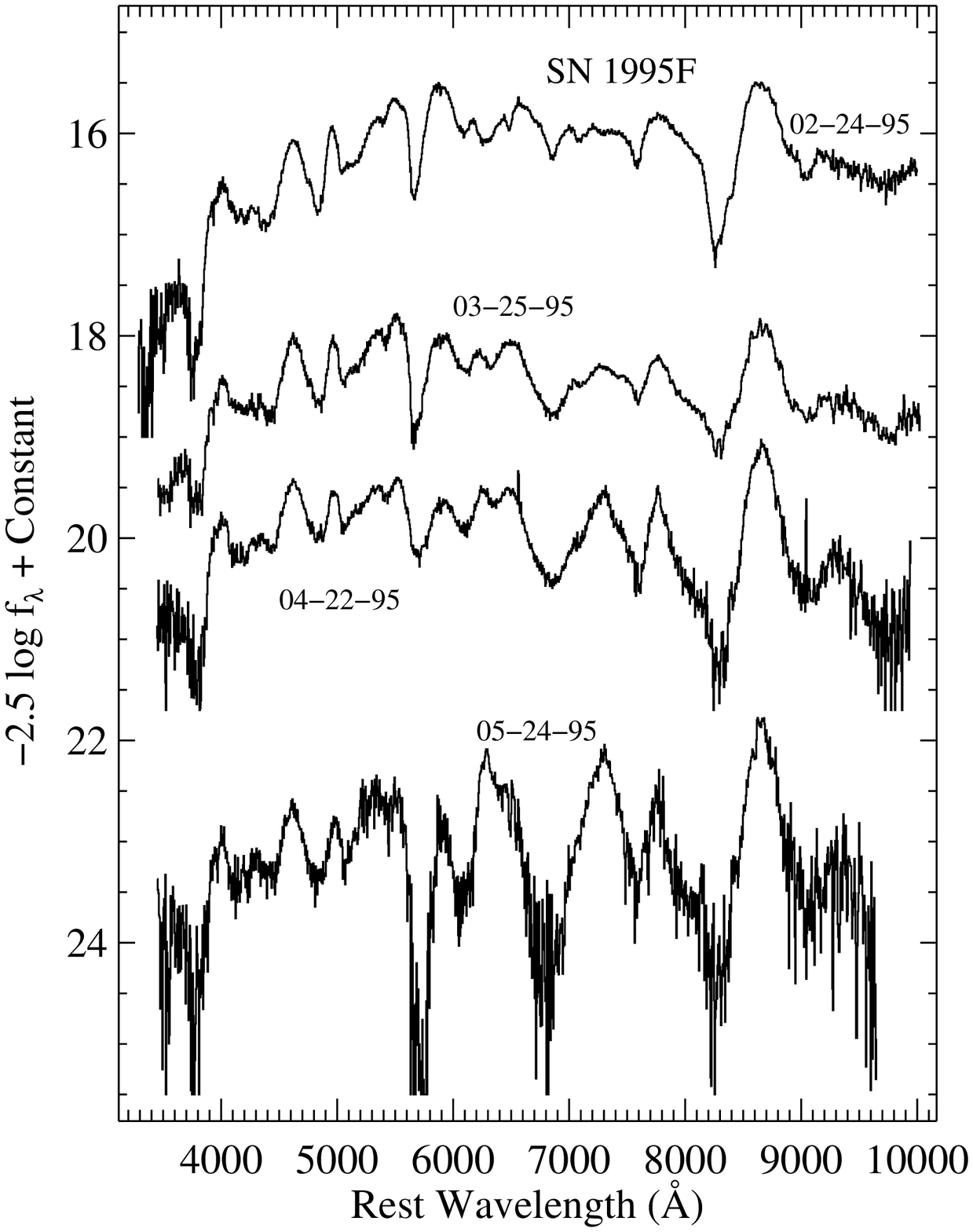}

\caption{Spectra of SN Ic 1995F with flux units as in Figure
\ref{sn1988l-mont}.  The following constants have been added to the
individual spectra (from top to bottom): 0.0, 1.8, 3.0, and 5.0.  The
recession velocity of the SN has been removed as described in the
introduction to \S 3.\label{sn1995f-mont}}
\end{figure}
\clearpage
\begin{figure}[ht!]
\rotatebox{180}{
        \plotone{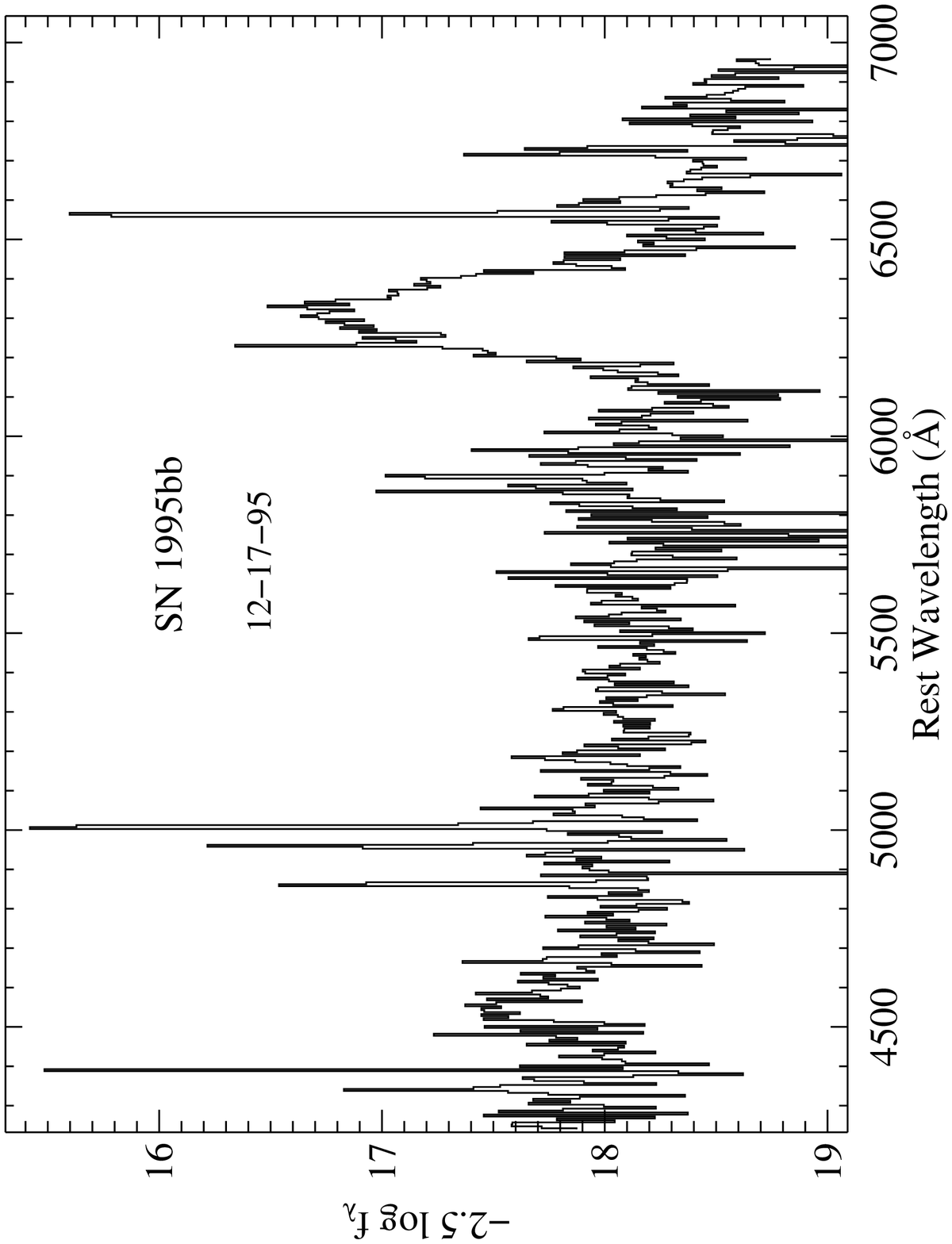}
}

\caption{Spectrum of SN Ib/c 1995bb with flux units as in Figure
\ref{sn1988l-mont}.  The recession velocity of the SN has been removed
as described in the introduction to \S 3.\label{sn1995bb-mont}}
\end{figure}
\clearpage
\begin{figure}[ht!]

        \plotone{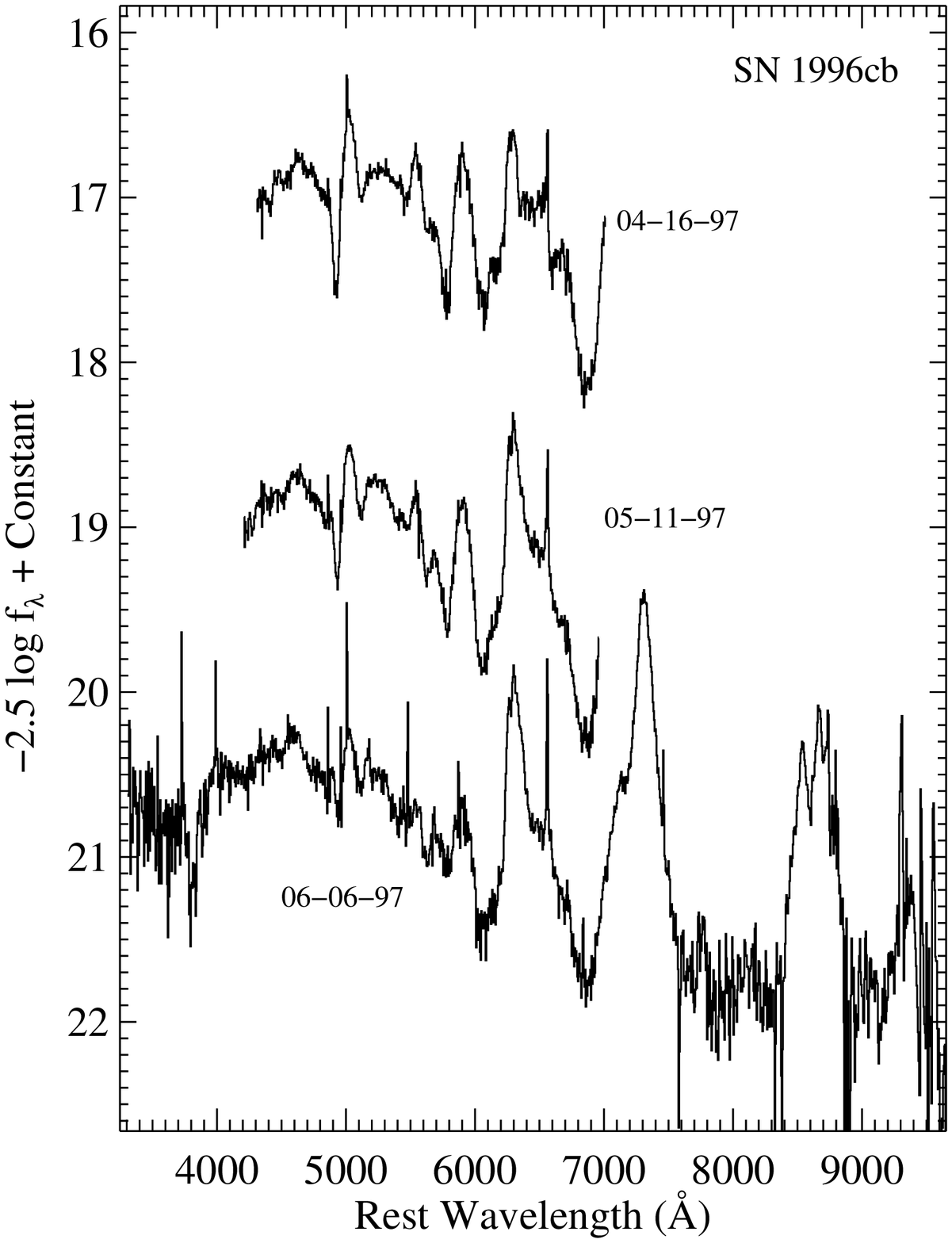}

\caption{Spectra of SN IIb 1996cb with flux units as in Figure
\ref{sn1988l-mont}.  The following constants have been added to the
individual spectra (from top to bottom): 0.0, 1.3, and 2.5.  The
recession velocity of the SN has been removed as described in the
introduction to \S 3.\label{sn1996cb-mont}}
\end{figure}
\clearpage
\begin{figure}[ht!]
\rotatebox{180}{
        \plotone{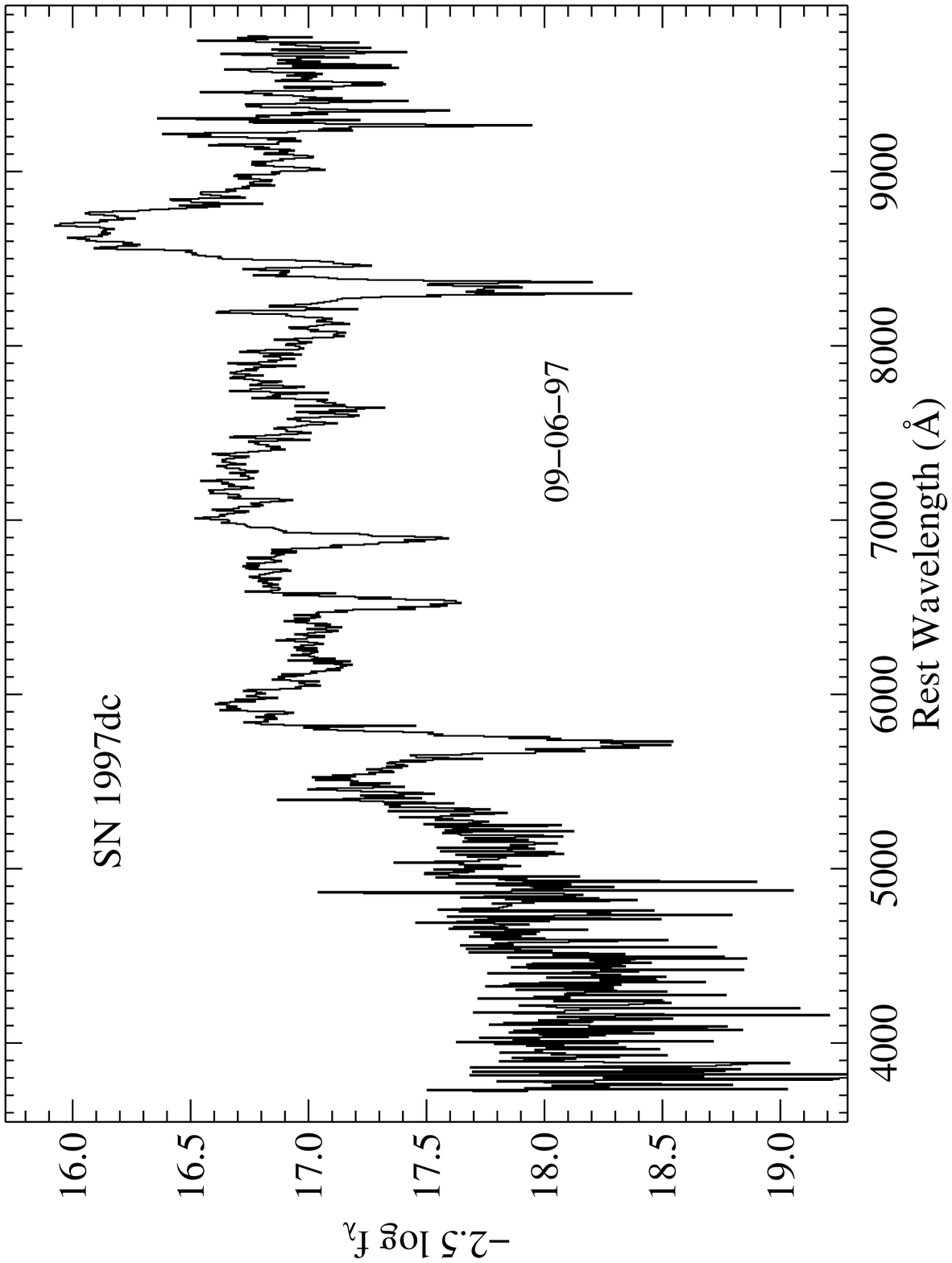}
}

\caption{Spectrum of SN Ib 1997dc with flux units as in Figure
\ref{sn1988l-mont}.  The recession velocity of the SN has been removed
as described in the introduction to \S 3.\label{sn1997dc-mont}}
\end{figure}
\clearpage
\begin{figure}[ht!]
\rotatebox{180}{
        \plotone{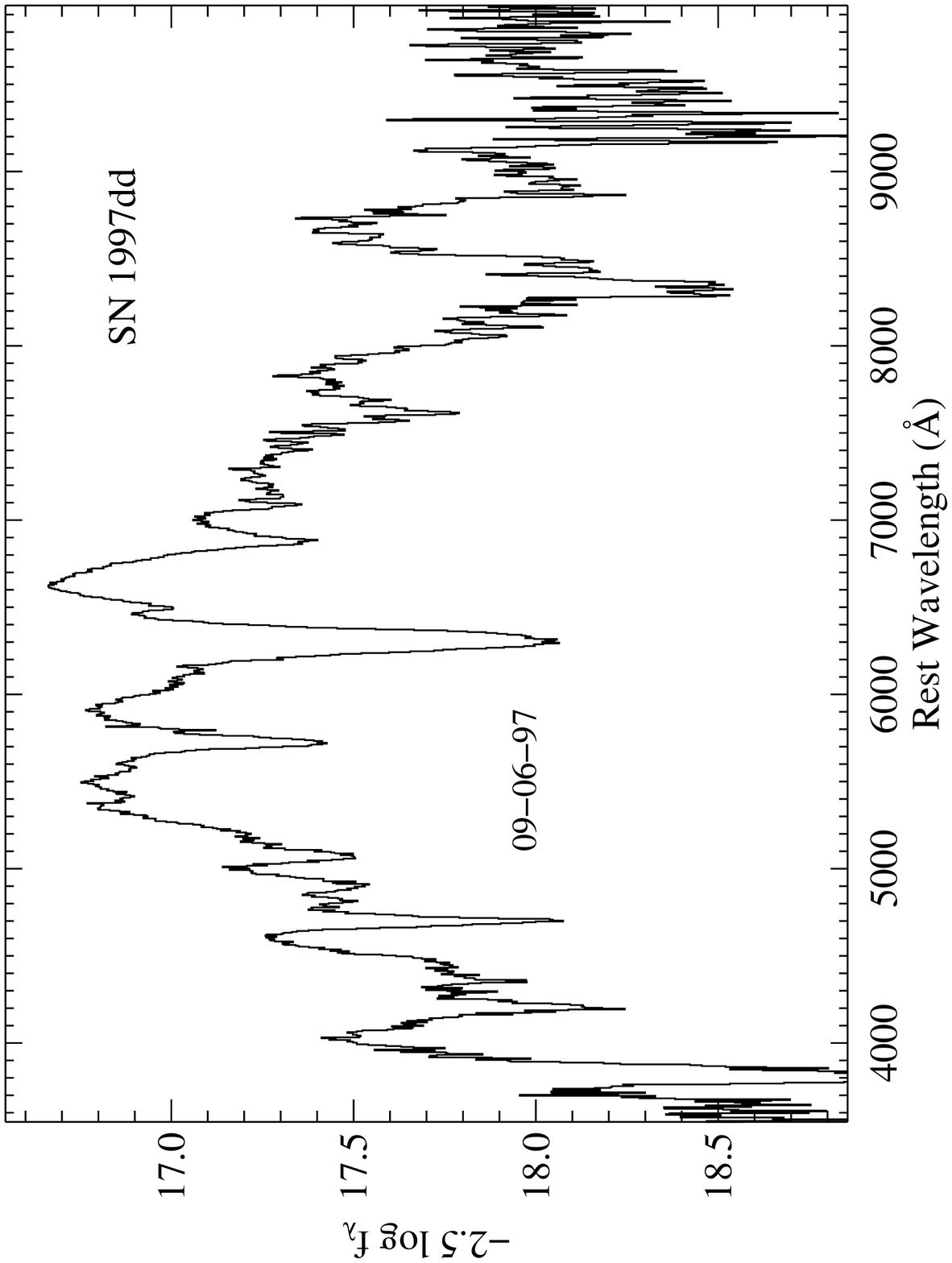}
}

\caption{Spectrum of SN IIb 1997dd with flux units as in Figure
\ref{sn1988l-mont}.  The recession velocity of the SN has been removed
as described in the introduction to \S 3.\label{sn1997dd-mont}}
\end{figure}
\clearpage
\begin{figure}[ht!]

        \plotone{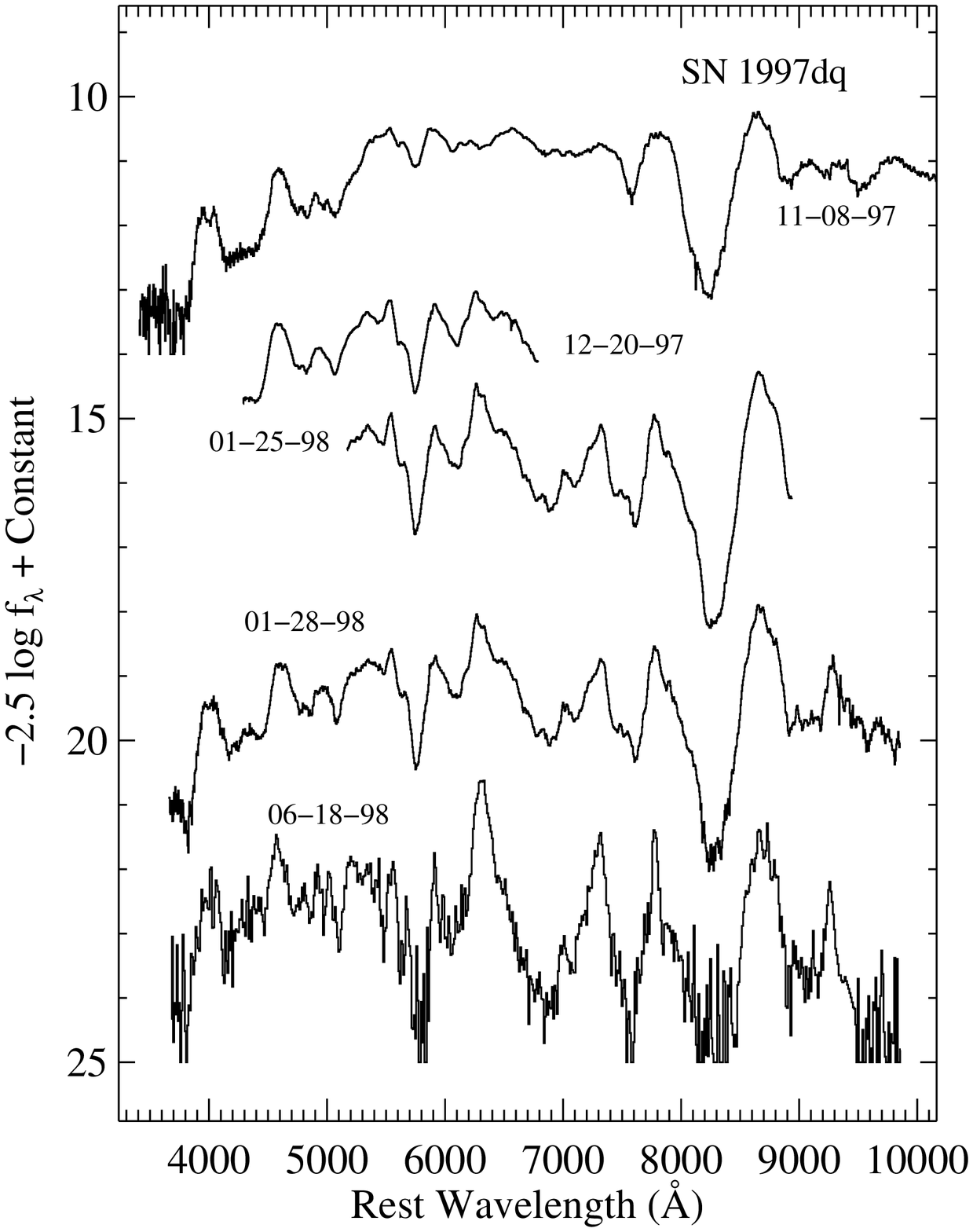}

\caption{Spectra of SN Ic 1997dq with flux units as in Figure
\ref{sn1988l-mont}.  The following constants have been added to the
individual spectra (from top to bottom): $-$4.0, $-$2.0, 8.0, 1.8, and
3.0.  The recession velocity of the SN has been removed as described
in the introduction to \S 3.\label{sn1997dq-mont}}
\end{figure}
\clearpage
\begin{figure}[ht!]

        \plotone{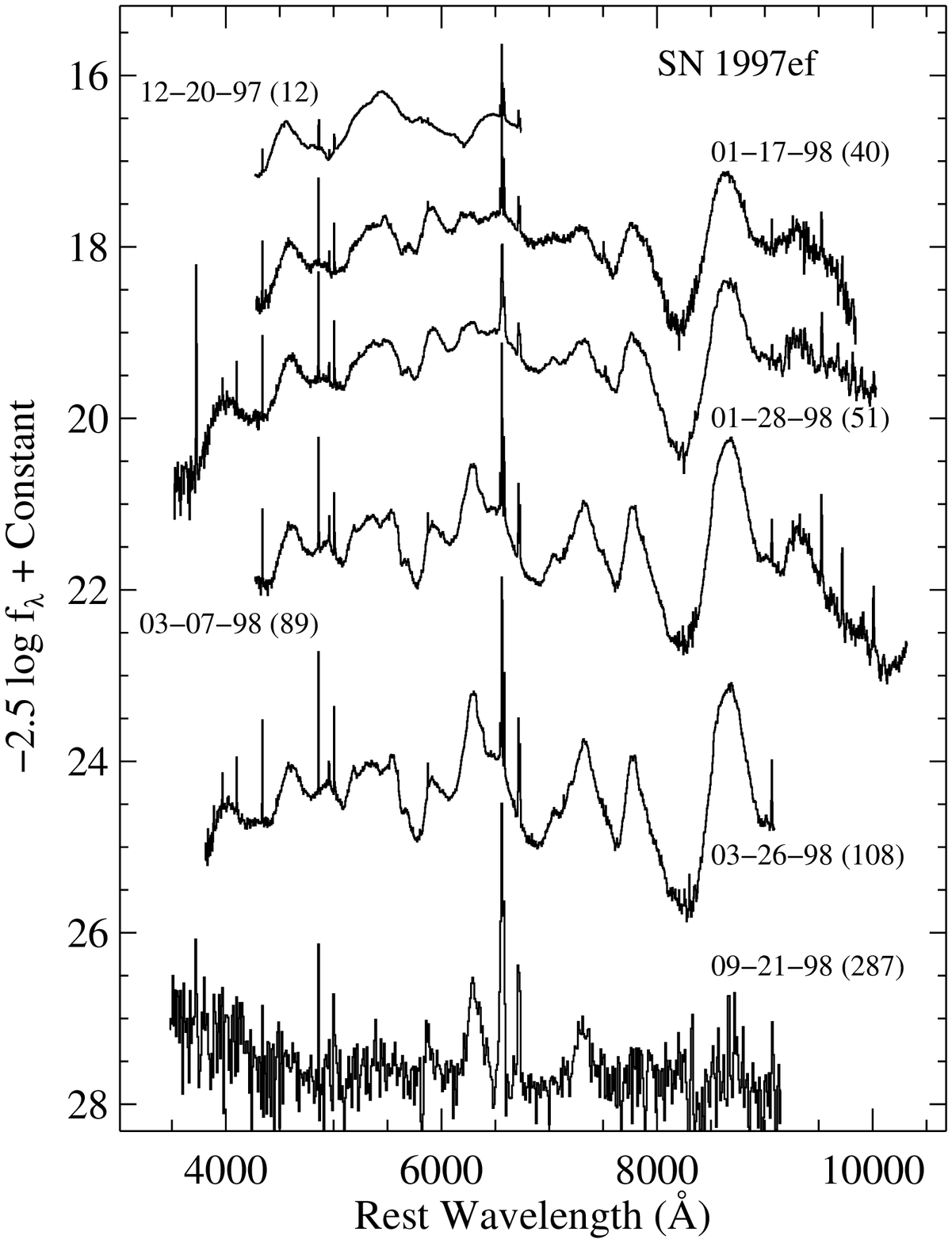}

\caption{Spectra of SN Ic 1997ef with flux units as in Figure
\ref{sn1988l-mont}.  The following constants have been added to the
individual spectra (from top to bottom): 0.0, $-$0.5, 1.5, 3.0, 6.0,
and 7.5.  The recession velocity of the SN has been removed as
described in the introduction to \S 3.  The numbers after the date of
observation indicate the approximate number of days past $V$-band
maximum brightness based upon the light curve of Iwamoto et
al. (2000).\label{sn1997ef-mont}}
\end{figure}
\clearpage
\begin{figure}[ht!]

        \plotone{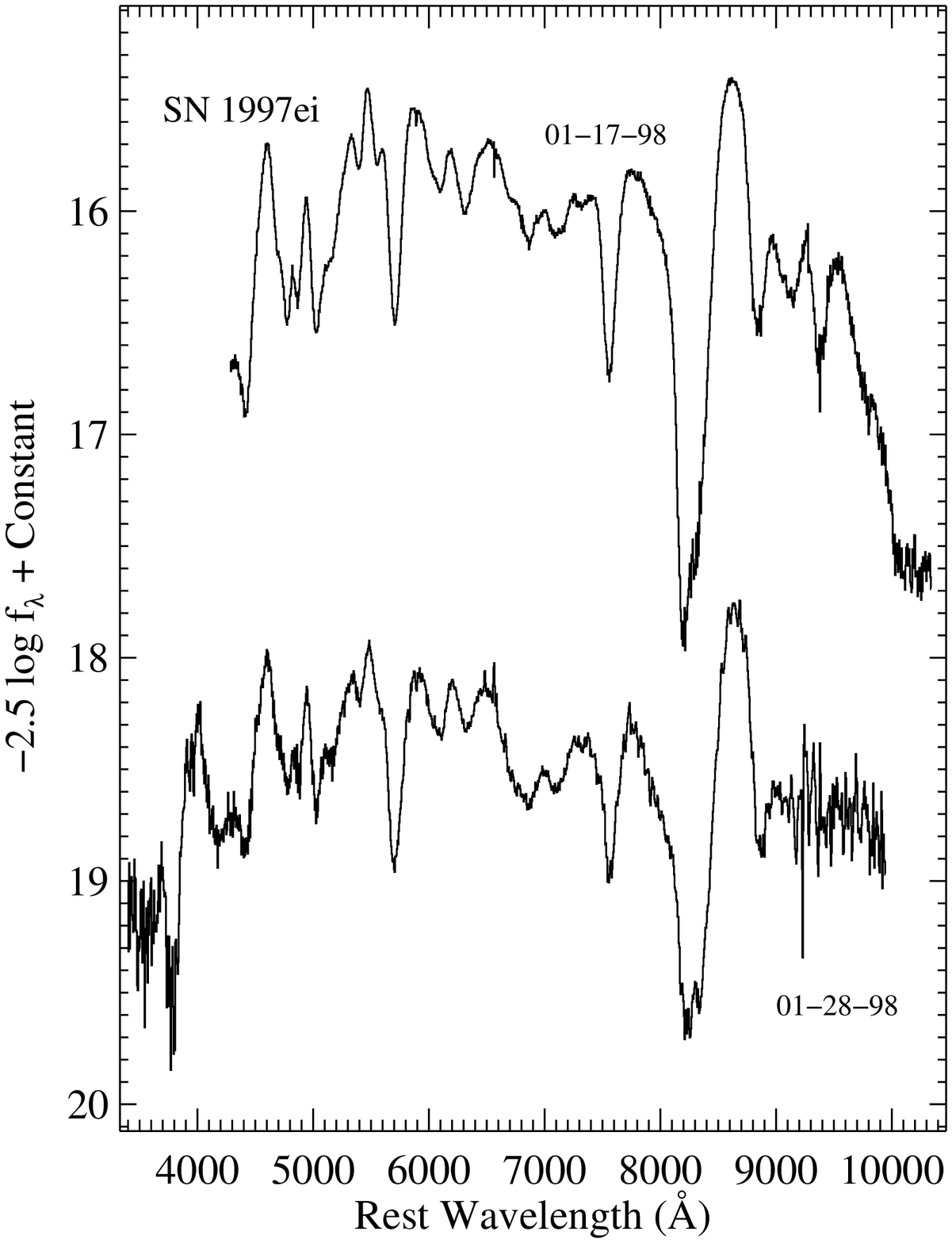}

\caption{Spectra of SN Ic 1997ei with flux units as in Figure
\ref{sn1988l-mont}.  The following constants have been added to the
individual spectra (from top to bottom): 0.0 and 0.5.  The recession
velocity of the SN has been removed as described in the introduction
to \S 3.\label{sn1997ei-mont}}
\end{figure}
\clearpage
\begin{figure}[ht!]

        \plotone{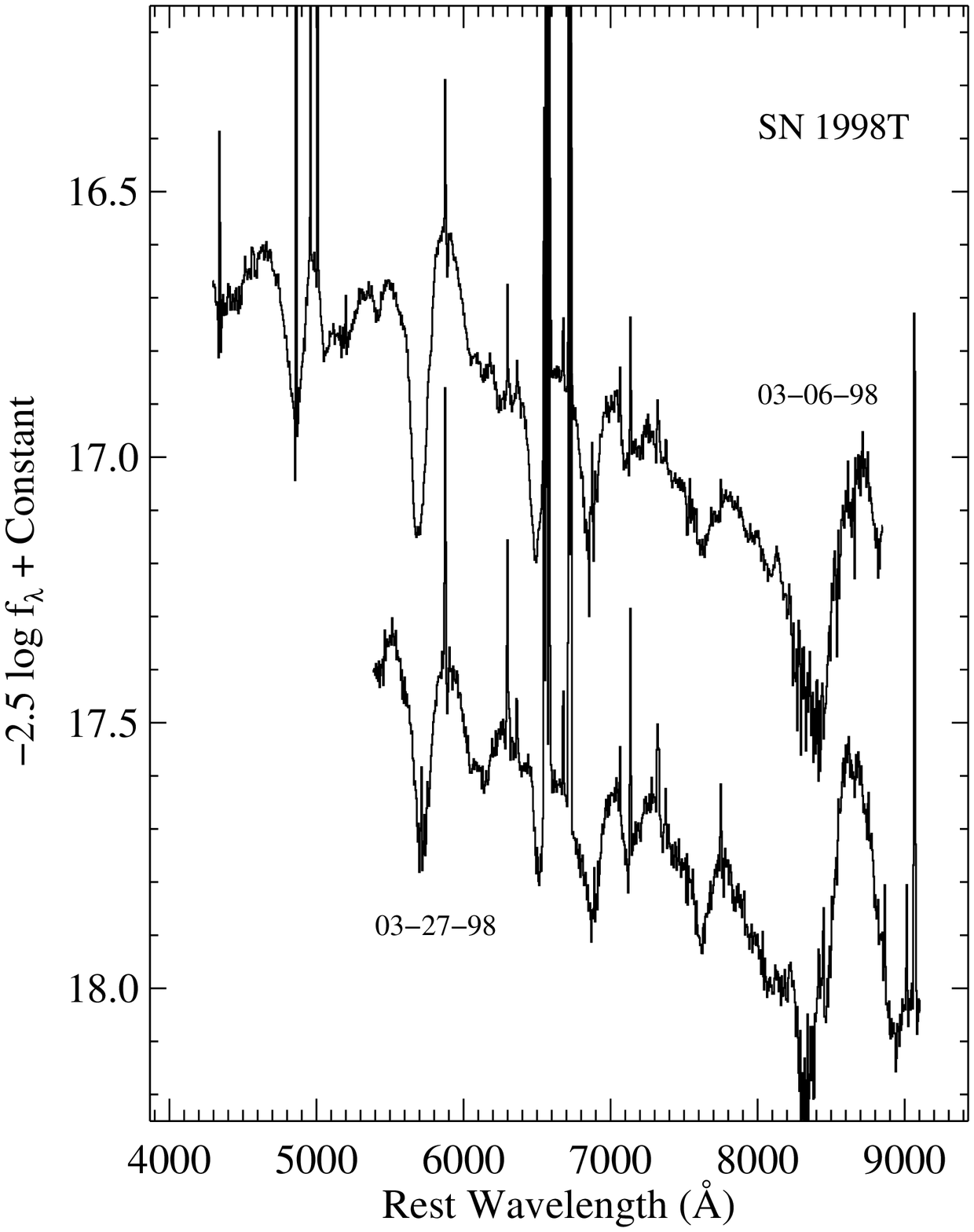}

\caption{Spectra of SN Ib 1998T with flux units as in Figure
\ref{sn1988l-mont}.  The following constants have been added to the
individual spectra (from top to bottom): 0.0 and 1.3.  The recession
velocity of the SN has been removed as described in the introduction
to \S 3.\label{sn1998t-mont}}
\end{figure}
\clearpage
\begin{figure}[ht!]

        \plotone{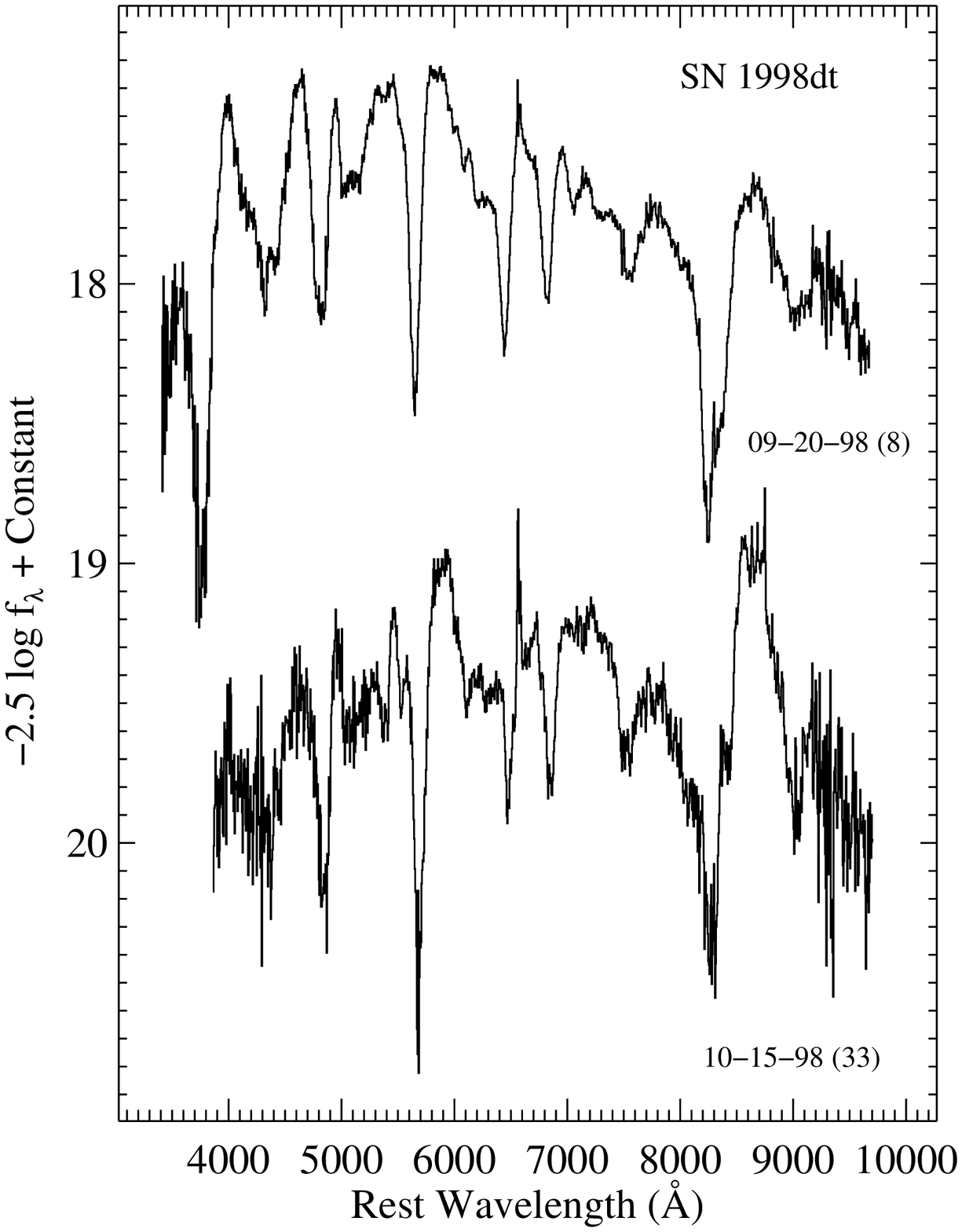}

\caption{Spectra of SN Ib 1998dt with flux units as in Figure
\ref{sn1988l-mont}.  The following constants have been added to the
individual spectra (from top to bottom): 0.0 and 1.0.  The recession
velocity of the SN has been removed as described in the introduction
to \S 3.  The numbers after the date of observation indicate the
approximate number of days past $R$-band maximum brightness based upon
the light curve shown in Figure
\ref{iblightcurve}.\label{sn1998dt-mont}}
\end{figure}
\clearpage
\begin{figure}[ht!]

        \plotone{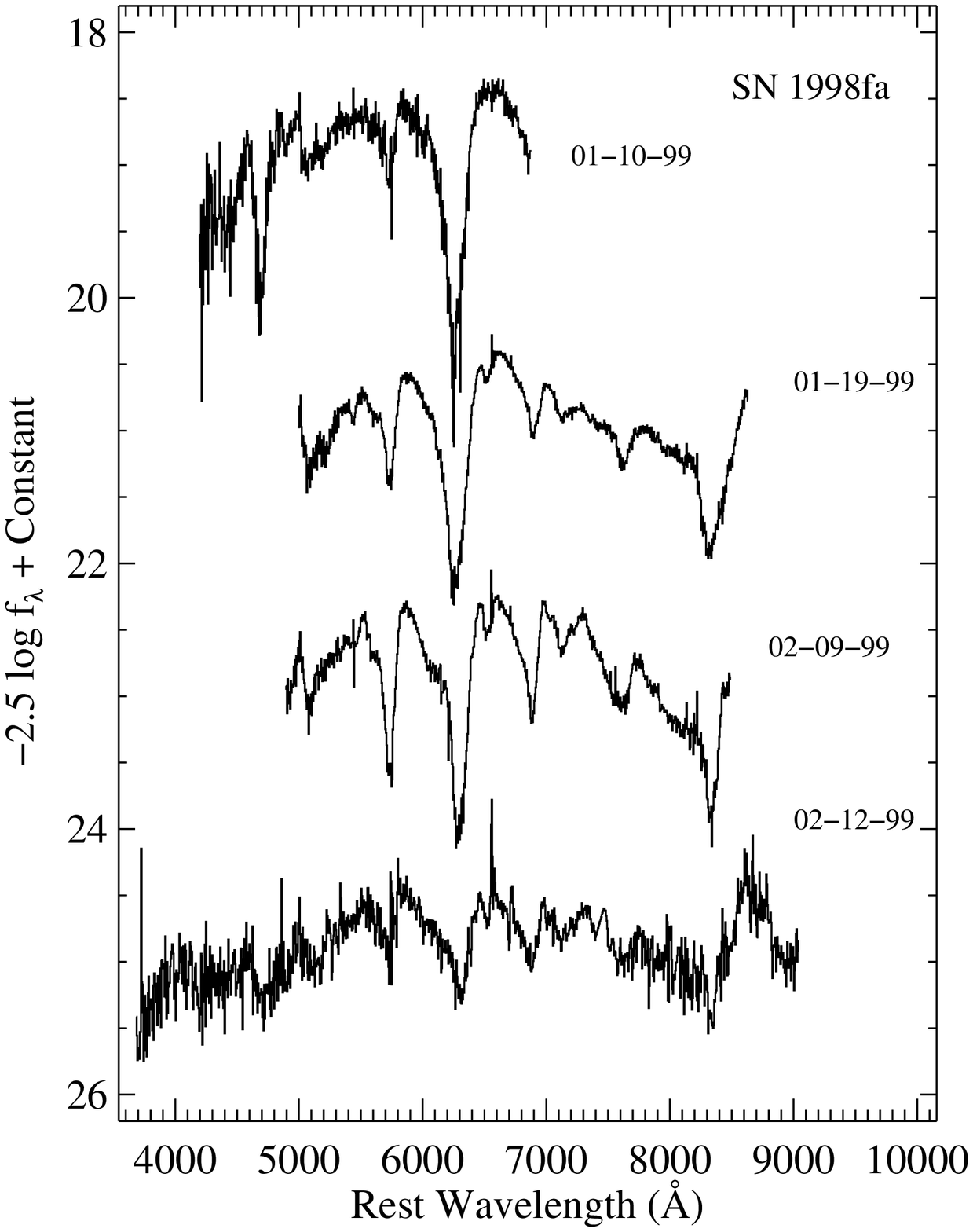}

\caption{Spectra of SN IIb 1998fa with flux units as in Figure
\ref{sn1988l-mont}.  The following constants have been added to the
individual spectra (from top to bottom): 0.0, 1.8, 3.0, and 6.0.  The
recession velocity of the SN has been removed as described in the
introduction to \S 3.\label{sn1998fa-mont}}
\end{figure}
\clearpage
\begin{figure}[ht!]
\rotatebox{180}{
        \plotone{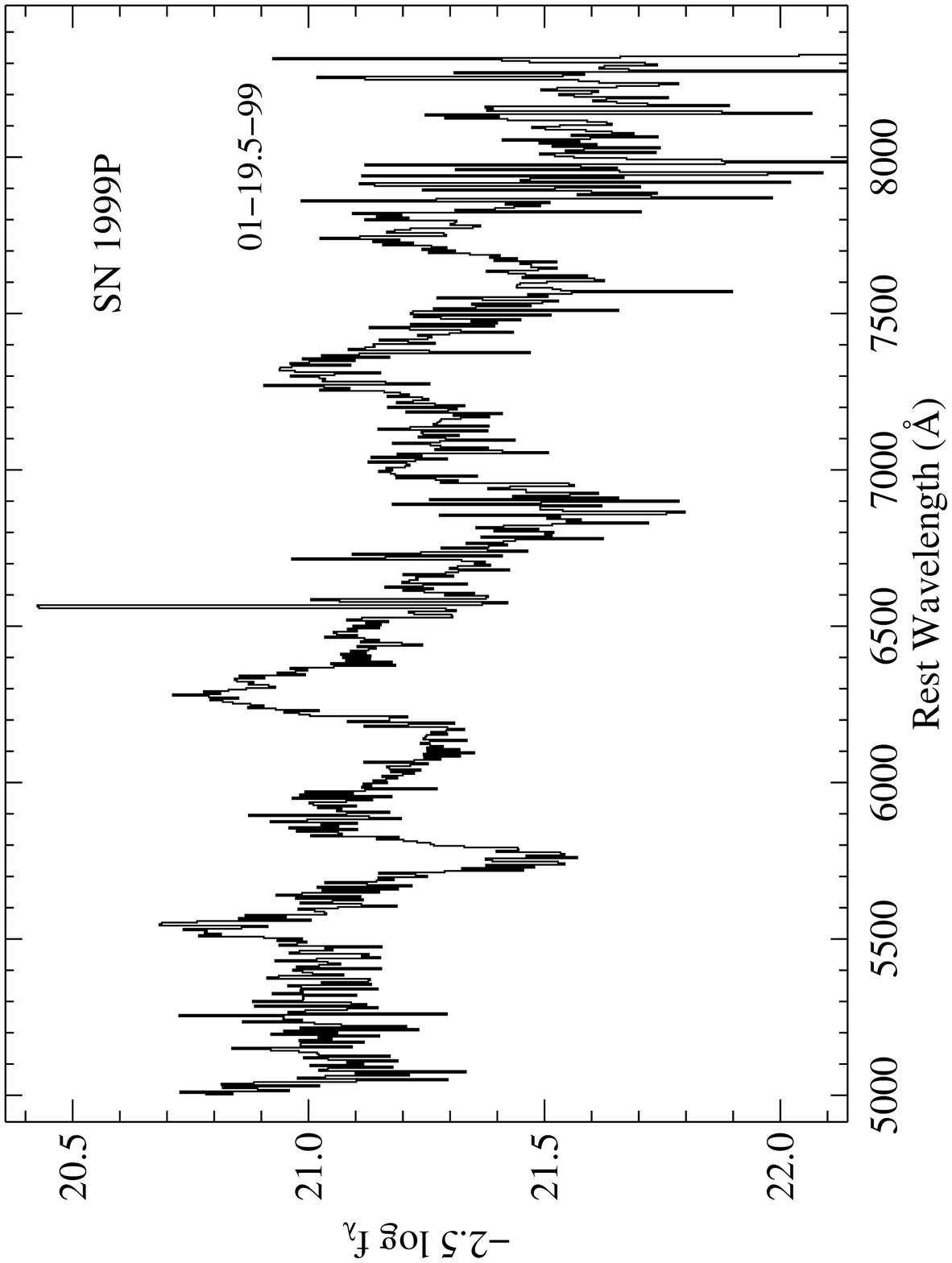}
 }

\caption{Spectrum of SN Ib/c 1999P with flux units as in Figure
\ref{sn1988l-mont}.  The recession velocity of the SN has been removed
as described in the introduction to \S 3.\label{sn1999p-mont}}
\end{figure}
\clearpage
\begin{figure}[ht!]

        \plotone{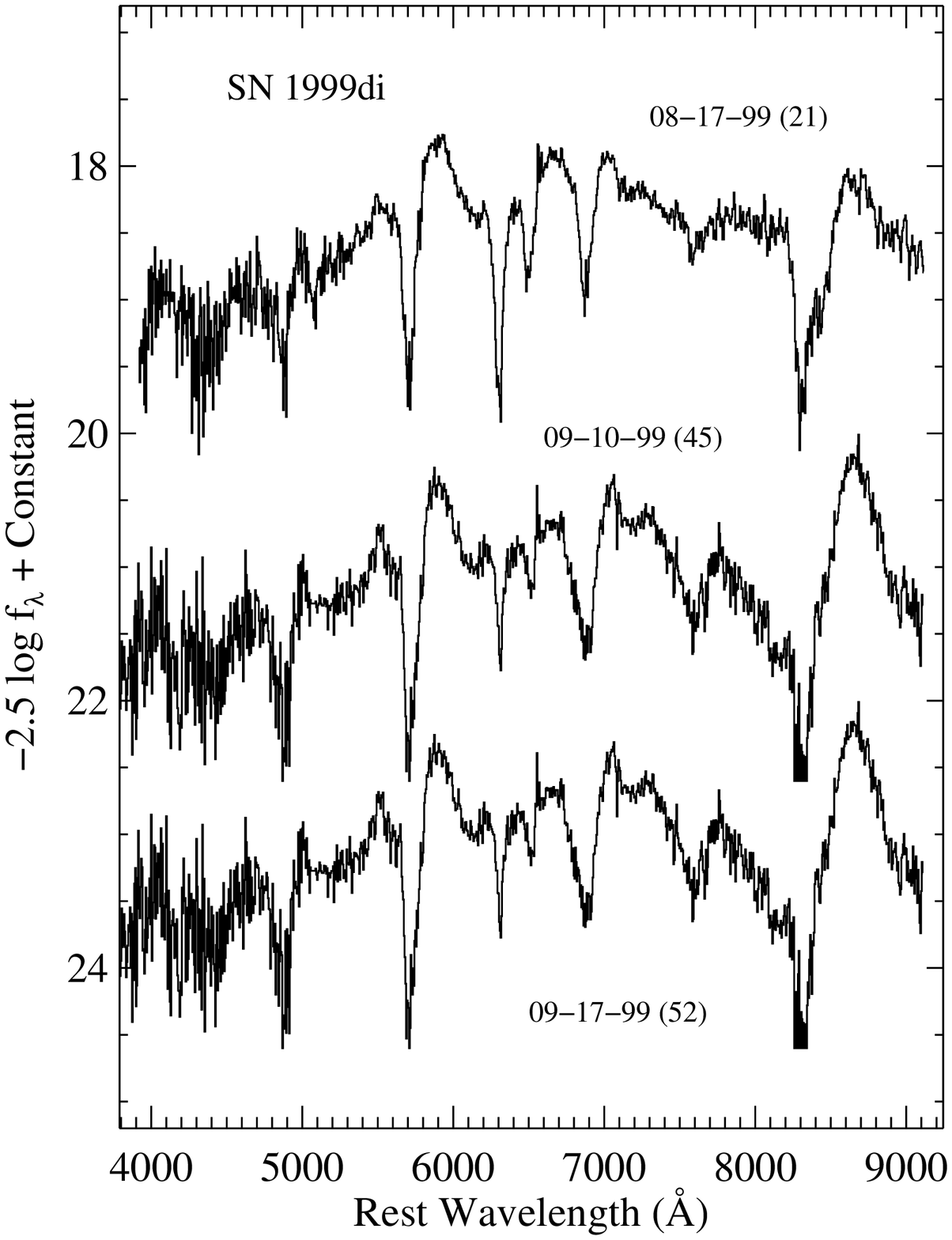}

\caption{Spectra of SN Ib 1999di with flux units as in Figure
\ref{sn1988l-mont}.  The following constants have been added to the
individual spectra (from top to bottom): 0.0, 1.5, and 3.5.  The
recession velocity of the SN has been removed as described in the
introduction to \S 3.  The numbers after the date of observation
indicate the approximate number of days past $R$-band maximum
brightness based upon the light curve shown in Figure
\ref{iblightcurve}.\label{sn1999di-mont}}
\end{figure}
\clearpage
\begin{figure}[ht!]

        \plotone{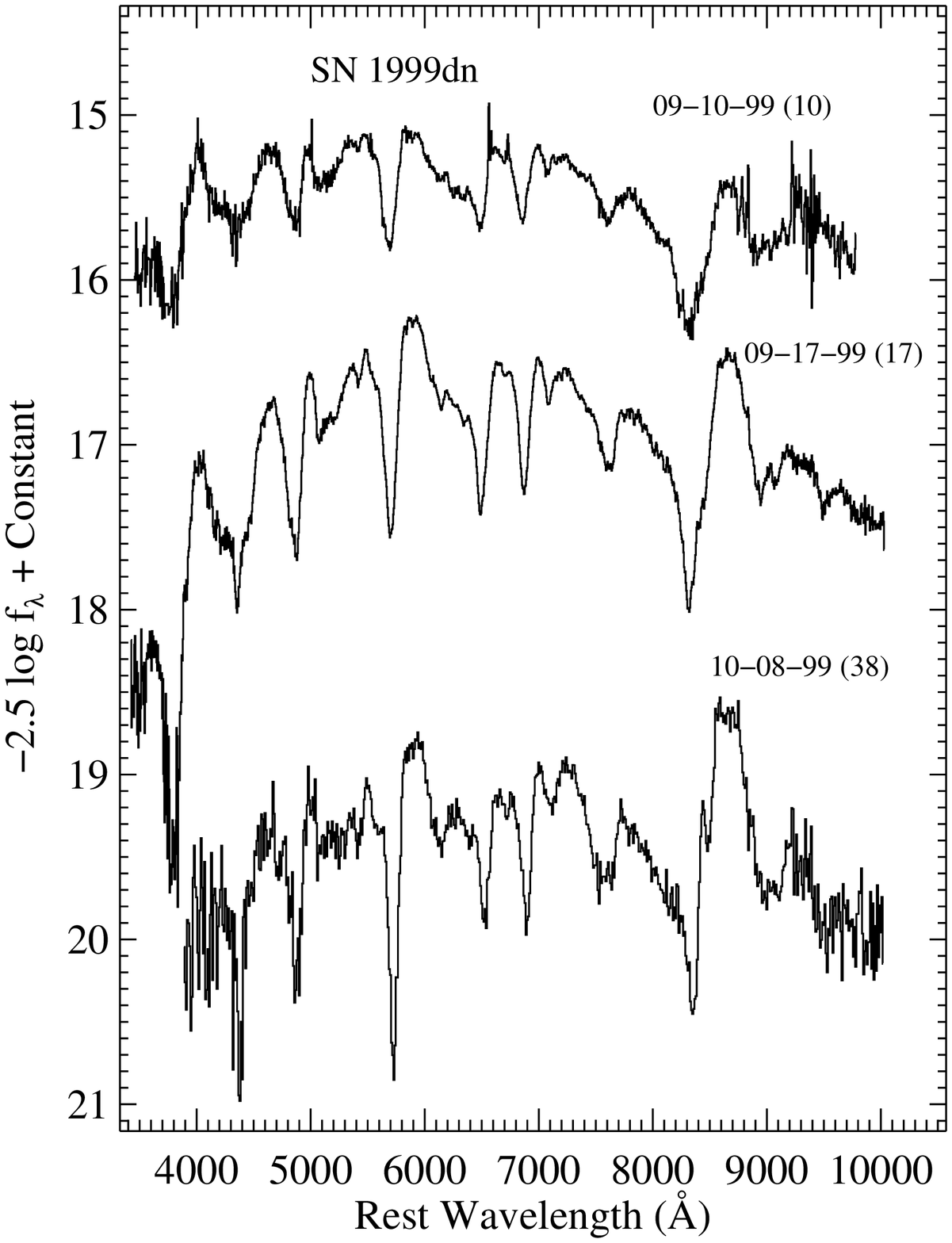}

\caption{Spectra of SN Ib 1999dn with flux units as in Figure
\ref{sn1988l-mont}.  The following constants have been added to the
individual spectra (from top to bottom): $-$2.0, 0.0, and 1.5.  The
recession velocity of the SN has been removed as described in the
introduction to \S 3.  The numbers after the date of observation
indicate the approximate number of days past $R$-band maximum
brightness based upon the light curve shown in Figure
\ref{iblightcurve}.\label{sn1999dn-mont}}
\end{figure}
\clearpage
\begin{figure}[ht!]
\rotatebox{180}{
        \plotone{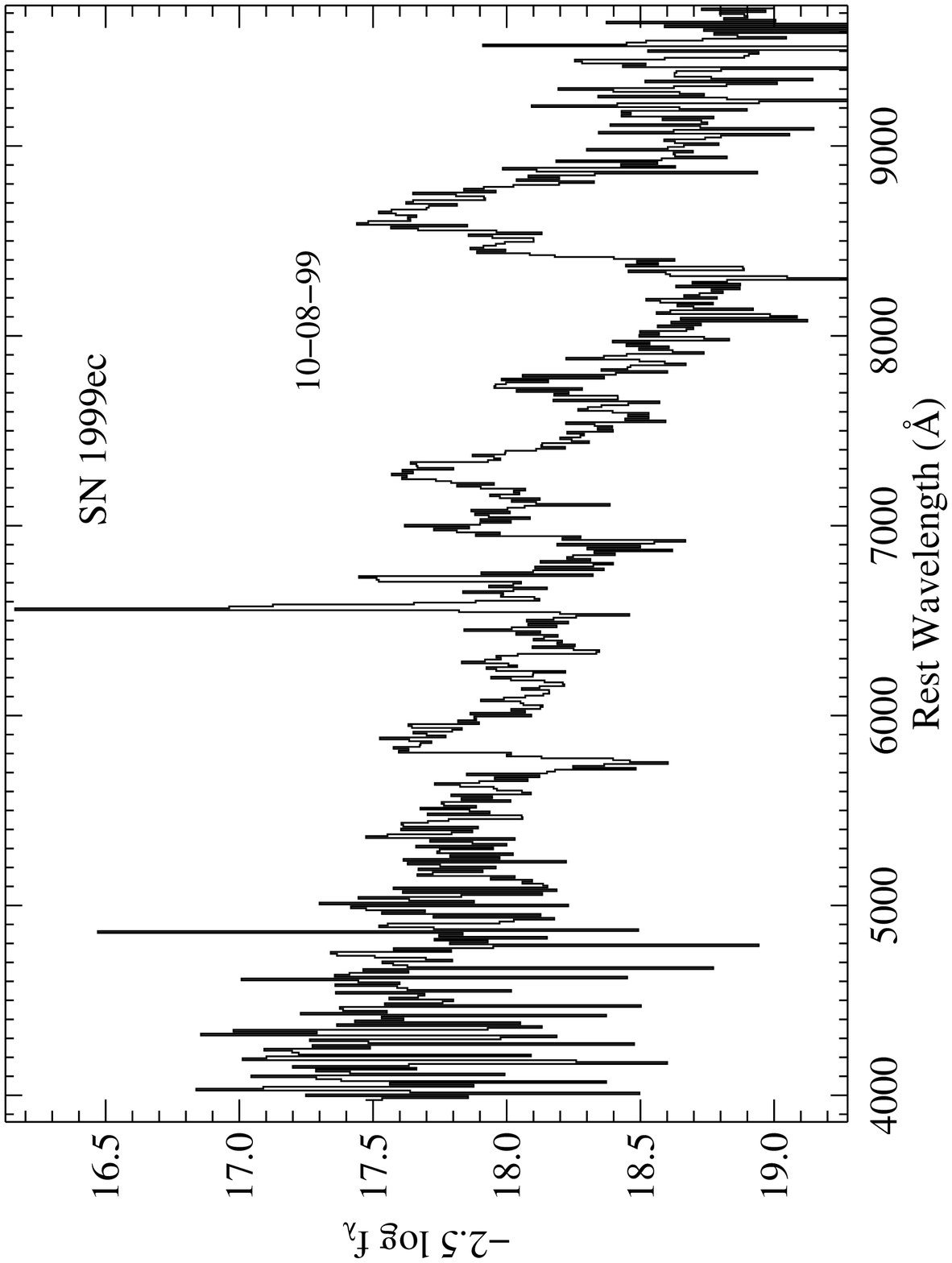}
}
\caption{Spectrum of SN Ib/c(?) 1999ec with flux units as in Figure
\ref{sn1988l-mont}.  The recession velocity of the SN has been removed
as described in the introduction to \S 3.\label{sn1999ec-mont}}
\end{figure}
\clearpage
\begin{figure}[ht!]
\rotatebox{180}{
\scalebox{0.8}{
        \plotone{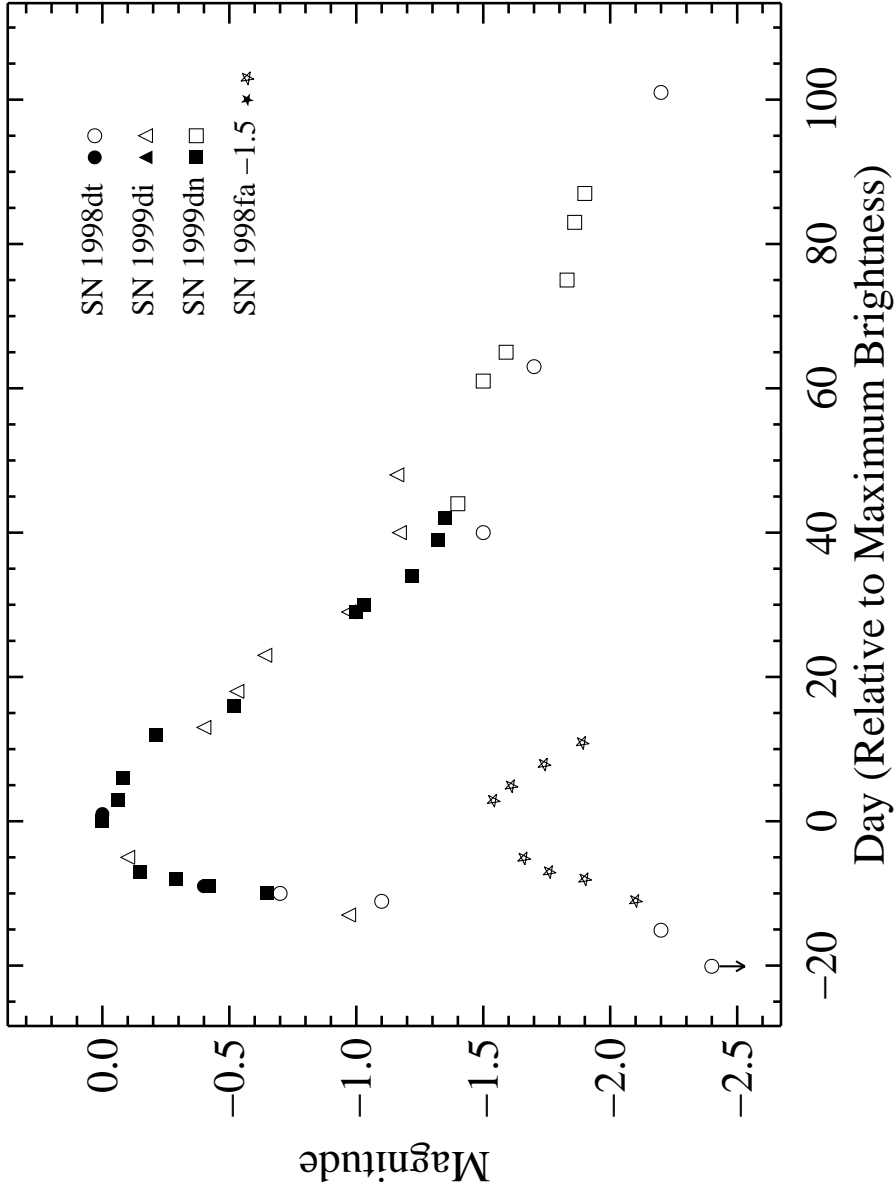}
}
}

\caption{KAIT $R$-band and unfiltered-magnitude light curves of the
SNe Ib 1998dt (\emph{circles}), 1999di (\emph{triangles}), and 1999dn
(\emph{squares}), along with the SN IIb 1998fa (\emph{stars}).  True
$R$-band values are indicated by solid symbols; unfiltered magnitudes
are shown as open symbols.  As described in \S 2, unfiltered
magnitudes for KAIT are similar to $R$-band values.  The points have
been shifted arbitrarily in magnitude to yield 0 at maximum (SN 1998fa
is offset by $-1.5$ mag).  The arrow on the first point of the SN
1998dt curve represents an upper limit.  The curves have been shifted
in time to match each other at $R$-band (or unfiltered) maximum.  The
dates of maxima are 1998 September 12, 1999 July 27, 1999 August 31,
and 1999 January 2 for SNe 1998dt, 1999di, 1999dn, and 1998fa,
respectively.\label{iblightcurve}}
\end{figure}
\clearpage
\begin{figure}[ht!]
%\rotatebox{180}{
\scalebox{0.9}{
        \plotone{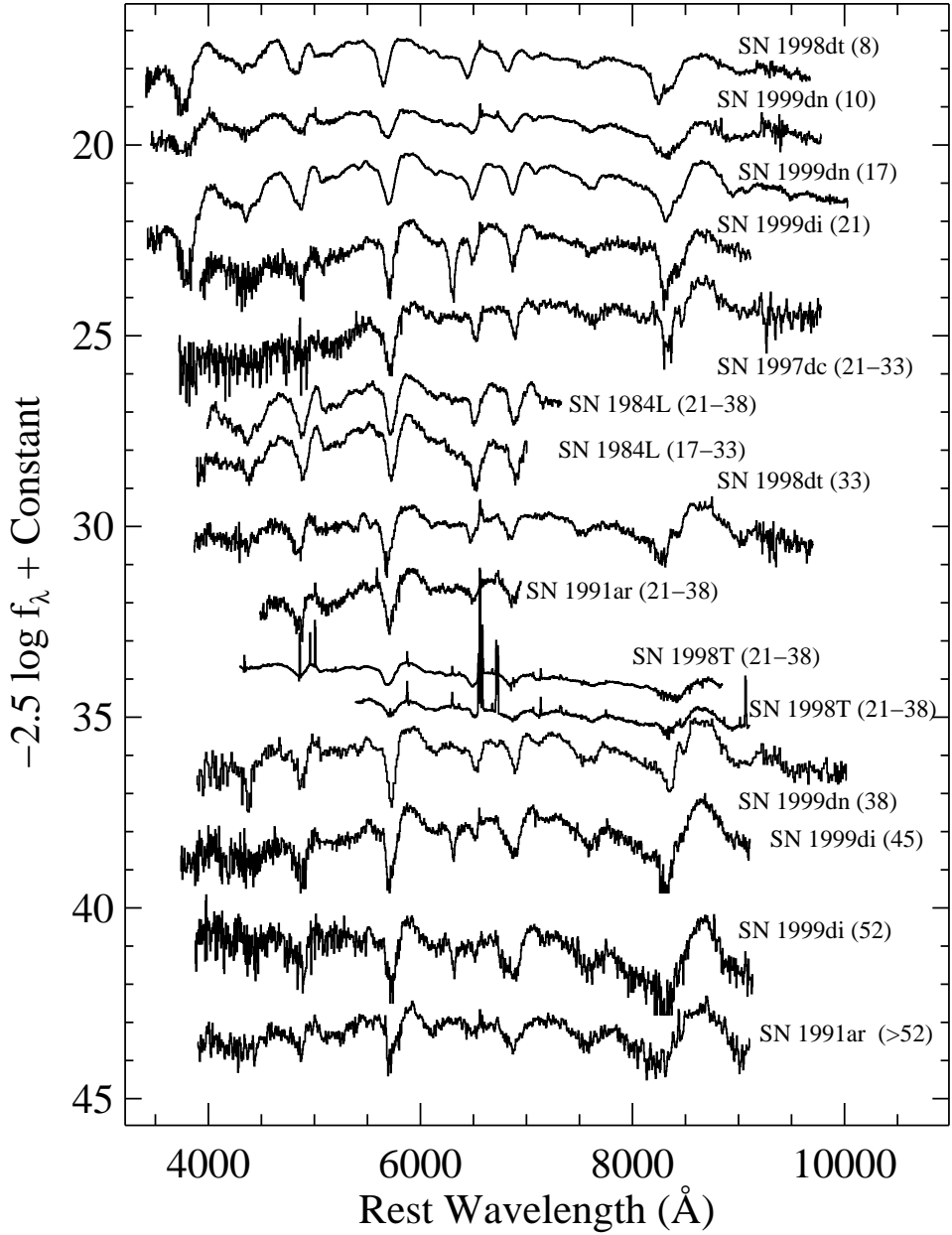}
%}
}
\caption{Spectra of the SNe Ib for which a phase is known from a
light curve or has been estimated from the helium-line depths, in
approximate temporal order.  Flux units as in Figure
\ref{sn1988l-mont}.  The recession velocities of the SNe have been
removed as described in the introduction to \S 3.  The number after
each SN name indicates the probable number of days after maximum
$R$-band brightness for each spectrum.\label{ibmontage}}
\end{figure}
\clearpage

\begin{figure}[ht!]
\rotatebox{180}{
\scalebox{0.9}{
        \plotone{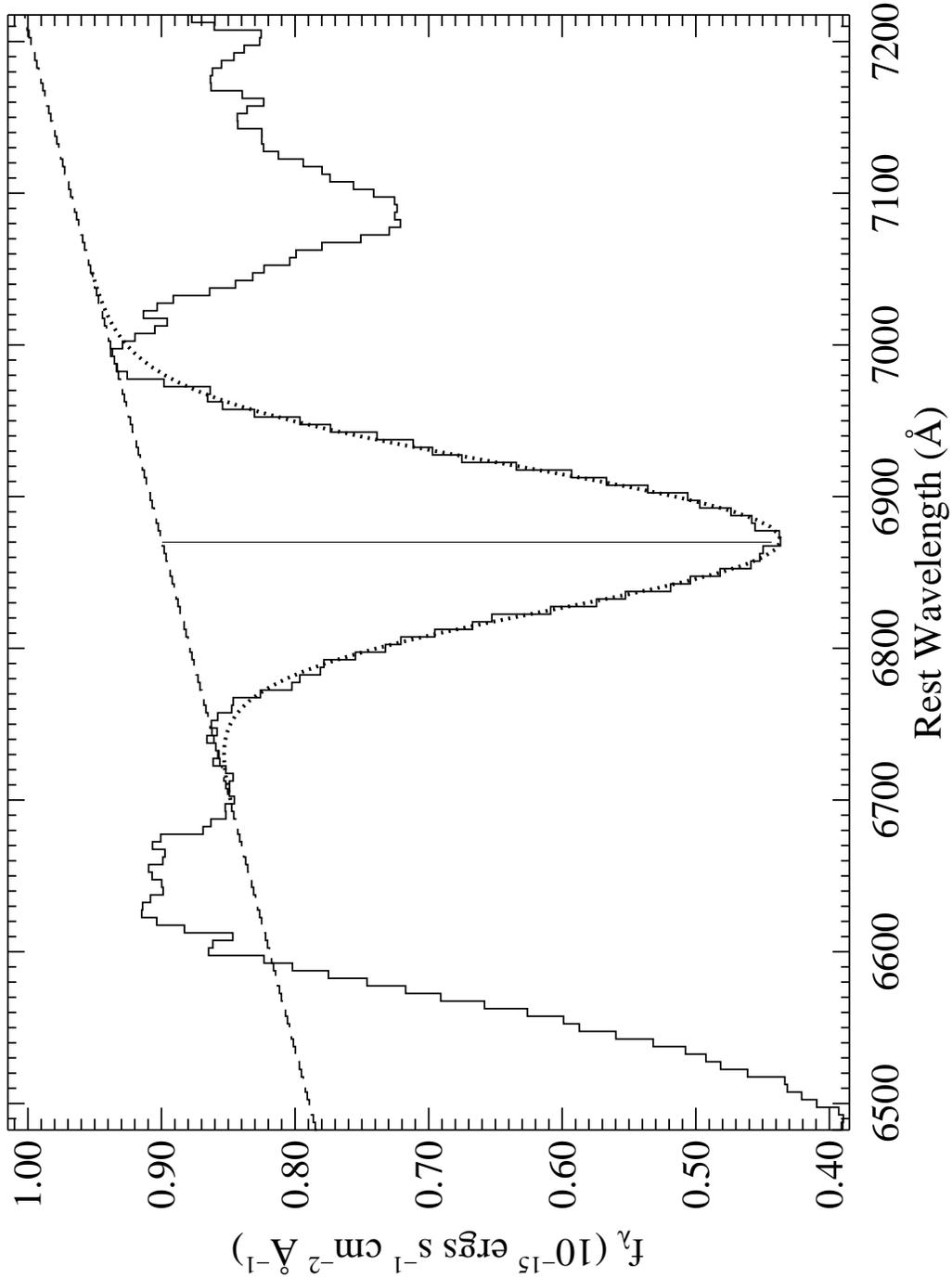}
}
}
\caption{Illustration of
the line-measurement technique.  Shown is the region of the 1999
September 17 spectrum of SN Ib 1999dn near the absorption caused by
\protect\ion{He}{1} $\lambda$7065.  The continuum determined from the
median value of the regions adjacent to the edges of the line is
represented by the dashed line.  The dotted line shows the Gaussian
fit to the absorption.  The vertical line in the middle of the
absorption marks the wavelength of the measured minimum of the line
and delimits the two flux values (continuum at minimum, spectrum flux
at minimum) that are used to calculate the fractional line depth as
described in the text.\label{depthfig}}
\end{figure}
\clearpage
\begin{figure}[ht!]
\rotatebox{180}{
\scalebox{0.8}{
        \plotone{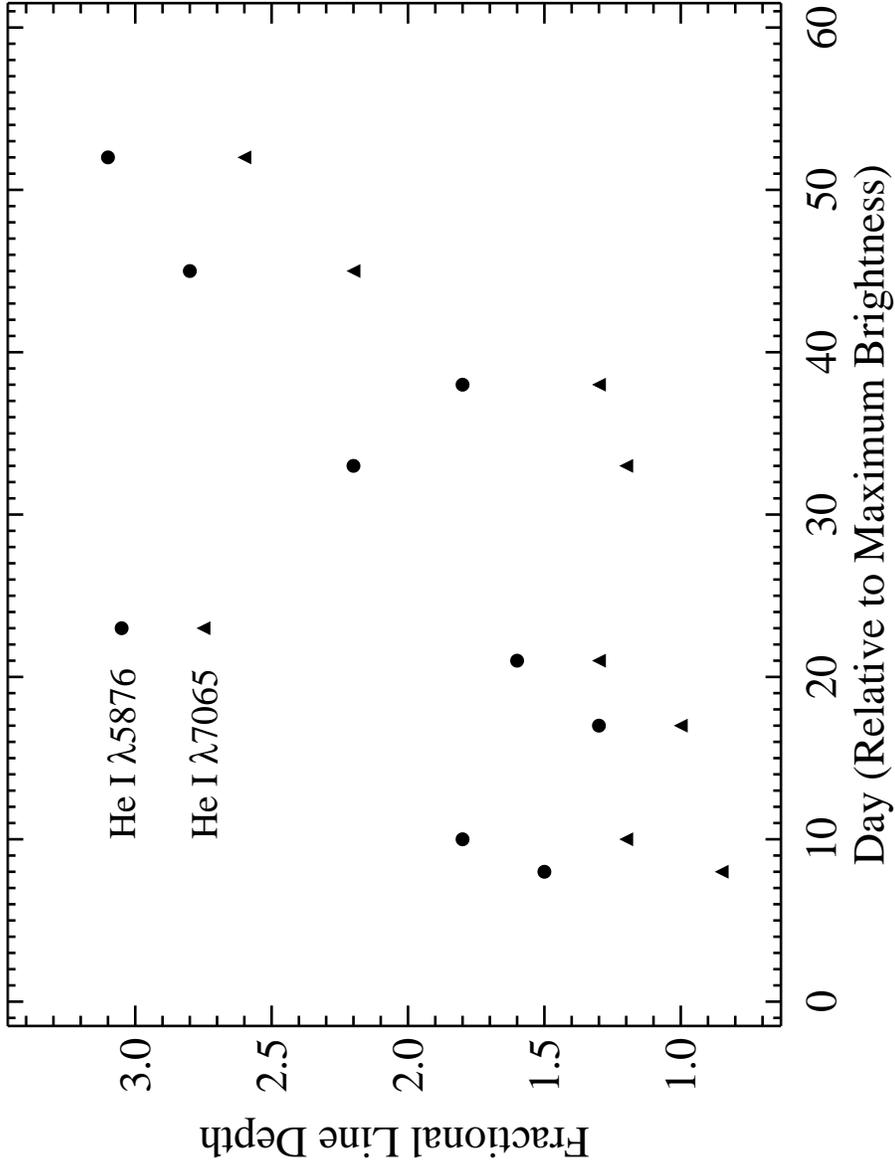}
}
}
\caption{Temporal evolution of fractional line depths of
\protect\ion{He}{1} $\lambda$5876 (\emph{circles}) and $\lambda$7065
(\emph{triangles}) normalized to the fractional line depth of
$\lambda$6678 (i.e., the fractional line depth of \protect\ion{He}{1}
$\lambda$6678 is set to unity) for SNe Ib 1998dt, 1999di, and 1999dn
(cf. Tables 2 and 3).  The number of days past maximum is determined
from the light curves in Figure \ref{iblightcurve}.  While there is
some scatter in the points, a trend is apparent.\label{linedepthrat}}
\end{figure}
\clearpage
\begin{figure}[ht!]
\scalebox{0.9}{
        \plotone{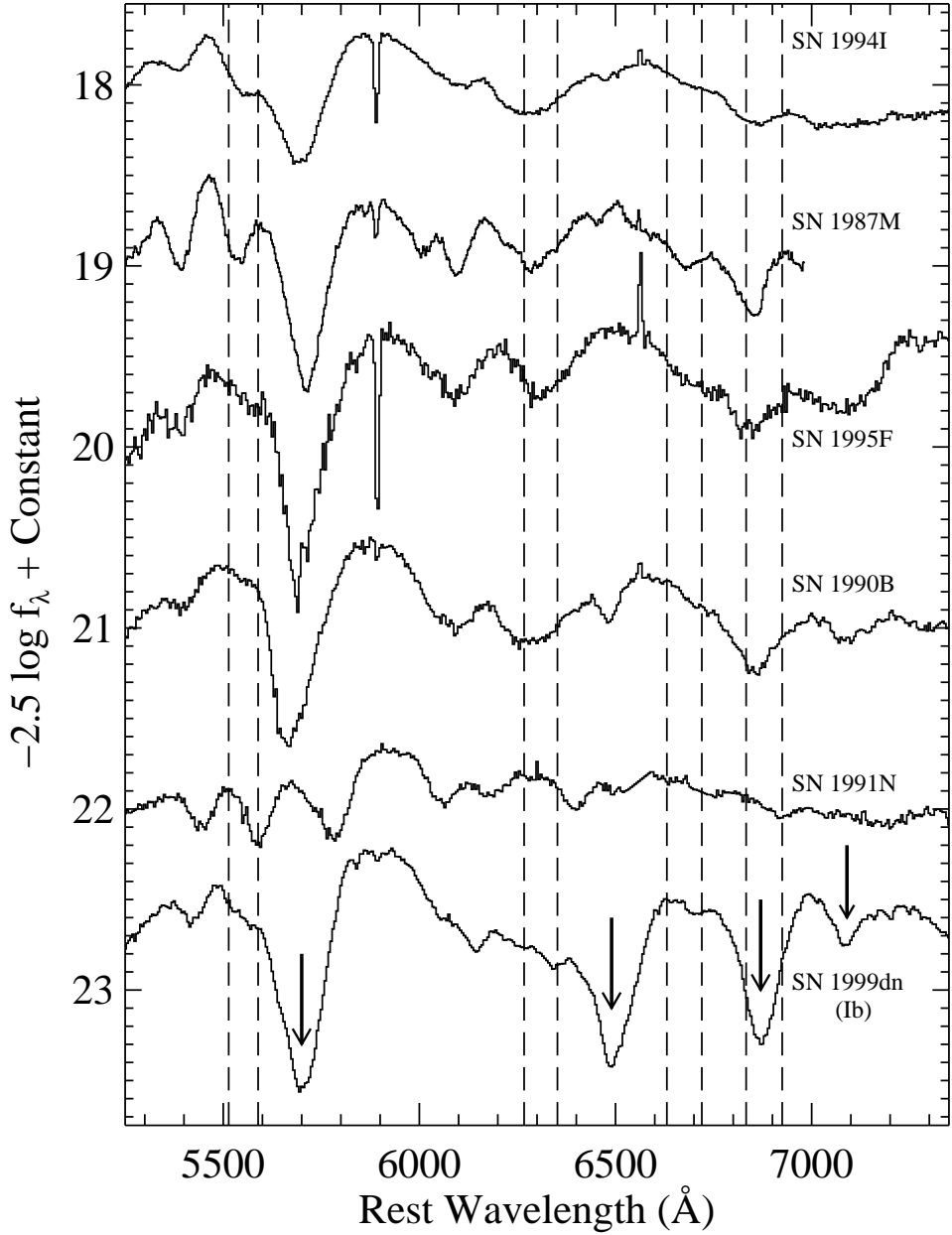}
}

\caption{Spectra of various SNe Ic and one SN Ib to illustrate the
potential presence of high-velocity helium absorption lines in SNe Ic.
The recession velocities of the SNe have been removed as described in
the introduction to \S 3.  (The SN Ib has strong helium absorption
lines at low expansion velocities of $-8000$ to $-9000$ km~s$^{-1}$,
marked with arrows.)  The dashed lines delimit the expected positions
of the main optical \ion{He}{1} transitions at high expansion
velocities in the range $-15000$ to $-19000$ km~s$^{-1}$.  Note that
while some features do line up, especially in SN Ic 1987M (and
possibly SN Ic 1994I), the overall pattern is inconsistent with the
identification of the features in most SN Ic spectra with helium
lines.  In fact, SN Ic 1990B appears to show weak helium lines at
\emph{low} velocity that match up with those in the SN Ib 1999dn,
albeit at unusual relative line-depth ratios.\label{ichelines}}
\end{figure}
\clearpage
\begin{figure}[ht!]
\rotatebox{180}{
\scalebox{0.8}{
        \plotone{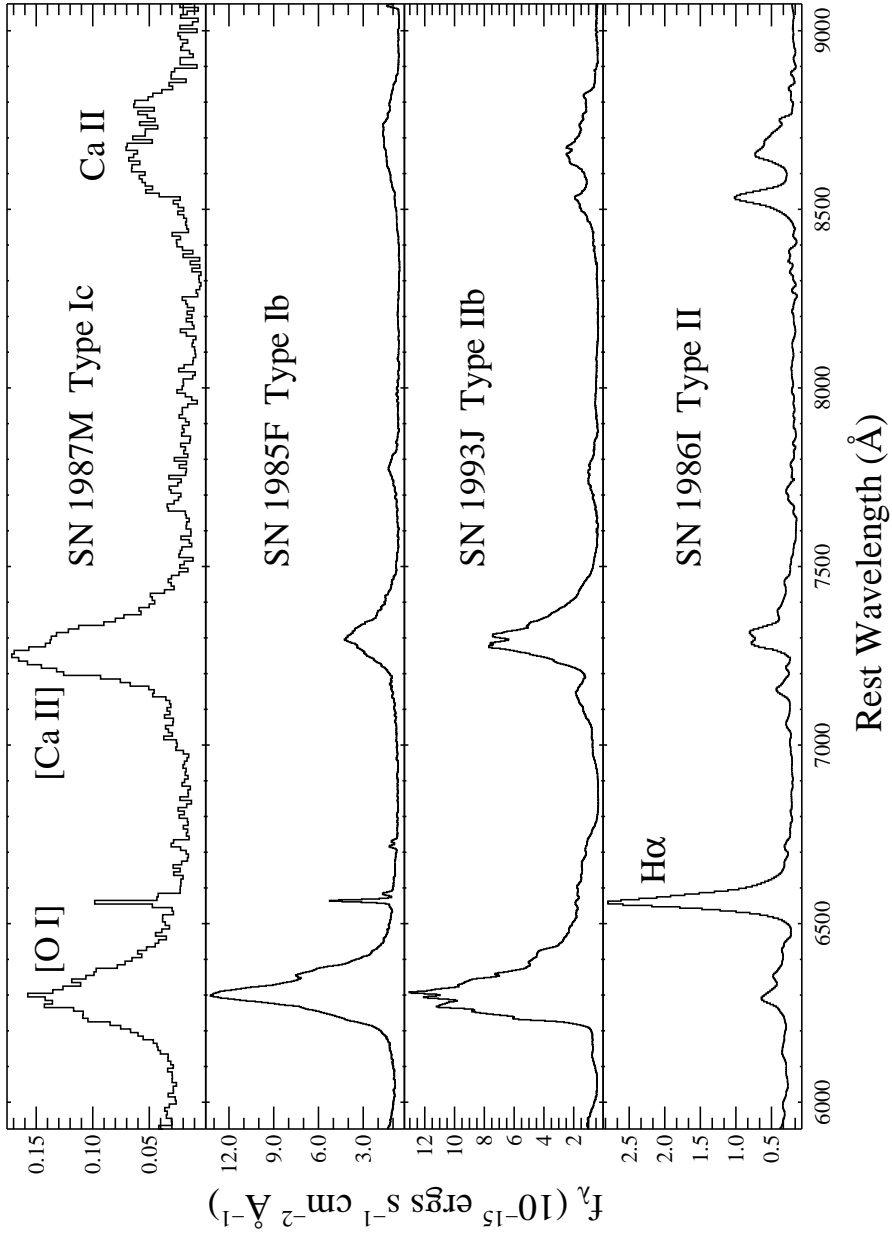}
}
}
\caption{Representative nebular-phase spectra of each of the
core-collapse SN Types (1986I, Type II, t $\approx$ 8 months; 1993J,
Type IIb, t $\approx$ 8 months; 1985F, Type Ib, t $\approx$ 9 months;
1987M, Type Ic, t $\approx$ 5 months).  Note the increase in line
width from Types II through IIb, Ib, and finally, Ic.  If the explosion
energies are comparable, then decreasing envelope mass could explain
increasing velocity of expansion.\label{linewidthfig}}
\end{figure}
\clearpage
\begin{figure}[ht!]
\rotatebox{180}{
\scalebox{0.8}{
        \plotone{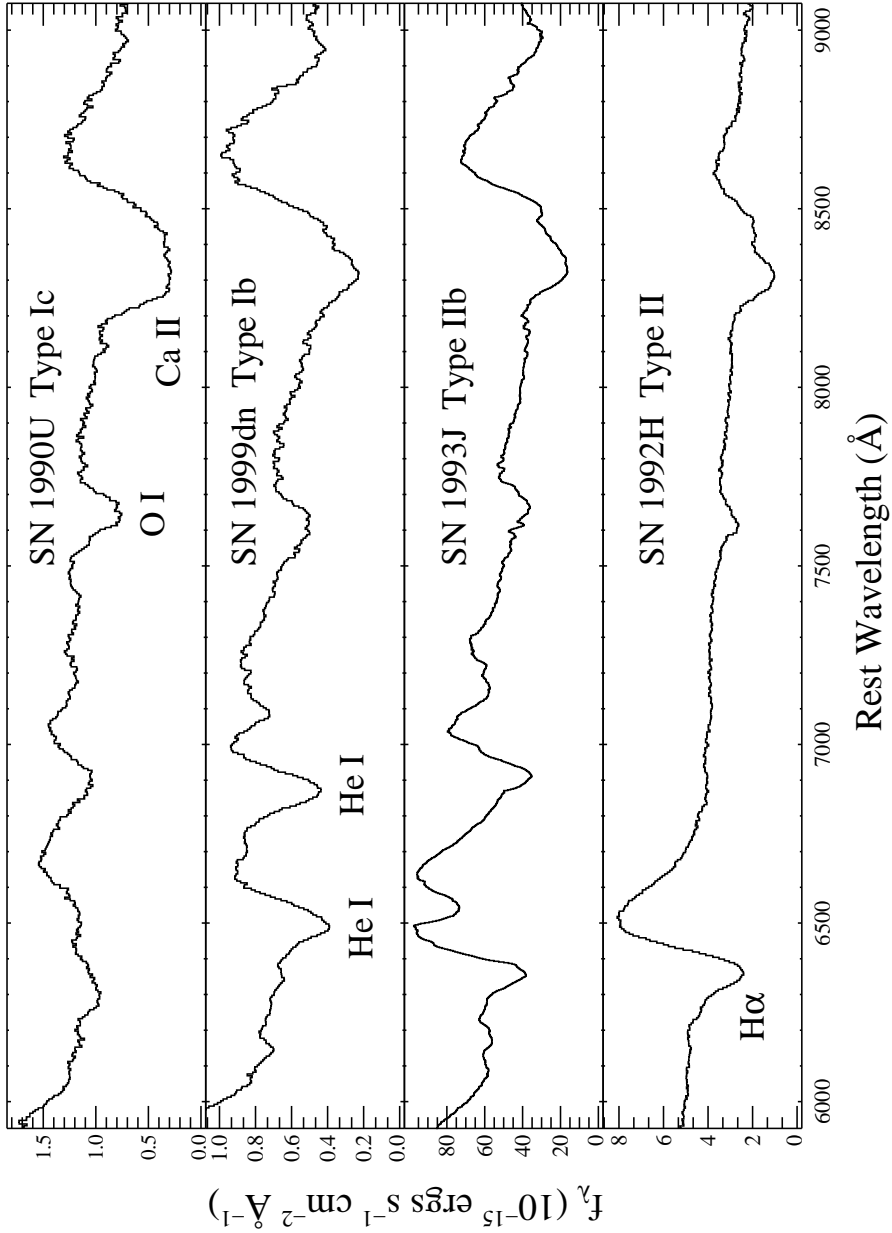}
}
}
\caption{Representative spectra of each of the core-collapse SN
Types from approximately $2-3$ weeks past maximum brightness (1992H,
Type II; 1993J, Type IIb; 1999dn, Type Ib; 1990U, Type Ic).  Narrow
lines from a superposed \protect\ion{H}{2} region have been removed
from the spectrum of SN 1990U for clarity.  There is a gradual
increase in the relative strength of the \protect\ion{O}{1}
$\lambda$7774 line from Types II through IIb, Ib, and finally, Ic.  If
a larger-mass envelope (of hydrogen and/or helium) dilutes the
strength of the oxygen line, then this increase in strength could
indicate a decreasing envelope mass from Types II to
Ic.\label{oidepth}}
\end{figure}
\clearpage
\begin{figure}[ht!]

        \plotone{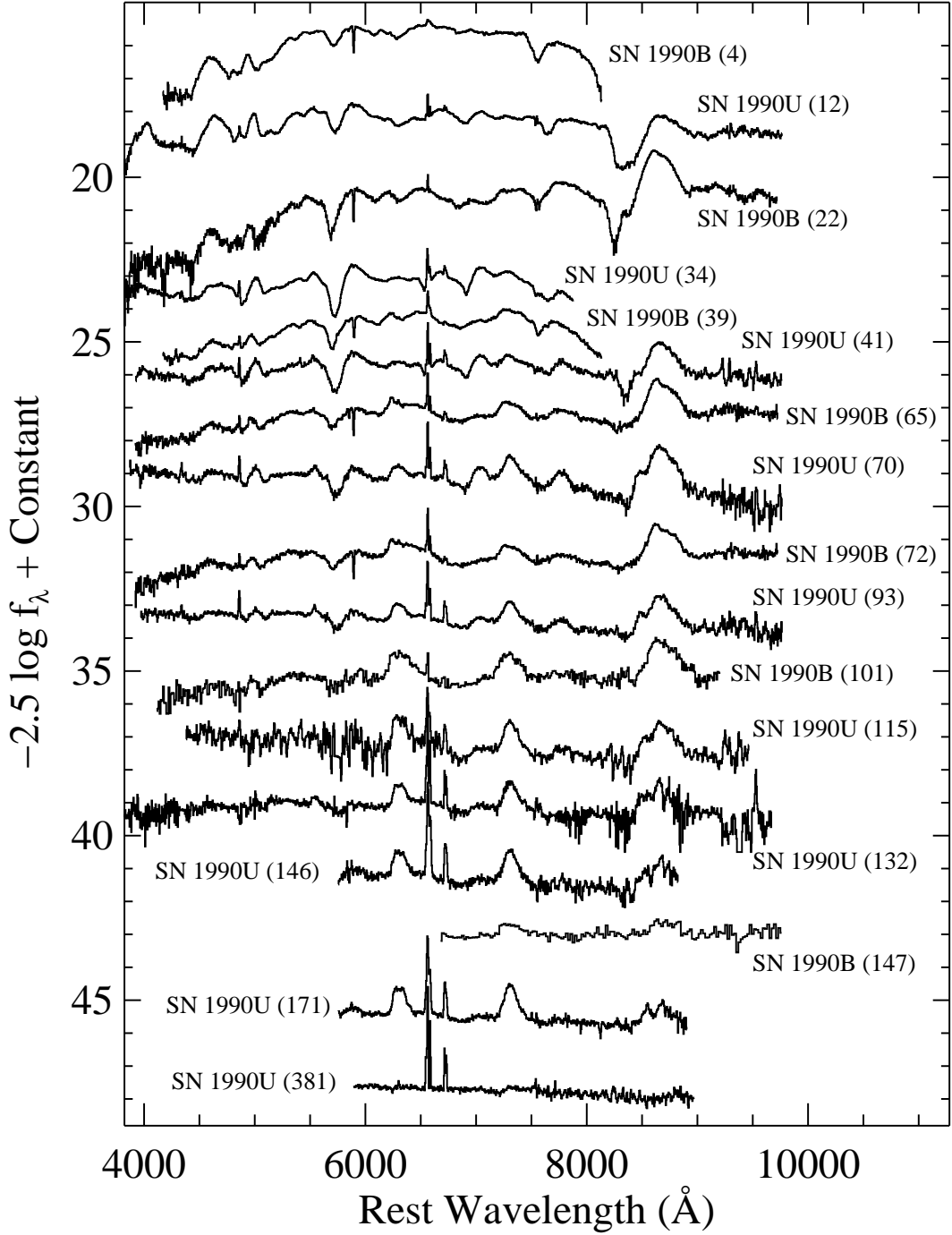}

\caption{Spectra of the SNe Ic 1990B and 1990 U in temporal order
based upon age relative to an estimated maximum of the $R$-band light
curve.  Flux units as in Figure \ref{sn1988l-mont}.  The recession
velocities of the SNe have been removed as described in the
introduction to \S 3.  The number after each SN name indicates the
probable number of days after $R$ maximum for each
spectrum.\label{icmontage}}
\end{figure}
\clearpage
\begin{figure}[ht!]

        \plotone{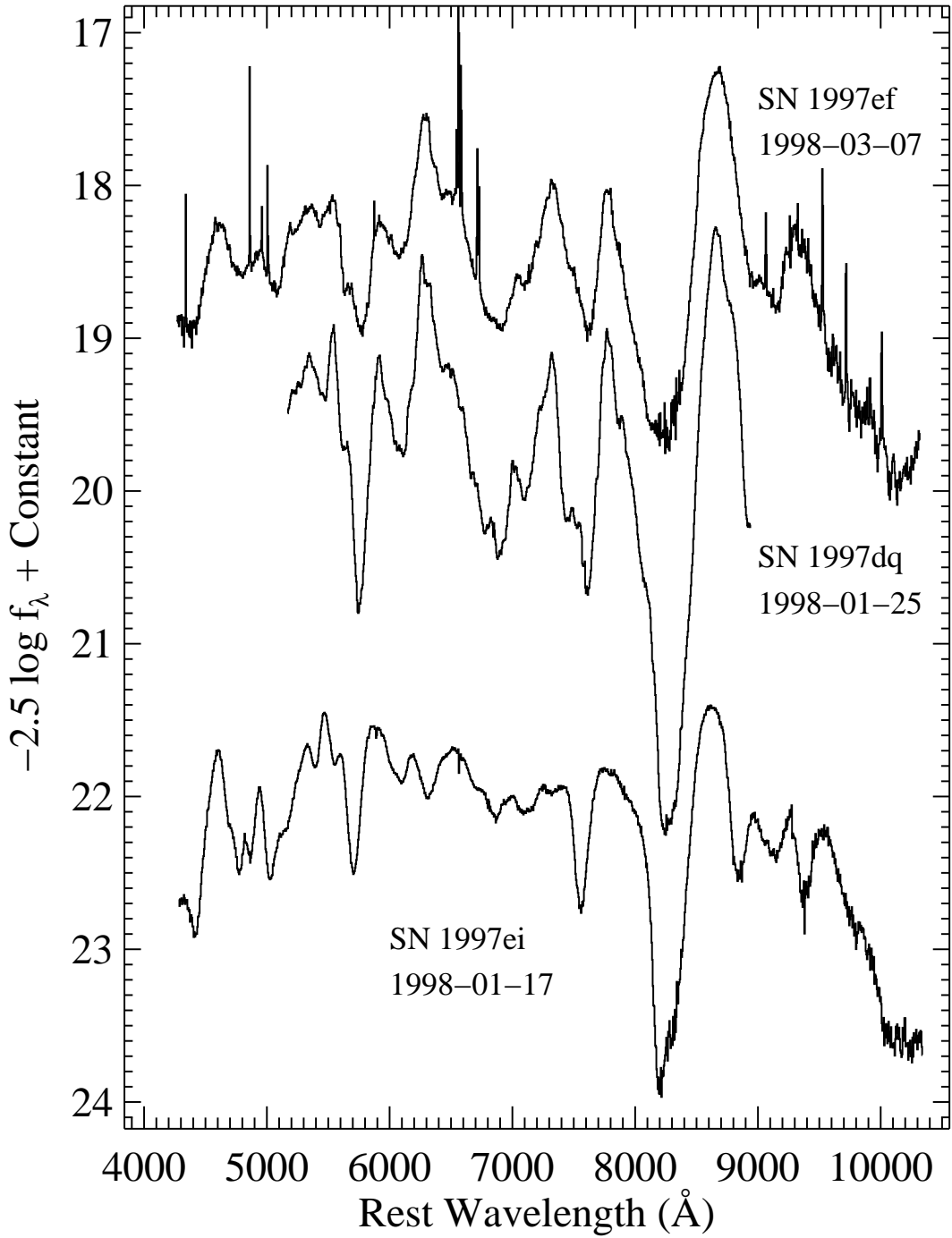}

\caption{Spectra of SNe Ic 1997dq, 1997ef, and 1997ei.  Flux units
as in Figure \ref{sn1988l-mont}.  The recession velocities of the SNe
have been removed as described in the introduction to \S 3.  The
spectrum of SN 1997ef is at approximately day 89 past maximum light,
based on the $V$-band light curve of Iwamoto et al. (2000).  All three
of these SNe may have been associated in time and location with a
gamma-ray burst.  SNe 1997dq and 1997ef are strikingly similar.  SN
1997ei more closely resembles a normal SN Ic.\label{dq-ef-comp}}
\end{figure}
\clearpage

%% The following command ends your manuscript. LaTeX will ignore any text

\end{document}